\documentclass[aps,reprint,superscriptaddress,footinbib,longbibliography]{revtex4-2}
\usepackage{graphicx}
\usepackage{dcolumn}
\usepackage{bm}
\usepackage{hyperref}
\hypersetup{colorlinks,
	linkcolor={blue!75!black!80!yellow},
	citecolor={blue!75!black!80!yellow},
	urlcolor={blue!75!black!80!yellow}
}

\usepackage{comment}
\usepackage{physics}
\usepackage{xcolor}
\usepackage{amssymb}
\usepackage{braket}
\usepackage[english]{babel}
\newcommand{\ii}{{\mathrm{i}}}
\newcommand{\rv}{{\bf r}}
\newcommand{\kv}{{\bf k}}
\newcommand{\bv}{{\bf b}}
\newcommand{\Rv}{{\bf R}}
\newcommand{\tR}{{\tilde{R}}}
\newcommand{\tX}{{\tilde{X}}}
\newcommand{\tY}{{\tilde{Y}}}
\newcommand{\tZ}{{\tilde{Z}}}

\newcommand{\hs}{{\mathcal{H}}}

\newcommand{\rextract}{{\rv^*}}
\newcommand*{\texpval}[1]{{\langle #1 \rangle}}
\newcommand{\transvec}{{\mathbf{R}_\ell}}
\newcommand{\psch}{p}
\newcommand{\esch}{\xi}
\newcommand{\halfproj}{V_{\mathrm{occ}}}
\newtheorem{assumption}{Assumption}
\newcommand{\satresidual}{\mathcal{S}}

\usepackage{array} 
\newcolumntype{C}[1]{>{\centering\arraybackslash}p{#1}} 

\begin{document}

\preprint{APS/123-QED}

\title{A Spatial Localizer for Electrons in Insulators}

\author{Haylen Gerhard} 
\affiliation{Department of Physics, Emory University, Atlanta, Georgia, USA}%
\author{Yifan Wang}
\affiliation{Department of Physics, Emory University, Atlanta, Georgia, USA}%
\author{Alexander Cerjan}%
\affiliation{Center for Integrated Nanotechnologies, Sandia National Laboratories, Albuquerque, New Mexico 87185, USA}%
\author{Wladimir A. Benalcazar}%
\email{benalcazar@emory.edu}
\affiliation{Department of Physics, Emory University, Atlanta, Georgia, USA}%

\date{\today}%

\begin{abstract}
The location of electrons governs phenomena ranging from chemical bonding and electric polarization to the topological classification of band insulators and the emergence of correlated states in quantum matter. While a prescription exists for finding local state representations of electrons in one-dimensional insulators, no comparably general theory exists in higher dimensions. Here, we introduce a general framework for finding the location of electrons in insulators in two and three dimensions based on the spectral properties of quantum-mechanical operators that we term \emph{Spatial Localizers}. This framework naturally extends the notion of Wannier centers to insulators with boundaries, defects, and disorder, which we use to establish a position-space formulation of the bulk–defect correspondence for electronic charge. This framework also yields maximally localized electronic states. As two representative examples, we show that these states reduce to maximally localized Wannier functions in atomic insulators, whereas in Chern insulators they form coherent states that mirror the coherent-state structure of Landau levels in the quantum Hall effect.

\end{abstract}

\maketitle

The locations of electrons play a central role in various properties of materials. In molecules, they underlie chemical bonding and catalytic activity~\cite{boys1960,foster1960,edmiston1963}, while in extended systems, they govern charge polarization, orbital magnetization~\cite{zak1989,king-smith_1993}, and provide the conceptual foundation for classifying topological band insulators and their bulk responses~\cite{qi2008,kitaev2009,ryu2010,hasan2010,qi2011}. Local state representations of electrons are also essential for constructing effective models of correlated phases, bridging first-principles electronic structure and the identification of candidate materials for strongly interacting quantum states~\cite{marzari2012,rosner2015,koshino2018,xie2024}.

In one-dimensional insulators, finding local state representations of electronic energy bands admits a complete solution. Polarization, Wannier centers, and maximally localized Wannier functions (MLWFs) can be obtained from Wilson loops in momentum space~\cite{zak1989} or, more generally (e.g., in the presence of disorder), from the many-body formulation of the position operator introduced by Resta~\cite{resta1998,kielson1982}. These approaches are gauge-invariant and have become standard tools across condensed matter physics, chemistry, and materials science~\cite{resta1994,vanderbilt2018book}.

By contrast, no comparably general framework exists in higher dimensions. In two and three dimensions, symmetry-based band-representation methods~\cite{hughes2011,bradlyn2017topological,cano2018,po2017symmetry-based,kruthoff2017,song2018,benalcazar2019} apply to a subset of insulators with crystal symmetries. While useful in the discovery of topological materials~\cite{catalogue1,catalogue2}, this framework leaves out many insulators whose Wannier representations cannot be assessed via symmetry indicators~\cite{cano2022}, including those with rotation anomalies~\cite{rotationAnomaly}, quadrupole and octupole moments~\cite{benalcazar2017_science,benalcazar2017}, boundary obstructions~\cite{botp}, and many atomic insulators with spin-orbit coupling~\cite{schindler2019}. Furthermore, strong topological insulators are independent of crystal symmetries. To complement symmetry-based approaches, hybrid Wannier functions, Wannier bands, and nested Wilson-loop constructions~\cite{sgiarovello2001,marzari2012,alexandradinata2014,benalcazar2017,zeng2023} further diagnose some insulators, but find states localized only along one dimension at a time. In practice, MLWFs are obtained through variational optimization over gauge choices in Bloch space~\cite{marzari1997,marzari2012,souza2001,wannier90,wannier90_v2}, a procedure that relies on user-defined ansätze and does not guarantee convergence to globally optimal solutions. Complementary real-space marker approaches can diagnose local insulating and topological character in inhomogeneous systems, but they do not directly inform Wannier center locations or construct localized state representations~\cite{bianco2011localchern,marrazzo2019}.

The absence of a general framework beyond one dimension is rooted in fundamental geometric constraints. First, position operators must be reconciled with periodic boundary conditions in crystalline systems~\cite{resta1998}. Second, when projected onto the occupied fermionic bands, position operators along different spatial directions generally fail to commute~\cite{neupert2012}, obstructing the simultaneous localization of electronic states in all directions. This obstruction is particularly severe in strong topological insulators, where topological invariants prevent the construction of exponentially localized Wannier functions altogether~\cite{brouder2007exponential}. Despite these constraints, the discovery of interaction-driven phases in highly tunable moiré and graphene-based materials~\cite{cao2018,sharpe2019,serlin2020,xie2021,zhou2021,cai2023,lu2024,xia2025,guo2025} has renewed interest in local state representations of electronic bands because, in the flat-band limit, where dispersion is suppressed, local electronic states form the natural basis for describing local interactions~\cite{kang2018,koshino2018,seo2019,hejazi2021,herzog2024,okuma2026}.

In this work, we introduce a general framework for finding local state representations of electrons in insulators beyond one dimension. Our approach is based on the spectral properties of quantum-mechanical operators that we term \emph{Spatial Localizers}, which address the geometric constraints described above by being compatible with periodic boundary conditions and simultaneously preserving the non-commutative structure of projected position operators across multiple directions. In contrast to variational approaches, our prescription consists of solving an eigenvalue problem; therefore, it is ansatz-free and non-iterative. It provides a gauge-invariant characterization of composite bands and applies to systems with crystalline symmetry or disorder, under periodic or open boundary conditions. A central result of this framework is the extension of the notion of Wannier centers to insulators with open boundaries, lattice defects, or disorder, which we use to establish a general bulk--defect correspondence for electronic charge. A second central result is that, beyond identifying the centers of electronic charge, Spatial Localizers also yield their associated wavefunctions, maximally localized in a rigorous sense. In atomic insulators, these states lead to MLWFs, whereas in Chern insulators they take the form of coherent states, directly mirroring the coherent-state structure of Landau levels in the quantum Hall effect (QHE)~\cite{fradkin2013field,thouless1984_WF_OC,girvin1984}.

To motivate the construction of Spatial Localizers, we briefly review electron localization in one dimension.
In insulators, the Fermi energy lies within an energy gap, so that their energy bands are either filled or empty at zero temperature. 
Electronic phenomena are the result of collective effects of $N_\textrm{occ}$ states below the Fermi level that define the Hilbert subspace $\hs_\mathrm{occ}$. For example, the position of the electrons in 1D can be determined by diagonalizing $P X P$, where $P=\sum_{E_i < E_F} \ketbra{\Psi_i}$ is the projector into occupied bands ($\ket{\Psi_i}$ is the i$th$ energy eigenstate, obeying $\hat{H} \ket{\Psi_i}=E_i \ket{\Psi_i}$) and $X=\sum_{x_i} x_i \ketbra{x_i}$ is the position operator of a crystal with $N$ unit cells of unit length, so that $x_i \in [0,1,\ldots, N-1]$. 
We will discard the null space of $P$ by using instead $\halfproj^\dagger X \halfproj$ with $\halfproj=\sum_{E_i < E_F} \ketbra{\Psi_i}{i}$ (and $P=\halfproj\halfproj^\dagger$). 
To eliminate spurious boundary effects and study the bulk properties of materials, condensed matter systems are often considered under periodic boundary conditions (PBC), so that the discrete coordinates form the torus $T^1=S^1=\{x_i \in \mathbb{Z}, \text{ s.t. } x_i \equiv x_i+N \}$. 
Under these conditions, the position operator needs to be reformulated, as done originally by  Resta~\cite{resta1998}, to $X_\mathrm{R} = e^{\ii\frac{2\pi}{N}X}$. 
The exponentiation of $X$ maps the coordinates of the crystal $\{x_i\}$ to the complex unit circle, and makes the Resta position operator $X_\mathrm{R}$ compatible with PBC. 
The eigenvalues of $\halfproj^\dagger X_\mathrm{R}\halfproj$ are $U(1)$ numbers $e^{\ii \frac{2\pi}{N}x^W}$ that nontrivially solve the eigenvalue equation $(\halfproj^\dagger X_\mathrm{R}\halfproj - e^{i\frac{2\pi}{N}x^W}I)\ket{\psi_{x^W}} = 0$, where $\ket{\psi_{x^W}} \in \hs_\mathrm{occ}$ give rise to the so-called \emph{Wannier functions} (WFs) centered around their \emph{Wannier centers} (WCs) $x^W$~\cite{Wannier1937,marzari1997,marzari2012}. There are $N_\textrm{occ}$ such WCs and WFs. The collection of WFs forms an alternative basis to the Bloch basis for $\hs_\textrm{occ}$, collectively referred to as a \emph{Wannier representation} of $\hs_\textrm{occ}$. 
In crystalline insulators, the WCs can be equivalently obtained in crystal momentum space (i.e., in $k$ space), by diagonalizing the Wilson loop over the occupied bands across the 1D Brillouin zone of the crystal~\cite{benalcazar2017,yu2011,alexandradinata2014} [Appendix A]. In 1D, the two formulations are equivalent and related by the Fourier transform~\cite{yu2011}.

Let us now define the operator
\begin{equation}
    L_X(x) = \halfproj^\dagger(X_\mathrm{R}e^{-\ii \frac{2\pi}{N}x} - I)\halfproj,
    \label{eq:RestaLocalizer}
\end{equation}
which we refer to as the \textit{Resta Spatial Localizer}. We define its \emph{localizer indicator function} (LIF), 
\begin{equation}
    \mu(x) = \min[|\sigma(L_X(x))|],
    \label{eq:WannierIndicatorFunction}
\end{equation}
where $\sigma(\hat{O})$ represents the spectrum of an operator $\hat{O}$. Since setting $\det[L_X(x)]=0$ recovers the eigenvalue problem for $X_\textrm{R}$ projected into $\hs_\textrm{occ}$~\footnote{Note that the Resta Spatial Localizer in~\eqref{eq:RestaLocalizer} is equivalent to the standard operator used in eigenvalue equations, $O - \lambda I$, scaled by $\lambda^{-1}$.}, the zeros of the LIF $\{ x^* | \mu(x^*)=0\}$ indicate the presence of WFs centered at the locations $\{x^*\}$. We therefore identify the positions $\{x^*\}$ 
with the WCs of $\hs_\textrm{occ}$, i.e., $\{x^*\}=\{x^W\}$.

The Resta position operator $X_\mathrm{R}$ embeds the position coordinates with PBC within $S^1=\{e^{\ii \phi}| \phi \in [0,2\pi)\}$. This embedding is not unique. To elucidate the generalization to higher dimensions, we first embed the 1D position operator onto the equator of the Bloch sphere, $S^1=\{\cos (\phi) \sigma_1 + \sin (\phi) \sigma_2| \phi \in [0,2\pi)\}$, and use it to define the 1D Spatial Localizer
\begin{align}
        L^\mathrm{PBC}_\mathrm{1D}(x) &= \tX_C(x)  \otimes \sigma_1 + \tX_S(x) \otimes \sigma_2
        \label{eq:localizer1D}\\
        & =\left(\begin{smallmatrix} 0 & L_X(x) \\ L^\dagger_X(x) & 0 \end{smallmatrix}\right),\nonumber
\end{align}
where $\tX_C(x) = \halfproj^\dagger(\cos(2 \pi (X- xI) / N) - I)\halfproj$ and $\tX_S(x) =  \halfproj^\dagger \sin(2 \pi (X-xI) / N)\halfproj$. The Spatial Localizer \eqref{eq:localizer1D} is Hermitian and obeys chiral symmetry, $\{L^\mathrm{PBC}_\mathrm{1D}(x),\Pi\}=0$, with $\Pi=I \otimes \sigma_3$. 
This renders its spectrum $\sigma(L_\mathrm{1D}(x))$ symmetric about zero; it comes in pairs $\{-|\sigma(L_X(x))|, +|\sigma(L_X(x))|\}$. Thus, the LIF is the same when calculated using either \eqref{eq:RestaLocalizer} or \eqref{eq:localizer1D}, and therefore the WCs are identical. 

The embedding of the position operator within the Clifford algebra $\mathrm{Cl}_{3,0}(\mathbb{R})$ in \eqref{eq:localizer1D} paves the way for its generalization. In higher dimensions, larger Clifford algebras allow the simultaneous embedding of multiple position operators, one per physical dimension of the lattice, all this while simultaneously preserving PBC.

\begin{figure*}
    \centering
    \includegraphics[width=0.99\linewidth]{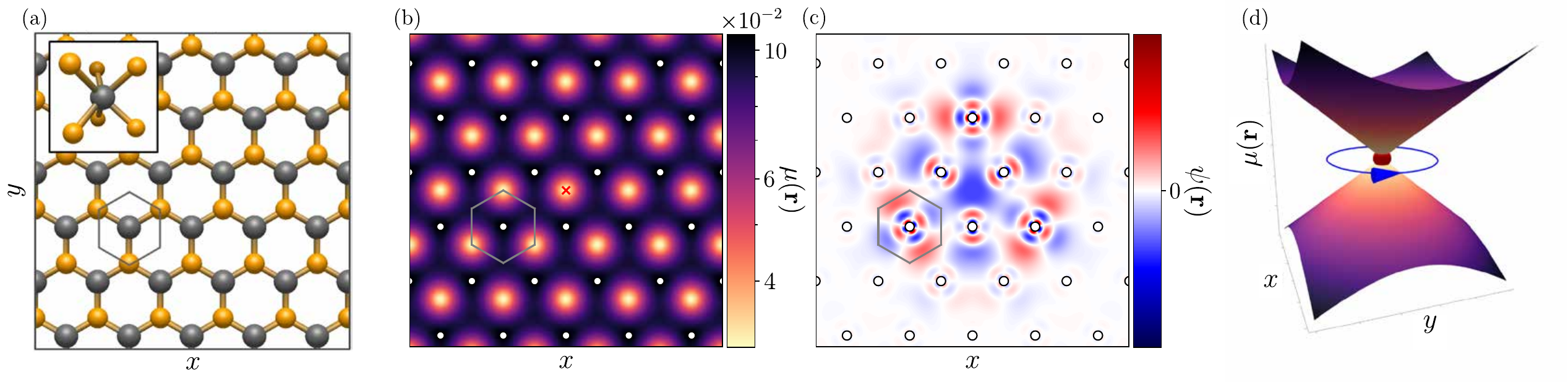}
    
    \caption{{\bf Spatial Localizer in the obstructed atomic insulator $\mathrm{WSe_2}$}. (a) Atomic structure of $\mathrm{WSe_2}$; the grey hexagon denotes a unit cell centered at a W atom. The W and Se atoms are colored grey and yellow, respectively. (b) LIF for the effective $\mathrm{WSe_2}$ model. The red cross denotes the extraction point for the WF in (c). (c) WF constructed from the eigenstate of the Spatial Localizer $L(\rv^W)$ with minimal $\mu(\rv^W)$ at the center of the plot. The WF is purely real. The color bar uses a linear color scaling. White dots with black edges in (b) and (c) indicate $\mathrm{W}$ atoms for reference. (d) LIF around a WC and its Wannier monopole as a topological charge (sphere).
    }
    \label{fig:wse2}
\end{figure*}

In 2D, the lattice of a crystal under PBC is covered by the discrete torus of coordinates $T^2=\{ (x_i,y_i)  \mid x_i,y_i \in \mathbb{Z} \text{, s.t. } x_i \equiv x_i + N_x, \; y_i \equiv y_i + N_y\}$. To embed the associated position operators $X$ and $Y$ in $T^2$, we use the Clifford algebra $\mathrm{Cl}_{4,0}(\mathbb{R})$. We define the 2D Spatial Localizer
\begin{equation}\label{eq:2D_loc_pbc}
    \begin{split}
        L^{\mathrm{PBC}}_\mathrm{2D}(\rv) &= \tX_C(x) \otimes \Gamma_1
         + \tX_S(x) \otimes \Gamma_2
         \\ &+ \tilde{Y}_C(y) \otimes \Gamma_3
         + \tilde{Y}_S(y) \otimes \Gamma_4,
    \end{split}
\end{equation}
where $\rv=(x,y)$, $\tilde{Y}_C(y)$ and $\tilde{Y}_S(y)$ have the same functional form as $\tX_C(x)$ and $\tX_S(x)$ in \eqref{eq:localizer1D}, with substitution $X \rightarrow Y$. For concreteness, $\Gamma_j = -\sigma_2 \otimes \sigma_j$ for $j=1,2,3$, and $\Gamma_4 = \sigma_1 \otimes I_2$ (although any other representation of $\mathrm{Cl}_{4,0}(\mathbb{R})$ is equally valid).
The 2D PBC Spatial Localizer also obeys chiral symmetry, $\{I \otimes \Gamma_0, L^{\mathrm{PBC}}_\mathrm{2D}({\bf r})\}=0$, with $\Gamma_0 = \sigma_3 \otimes I_2$. Due to the absence of $\Gamma_0$, the Spatial Localizer \eqref{eq:2D_loc_pbc} embeds the projected position operators into the equatorial manifold of $S^3$, i.e., the Clifford torus, which is locally flat everywhere~\footnote{There are other embeddings for localizers under PBC. For example, see Ref.~\cite{doll2024}, where the embedding is into $S^2$ using Pauli matrices. We have chosen an embedding into the Clifford torus instead because it is locally flat everywhere, resembling both the periodicity and flat nature of the physical space for a translationally invariant system.}. The existence of WFs and WCs can then be obtained by inspecting the LIF $\mu(\rv)=\min[|\sigma(L^\textrm{PBC}_\textrm{2D}(\rv))|]$ for positions $\{\rv^W\}$ where $\mu(\rv^W)=0$. In insulators with discrete translation invariance (i.e., pristine crystals), scanning $\mu(\rv)$ over one unit cell suffices to find the WC configuration of the crystal. We note, however, that discrete translation invariance is not necessary in this formulation. The Spatial Localizer \eqref{eq:2D_loc_pbc} also finds WCs in the presence of disorder.

For open boundary conditions (OBC), there is no need to embed the crystal's unit-cell coordinates on a closed manifold. Still, Clifford algebras can be used to embed various position operators. In 2D, we define
 \begin{equation}\label{eq:2D_loc_obc}
     L^{\mathrm{OBC}}_{\mathrm{2D}}(\rv) = \tX(x) \otimes \sigma_1 + \tilde{Y}(y) \otimes \sigma_2,
 \end{equation}
where $\tX(x) = \halfproj^\dagger(X- xI)\halfproj$ and similarly for $\tilde{Y}(y)$. Equation~\eqref{eq:2D_loc_obc} obeys chiral symmetry, with $\Pi=I \otimes \sigma_3$. 
Contrary to the case of PBC, there is no translation invariance, and thus we generally need to consider the LIF over the entire lattice to get the full distribution of WCs. Formulations of Spatial Localizers under OBC and PBC and in arbitrary spatial dimensions are presented in Appendix B. 

The Spatial Localizer \eqref{eq:2D_loc_pbc} is constructed for crystals with orthogonal primitive lattice vectors. More generally, Spatial Localizers under PBC require an embedding of position operators that equally localize in all directions. For example, in $C_3$ symmetric lattices, the PBC Spatial Localizer is built from three Resta position operators along directions rotated by $2\pi/3$ rad from one another~\footnote{Similar considerations of symmetry are also used in optimization methods such as those in Ref.~\cite{marzari1997}}.

As an example, we apply the Spatial Localizer to the $C_3$ symmetric transition‐metal dichalcogenide $\mathrm{WSe_2}$~\footnote{The model itself is an effective model for the monolayer of the W sublattice, which takes into account both (i) direct W-to-W hoppings and (ii) W-to-W hoppings mediated by the Se sublattice, and which predominantly accounts for the conduction and valence bands of $\mathrm{WSe_2}$~\cite{wse2_model}.} under PBC [Fig.~\ref{fig:wse2}(a)]. This material has been shown to exhibit superconductivity in Moir\'e heterostructures~\cite{xia2025,guo2025}, and was recently inferred as an obstructed atomic limit (OAL) insulator by real‐space imaging of its band topology and analysis of its band representations at high‐symmetry points of the Brillouin zone~\cite{holbrook2024realspaceimagingbandtopology}. Crucially, its projected Resta position operators along different directions do not commute, allowing us to test the effectiveness of the Spatial Localizer in a nontrivial system. 

Figure~\ref{fig:wse2}(b) shows its LIF  $\mu(\rv)$. There are $N_\textrm{occ}$ positions $\{\rv^W\}$ that minimize the LIF to $\mu(\rv^W) \rightarrow 0$ in the thermodynamic limit. These positions correspond to the WCs, which match those expected from band representation analysis~\cite{cano2018}. Figure~\ref{fig:wse2}(c) shows a sample WF constructed from the Spatial Localizer's eigenstate whose eigenvalue is the LIF at the center of the plot [which coincides with a WC, red cross in Fig.~\ref{fig:wse2}(b)]. The WF is indeed centered around its WC, and has the trivial irrep of $C_3$ symmetry, matching the irrep of the occupied energy band at the $\Gamma$ point of the Brillouin zone, as expected. Furthermore, we find that the resulting WFs cannot be further localized by traditional spread-optimization techniques [Appendix F], confirming that these states are MLWFs.

\begin{figure}
    \centering
    \includegraphics[width=1\linewidth]{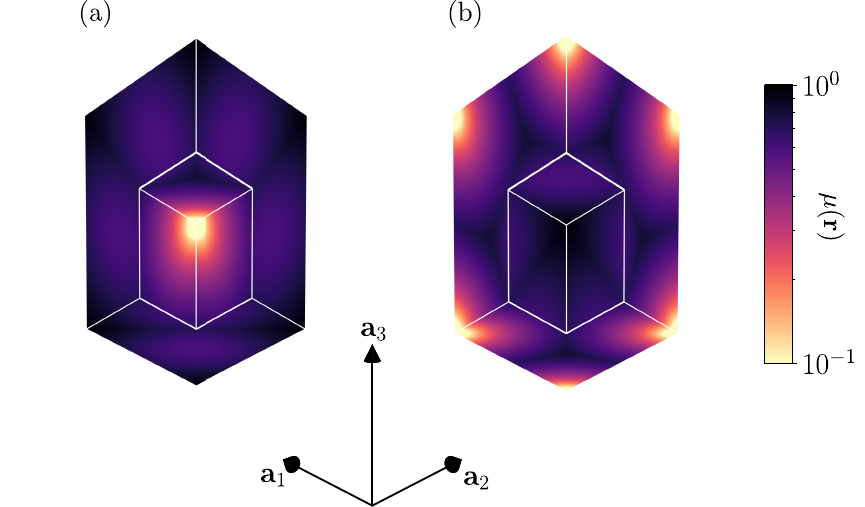}
    \caption{{\bf WCs in insulators with equal band representations}. LIF over the unit cell of an insulator in two phases protected by the space group F222. A trivial insulator (a) and an OAL (b). An octant of the unit cell has been removed to reveal its center, $\frac12 (\mathbf{a}_1 + \mathbf{a}_2 + \mathbf{a}_3)$. Both phases have identical band representations, and thus their WCs cannot be determined from symmetry considerations.}
    \label{fig:3D}
\end{figure}

As an illustration in a 3D insulator, we consider a symmetry-preserving variation of the three-dimensional F222 model of Ref.~\cite{cano2022}, which realizes a trivial and an OAL phase, with WCs at the maximal Wyckoff positions $4a$ and $4b$, respectively, where $\rv_\text{4a}=(0,0,0)$ and $\rv_\text{4b}=(0,0,\tfrac{1}{2})$ in Cartesian coordinates. In the minimal form presented in Ref.~\cite{cano2022}, this model is closely related to stacked Rice-Mele chains, the projected position operators commute, and the Wannier-center diagnosis is simple. To obtain a nontrivial 3D benchmark for the Spatial Localizer, we add a term that preserves the symmetry and phase, and consequently their WC configurations, while rendering the projected position operators noncommuting [Appendix F]. The corresponding Spatial Localizer correctly distinguishes the two WC configurations, with minima developing at the appropriate real-space locations in each phase [Figure~\ref{fig:3D}]. This example is significant because the two phases are irrep-equivalent at all high-symmetry momenta and, moreover, are not distinguished by the polarization along $z$. The Spatial Localizer therefore captures information beyond band representation and symmetry indicator frameworks and polarization.

By construction, the Spatial Localizer possesses chiral symmetry, and the WCs, which point to the zeros of the Spatial Localizer's spectrum, can be associated with the existence of ``Wannier monopoles''. In 2D, this enables counting the number of electrons $N_e$ within a region enclosed by a path $\mathcal{C}$ by performing the loop integral,
\begin{align}\label{eq:Ne}
N_e=\frac{1}{2\pi \ii} \oint_\mathcal{C} \dd \mathbf{r} \cdot \mathbf{\nabla}_r \log \det \tZ^\dagger(\mathbf{r}) \;\; \in \; \mathbb{Z},
\end{align}
where $\tZ(\rv)=\tX(x) + \ii \tY(y)$ is the off-diagonal component of the Spatial Localizer, which under chiral symmetry can always be written as $L=[0 \;\; \tZ^\dagger;\;\; \tZ \;\; 0]$. $N_e$ is thus given by the winding number generated by the Wannier monopoles enclosed within the loop $\mathcal{C}$ [Fig.~\ref{fig:wse2}(d)].

\begin{figure}
    \centering
    \includegraphics[width=1\linewidth]{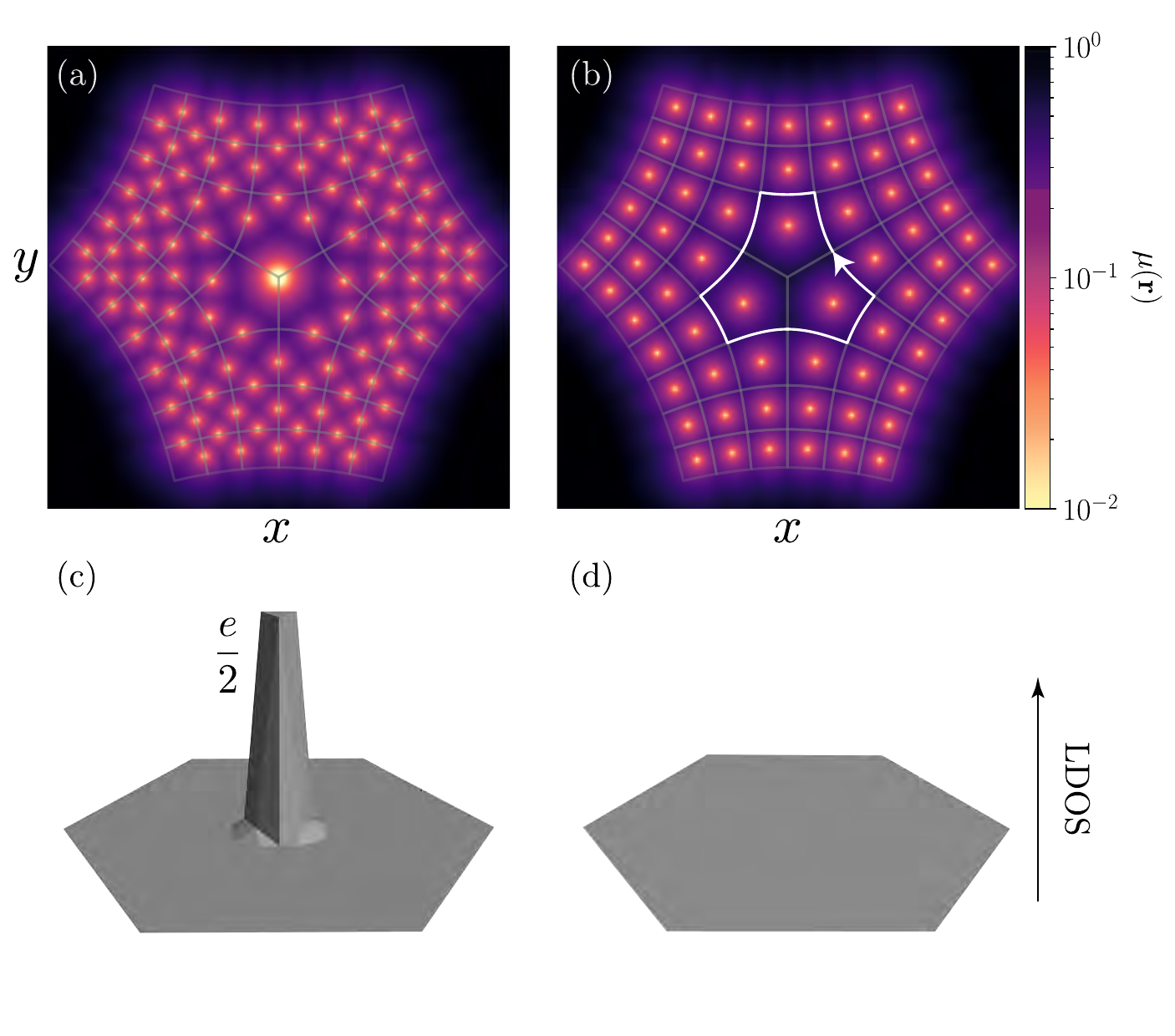}
    \caption{{\bf Bulk-defect correspondence for electronic charge around a topological defect}. LIF for a crystalline insulator with a disclination and two electrons per unit cell in an OAL phase (a) and a trivial phase (b). Both figures follow the color bar on the right. Grey lines denote unit cell boundaries. (c,d) Net charge densities (with $+2e$ ionic charges per unit cell) quantized to $e/2$ in (c) and zero in (d). Excess edge charge is not shown for clarity.}
    \label{fig:disclinations}
\end{figure}

Equation~\eqref{eq:Ne} enables a rigorous, position space formulation of the bulk-defect correspondence for electronic charge. Drawing from analogies with the bulk-boundary correspondence in polarized insulators, where boundary charge is proportional to the displacement of WCs from their trivial positions (at the centers of the unit cells), it follows that the WC configurations around defects determine any local electric charge bound to them. Of particular interest are topological defects, which can fractionally quantize that charge~\cite{benalcazar2019, li2020fractional} and bind midgap states~\cite{teo2010,teo2013,benalcazar2014classification,deng2022observation}. Figure~\ref{fig:disclinations} shows the WC configurations obtained with the Spatial Localizer \eqref{eq:2D_loc_obc} around a disclination with a Frank angle of $\pi$ in a $C_4$ symmetric crystalline insulator that can transition between an OAL phase and a trivial insulator phase [Appendix F]. Due to the curvature induced by the disclination, we have that $[\tX(x),\tY(y)] \neq 0$, even though $[\tX(x),\tY(y)] = 0$ in the absence of a defect. Any loop $\mathcal{C}$ in \eqref{eq:Ne} enclosing an integer number of unit cells and surrounding the defect's core [e.g., white path in Fig.~\ref{fig:disclinations}(b)] mod 1 will indicate the fractional charge (in units of $e$) bound to the defect. For the insulator in Fig.~\ref{fig:disclinations}(a), the OAL phase carries a fractional charge of $e/2$ at the disclination's core, as the path cuts through the middle of an odd number of Wannier monopoles. In contrast, paths $\mathcal{C}$ in the trivial phase always enclose an integer number of Wannier monopoles (Fig.~\ref{fig:disclinations}b), and thus, this phase is not expected to bind a fractional charge. Both predictions are confirmed by local density of state calculations [Fig.~\ref{fig:disclinations}(c) and \ref{fig:disclinations}(d)]. 

Spatial Localizers offer a gauge-invariant pathway for identifying WCs, their corresponding maximally localized Wannier states, and, more generally, sets of ``local state representations'' of energy bands. They find the best approximation to joint eigenstates of the projected position operators $\tX(x)$ and $\tY(y)$, which generally do not commute, $[\tX(x),\tY(y)]\ne 0$, as in the three examples described above.

When the projected position operators commute, $[\tX(x),\tY(y)]=0$, the Spatial Localizer simultaneously diagonalizes $\tX(x)$ and $\tY(y)$, and the magnitude of the LIF decreases the closer the ``extraction point'' $\rv=\sum_j r_j \hat{e}_j$ gets to a WC $\rv^W$. Specifically, the LIF near a WC $\rv^W$ takes the form
\begin{align}
    \mu(\rv) = \sqrt{\sum_j(r^W_j-r_j)^2}
    \label{eq:WannierCone}
\end{align}
as shown in Fig.~\ref{fig:wse2}(d). 

When the projected position operators do not commute, $[\tX(x),\tY(y)] \neq 0$, a crucial distinction arises depending on whether $\hs_\textrm{occ}$ is in a strong topological phase, such as a Chern insulator phase or a quantum spin Hall (QSH) phase, or not. In the absence of a strong topological phase, and despite $\tX$ and $\tY$ not commuting, the Spatial Localizer finds $N_\textrm{occ}$ WCs (more precisely, $N_\textrm{occ}$ positions $\rv^W$ where $\mu(\rv^W) \rightarrow 0$ as $N\rightarrow \infty$)~\footnote{Based on studies of operators with the same mathematical structure of our localizers, we conclude that the existence of $N_{\text{occ}}$ WCs, where $\mu(\rv^W)\rightarrow0$, is guaranteed to exist for localizers consisting of two position operators (e.g., our 1D PBC or 2D OBC Spatial Localizers). The general existence of zero spectra for localizers with three or more operators (spatial dimensions) has been conjectured, but not proven~\cite{kisil1996mobius,jefferies1998weyl}. However, we provide an argument for the existence of $N_\text{occ}$ points for the 2D PBC Spatial Localizer that converge to zero as $\mu(r^W)\propto N^{-2}$, while generic points scale as $N^{-1}$ [Appendix C]}, although the general functional form of the LIF around them does not necessarily follow \eqref{eq:WannierCone}. In contrast, in strong topological phases, in which $[\tX(x),\tY(y)] \neq 0$ always, we observe that the entire LIF approaches zero in the thermodynamic limit, and the maximally-localized states we extract from the Spatial Localizer have a coherent-state structure, as in the QHE~\cite{fradkin2013field,thouless1984_WF_OC}.

To support our assertions, we now show that the Spatial Localizer's eigenstates obtained from the minima of the LIF contain maximally localized state representations within $\hs_\mathrm{occ}$, regardless of the topological phase of the insulator. The spatial spread of a general state is naturally measured by its variance, e.g., $\Delta R^2 = \Delta X^2 + \Delta Y^2$ in 2D, with $\Delta X^2 = \expval{X^2} - \expval{X}^2$ and similarly for $\Delta Y^2$. To make a connection with the Bloch state decomposition of localized states within the occupied subspace $\hs_\textrm{occ}$, it is convenient to split $\Delta R^2$ into two contributions, $\Delta R^2=\Delta R_I^2+\Delta \tR^2$ with $\Delta R_I^2 = \expval{PXQXP + PYQYP}$, $\Delta\tilde{R}^2=\Delta \tX^2 + \Delta \tY^2$, and $Q=I-P$ [Appendix C].

For WFs, $\Delta R^2$ is equivalent to the Marzari-Vanderbilt localization functional $\Omega$ for a single band, which correspondingly splits as $\Omega=\Omega_I+\tilde{\Omega}$~\cite{marzari1997}. The first contribution, $\Omega_I$, is related to the gauge-invariant quantum metric via $\Omega_I=\int \Tr(g_\kv)d^2\kv$, where $[g_\kv]_{ij}=\frac12 ([\eta_\kv]_{ij} + [\eta_\kv]_{ij}^\dagger)$ and $[\eta_\kv]_{ij} = \sum_n \bra{\partial_{k_i}u_{n,\kv}} (I - P_\kv)\ket{\partial_{k_j}u_{n,\kv}}$ is the quantum geometric tensor over the unit-cell periodic components of occupied Bloch states $\ket{u_{n,\kv}}$~\cite{marzari1997}. The second contribution, $\tilde{\Omega}$, is the variance of the state with respect to the position operators projected into $\hs_\textrm{occ}$ [Appendix D]. Unlike the quantum geometric contribution, $\tilde{\Omega}$ is gauge-dependent and commonly minimized via the projection method and optimization algorithms~\cite{marzari1997,wannier90,wannier90_v2}. Thus, we expect states that minimize $\Delta \tR^2$ to serve as optimal seed/trial states for constructing Wannier bases in OALs.
 
 The Spatial Localizer framework we introduce in this work generates maximally localized states by minimizing $\Delta \tilde{R}^2$ via exact diagonalization of the eigenproblem $L(\rv)\ket{\psi^L(\rv)} = \mu(\rv)\ket{\psi^L(\rv)}$. It does so regardless of the presence of strong topological invariants, lack of discrete translation invariance, and without the need for ansätze or gauge-fixing procedures. The maximally localized states are obtained by separating the spatial and embedding degrees of freedom in the localizer eigenstates $\ket{\psi^L(\rv)}$ via the Schmidt decomposition
\begin{equation}
    \ket{\psi^L(\rv)} = \sum_i s_i \left( \ket{\psch_i(\rv)} \otimes \ket{\esch_i(\rv)} \right),
    \label{eq:SchmidtDecomposition}
\end{equation} 
where $s_i \geq 0 $ are the Schmidt values, obeying $\sum_i s_i^2=1$, and $\ket{\psch_i(\rv)}$ and $\ket{\esch_i(\rv)}$ are, respectively, the ``physical'' and ``embedding'' Schmidt vectors. We order the Schmidt values $\{s_i\}$ in descending order, and define the first physical Schmidt vector, $\ket{p_1(\rv)}$, as our ``localized state''. As we will now see, the states $\{\ket{p_1(\rv^*)}\}$ extracted from the $N_\textrm{occ}$ positions $\{\rv^*\}$ with minimal $\mu(\rv)$ are maximally localized. In particular, in OALs, the states $\{\ket{p_1(\rv^W)}\}$ extracted from the $N_\textrm{occ}$ WCs $\{\rv^W\}$ yield MLWFs, while in insulators in strong topological phases, the Spatial Localizer generates a set of $N_\textrm{occ}$ exponentially-localized coherent states, overcomplete in $\hs_\textrm{occ}$ by an amount given by the strong topological invariant.

For the OBC Spatial Localizer $L^\textrm{OBC}_\textrm{2D}(\rv)$ in~\eqref{eq:2D_loc_obc}, the spread $\Delta\tilde{R}^2$ of $\ket{p_1(\rv)}$ is bound by the relation
\begin{equation}\label{eq:main_text_variance bound}
        \Delta\tilde{R}_H^2 \leq \Delta\tilde{R}^2 \leq \Delta\tilde{R}_H^2+\satresidual,
\end{equation}
where
\begin{align}\label{eq:main_text_variance bound terms}
        &\Delta\tilde{R}_H^2 = \abs{\expval{\comm{\tX}{\tY}}}, \nonumber   \\
        &\satresidual = \mu^2(\rv) - \norm{\expval{\tilde{\mathbf{R}}(\rv)}}^2 +  \sum_{i\neq 1} \frac{s_i}{s_1} \abs{\bra{\psch_1(\rv)} \comm{\tilde{X}}{\tilde{Y}} \ket{\psch_i(\rv)}}, \nonumber
\end{align}
and the arguments in $\tX(x)$ and $\tY(y)$ are omitted to simplify the presentation, $\tilde{\mathbf{R}}(\rv)= \tX(x)\mathbf{\hat{x}} + \tY(y)\mathbf{\hat{y}}$, and all implicit expectation values are with respect to $\ket{\psch_1(\rv)}$. Finally, it is useful to note that ${\tilde{\mathbf{R}}}(\rv)={\tilde{\mathbf{R}}}({\bf 0})-V^\dagger_\mathrm{occ}\rv V_\mathrm{occ}$ [c.f. \eqref{eq:2D_loc_obc}]. The bound in \eqref{eq:main_text_variance bound} and its generalizations to arbitrary spatial dimensions for both OBC and PBC are demonstrated in Appendix C. The lower bound $\Delta\tilde{R}_H^2$ is the Heisenberg-Robertson uncertainty relation (HRUR)~\cite{robertson_1929}. The upper bound is determined by quantities that characterize the Spatial Localizer $L(\rv)$ and its $\mu(\rv)$ eigenstate at the extraction point $\rv$.

Let us start by considering OAL insulators under OBC. Regardless of whether or not $\comm{\tX}{\tY}=0$, there are always $N_\textrm{occ}$ WCs $\{\rv^W\}$ where $\mu(\rv^W)=0$, $\texpval{\tilde{\bf R}(\rv^W)}=0$, and $s_1=1$; thus $\satresidual = 0$ and the corresponding $\ket{\psch_1(\rv^W)}$ are eigenstates with minimum uncertainty [Appendix C]. Furthermore, if $\comm{\tX}{\tY}=0$, we have $\Delta \tR_H^2=0$ and the states $\{\ket{\psch_1(\rv^W)}\}$ are simultaneous eigenstates of $\tX,\tY$, and form a set of orthogonal MLWFs~\cite{kielson1982,marzari1997}. However, if $\comm{\tX}{\tY}\neq 0$, the states $\{\ket{\psch_1(\rv^W)}\}$ are generally non-orthogonal, but complete in $\hs_\text{occ}$ and a Wannier representation can be built using L\"{o}wdin orthogonalization~\cite{lowdinortho1950}\footnote{This is a standard orthogonalization method, widely used in numerical approaches that find Wannier bases~\cite{marzari1997}. For one band, this procedure amounts to simply setting equal magnitude to the coefficients in the expansion of $\ket{p_1(\rv)}$ in terms of occupied Bloch states}. 

In insulators with strong topological invariants, a system with OBC is necessarily gapless due to edge-localized topological states that cross the bulk band gaps, rendering it metallic. To consider fully insulating configurations, we study strong topological insulators only under PBC.

For the PBC Spatial Localizer $L^\textrm{PBC}_\textrm{2D}(\rv)$ in \eqref{eq:2D_loc_pbc}, the bound is presented in Appendix C. The upper bound converges to that in \eqref{eq:main_text_variance bound} in the thermodynamic limit upon the replacement $\mu \rightarrow \mu N/2\pi$
and under the assumption that $\ket{p_1(\rv)}$ is tightly confined. As we will see, this assumption is valid even in systems with topological obstructions, as the $\ket{p_1(\rv)}$ states are tightly localized coherent states.

\begin{figure}
    \centering
    \includegraphics[width=1\linewidth]{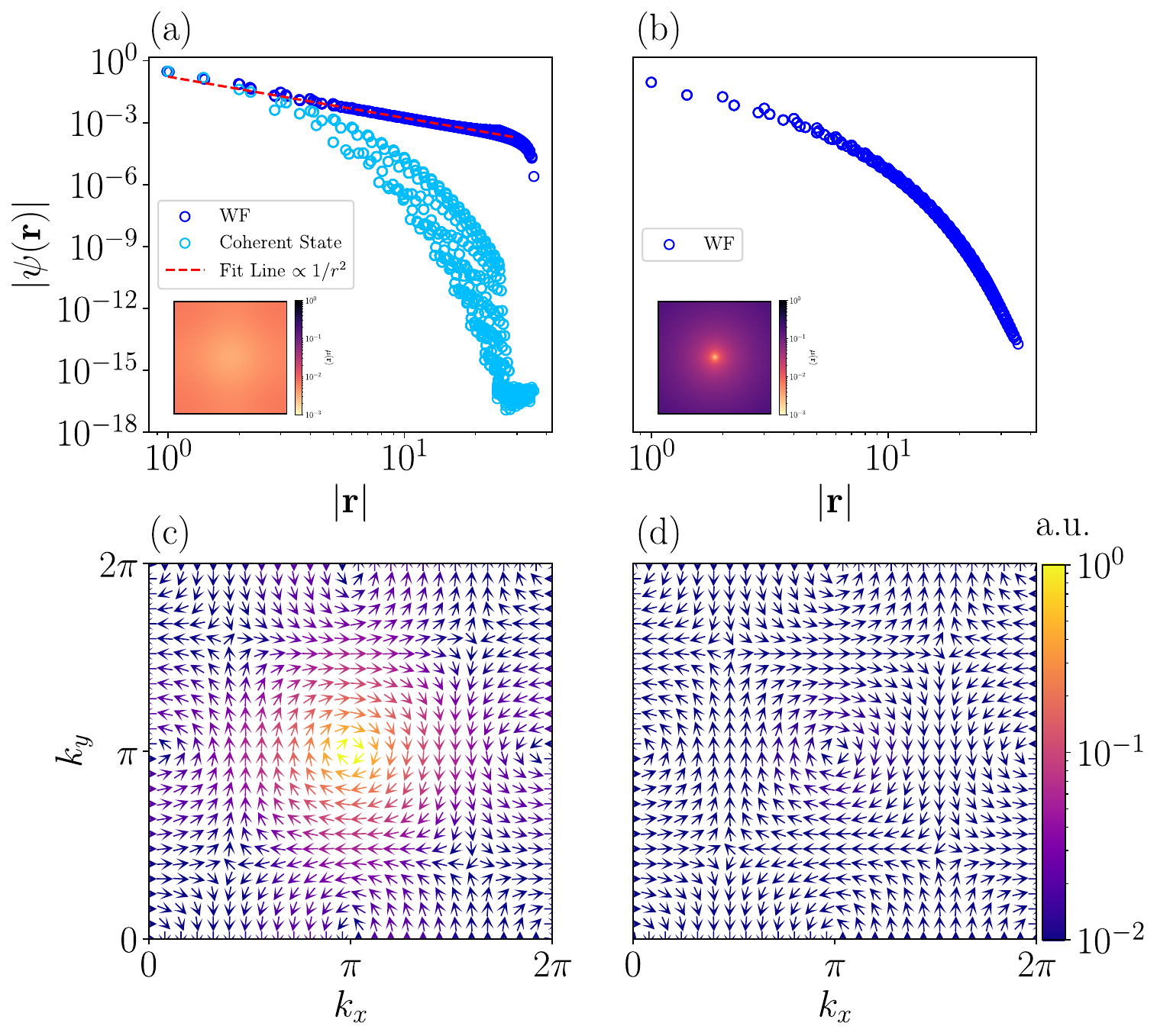}
    \caption{ {\bf The Spatial Localizer in a Chern insulator}.
    (a) and (b) Magnitude of WFs and coherent states as a function of distance from the mean, $|\mathbf{r}|$ for $|\mathbf{r}|>0.5$. In (a), the fit line is proportional to $1/r^{2}$. (c) and (d) Berry connection $\mathcal{A}(\kv)$ over the Brillouin zone obtained from the coherent state in (a) and WF in (b).
    The color on (c) and (d) corresponds to the magnitude of the Berry connection. The left (right) column corresponds to the topological (trivial) phase.
    All states are calculated on an $N\times N$ square lattice of unit cells with $N=50$. Insets: LIF for a single unit cell with $N=24$. Details on the insets are provided in Appendix F.}
    \label{fig:chern_QSH}
\end{figure}

Our choice for a strong topological insulator is a Chern insulator, known to exhibit a \emph{topological obstruction} to the existence of exponentially localized WFs and a divergent spatial variance for any WF in the thermodynamic limit~\cite{brouder2007exponential,panati2007,monaco2018}. In momentum space, the topological obstruction manifests in the impossibility of defining a smooth, periodic gauge for the Bloch states over the entire Brillouin zone and is quantified by the Chern number $C = \frac{1}{2\pi} \int_{\text{BZ}} \mathrm{Tr} \left[ \mathcal{F}(\mathbf{k}) \right] \, d^2k$, where $\mathcal{F}(\mathbf{k}) = d\mathcal{A}(\mathbf{k}) + \mathcal{A}(\mathbf{k}) \wedge \mathcal{A}(\mathbf{k})$ is the Berry curvature, and $\mathcal{A}_{mn}(\mathbf{k}) = -i \braket{u_{m,\mathbf{k}} | \nabla_{\mathbf{k}} | u_{n,\mathbf{k}}}$ is the Berry connection over the unit-cell periodic components of the Bloch eigenstates $\ket{u_{m,\kv}}$ of occupied energy bands.

We study the Qi--Wu--Zhang (QWZ) model under PBC, a minimal lattice Hamiltonian for a Chern insulator~\cite{qi2006}. Figure~\ref{fig:chern_QSH} summarizes our results. It shows the localized states obtained with the Spatial Localizer \eqref{eq:2D_loc_pbc} in the topological [$C=1$, Fig.~\ref{fig:chern_QSH}(a) and (c)] and trivial [$C=0$, Fig.~\ref{fig:chern_QSH}(b) and (d)] phases, along with the LIFs (insets). In the topological phase, the LIF is nearly flat and converges to zero everywhere in the unit cell in the thermodynamic limit [Appendix C]. The minimum $\rextract$ of the LIF points to a center with maximal localization of the local states $\ket{\psch_1(\rv)}$.
In contrast, the LIF in the trivial phase has a conical shape centered around the WC at the center of the unit cell.

In the $C=1$ phase, the $N_\textrm{occ}$ states $\{\ket{\psch_1(\rv^*)}\}$ indeed minimize the upper bound in \eqref{eq:main_text_variance bound}, and form a set of exponentially localized but overcomplete set of states within $\hs_\textrm{occ}$, spanning a Hilbert space of dimension $N_\textrm{occ}-C$. These states are \emph{coherent states}, as they saturate the HRUR and are zero modes of the operator~\cite{fradkin2013field,okuma2024} $\tZ^\dagger(\rv) = \tX(x) - \ii \tY(y)$ in the thermodynamic limit. Note that this resembles the coherent states in the QHE in 2D electron gases under strong magnetic fields~\cite{fradkin2013field,thouless1984_WF_OC,tao1986_WF_OC,okuma2024}.

The overcompleteness, a property common to coherent states~\cite{thouless1984_WF_OC,Perelomov1971}, implies the $N_\mathrm{occ}$ states are not linearly independent. Indeed, they obey the relation
\begin{equation}
    \sum_{\mathbf{R}} e^{\ii \mathbf{k_v} \cdot \transvec} \ket{p_1(\rv^*+\transvec)} = 0,
\end{equation}
where $\rextract=(0,0)$, $\mathbf{k_v}=(\pi,\pi)$ and the sum is over the $N_\mathrm{occ}$ coherent states related by lattice translations $\transvec=n_1 \mathbf{a}_1+n_2 \mathbf{a}_2$, for $n_1,n_2 \in \mathbb{Z}$ and primitive lattice vectors $\mathbf{a}_1$, $\mathbf{a}_2$. The phase factor $e^{\ii \mathbf{k_v} \cdot \transvec}$ reflects the fact that the null space of the coherent states within $\hs_\textrm{occ}$ corresponds to the Bloch state with crystal momentum $\mathbf{k_v}$ in the Brillouin zone (BZ) [Appendix F]. A decomposition of the coherent states into the Bloch states within $\hs_\textrm{occ}$ reveals the existence of a ``vortex gauge'', where the Berry connection manifests the topological discontinuities in the gauge via vortices, the largest one centered precisely at $\mathbf{k_v}$ [Fig.~\ref{fig:chern_QSH} (c)].

Such a gauge has recently been proposed as an optimal gauge for constructing WFs in Chern insulators~\cite{gunawardana2024,xie2024,li2024}. Our findings corroborate these proposals from the complete opposite perspective: the Spatial Localizer finds the most localized state within $\hs_\textrm{occ}$ via a gauge-invariant eigenvalue problem in position space, and this state --the coherent state-- is found to correspond to the vortex gauge in reciprocal space only a posteriori. To generate a Wannier basis from these coherent states, it suffices to restore support at the $\mathbf{k_v}$ point of the BZ. In our model, the second physical Schmidt vector, $\ket{\psch_2(\rv)}$, has support at the $\mathbf{k_v}$ point and, when considered in superposition with $\ket{\psch_1(\rv)}$, furnishes a gauge for localized WFs.
Specifically, we construct a state as $\ket{\psi^T(\rv)}=\alpha_1\ket{\psch_1(\rv)} + \alpha_2 \ket{\psch_2(\rv)}$ with $\alpha_1 \gg \alpha_2$ and utilize a L\"{o}wdin procedure~\cite{lowdinortho1950,marzari1997} to produce WFs. These WFs have amplitudes that decay from their centers as $1/r^2$, in accordance with the known lower bound~\cite{li2024,brouder2007exponential,panati2007,monaco2018} [Fig.~\ref{fig:chern_QSH}(a)]. 

We have shown that the Spatial Localizer framework solves the problem of electron localization in extended systems. In OALs, Spatial Localizers find MLWFs. Remarkably, these representations are unique; states pulled from positions $\rv \neq \rv^W$ are still centered around their nearest WC $\rv^W$, and possess the same spread. The collection of these unique MLWFs spans the space of occupied energy bands, forming ``Wannier representations'' of the bands. In Chern insulators, on the other hand, the Wannier representations are \emph{not} unique. The coherent state $\ket{\psch_1(\rv)}$ is always centered around the extraction position $\rv$, even for $\rv$ not corresponding to minima of the LIF, $\rv^*$. In other words, there is a continuum of coherent states centered anywhere in the unit cell, as one has in the QHE~\cite{fradkin2013field,girvin1984}  [Appendix F]. To any arbitrary extraction point $\rv$ corresponds a WF (constructed using $\ket{\psch_1(\rv)}$ and adding the necessary support at $\mathbf{k_v}$) centered at $\rv$ and with a gauge vortex in reciprocal space centered at $\mathbf{k_v}$, such that the polarization $\frac{\mathbf{P}}{e} = -\frac{\rv}{V_c} - \frac{C}{2\pi} \hat{z} \times \mathbf{k_v}$ remains invariant [Appendix F], where $V_c$ is the volume of the unit cell. This relation, recently derived in Ref.~\cite{gunawardana2025} and independently verified with our Spatial Localizer, sheds light on the problem of polarization and weak topology in Chern insulators~\cite{coh2009,gunawardana2025,vaidya2024,zhang2025}. We do find, however, that maximal localization of a coherent/Wannier state is obtained from the position $\rv^*$ that minimizes the LIF, $\mu(\rv^*)$. This is in agreement with recent work using the vortex gauge approach, in which maximally localized WFs are associated with an optimal vortex location, determined from Fourier components of the Berry curvature~\cite{gunawardana2024}.

\textit{Acknowledgements:} We thank Pedro Fittipaldi de Castro, Ajit Srivastava, and David Vanderbilt for discussions on this topic. We also thank Thomas Christensen for bringing to our attention OAL insulators beyond symmetry representations. This work was supported by the startup funds from Emory University and the Laboratory Directed Research and Development program at Sandia National Laboratories. This work was performed, in part, at the Center for Integrated Nanotechnologies, an Office of Science User Facility operated for the U.S. Department of Energy (DOE) Office of Science. Sandia National Laboratories is a multimission laboratory managed and operated by National Technology \& Engineering Solutions of Sandia, LLC, a wholly owned subsidiary of Honeywell International, Inc., for the U.S.\ DOE's National Nuclear Security Administration under contract DE-NA-0003525. The views expressed in the article do not necessarily represent the views of the U.S.\ DOE or the United States Government.

\bibliography{refs.bib}

\begin{thebibliography}{110}%
\makeatletter
\providecommand \@ifxundefined [1]{%
 \@ifx{#1\undefined}
}%
\providecommand \@ifnum [1]{%
 \ifnum #1\expandafter \@firstoftwo
 \else \expandafter \@secondoftwo
 \fi
}%
\providecommand \@ifx [1]{%
 \ifx #1\expandafter \@firstoftwo
 \else \expandafter \@secondoftwo
 \fi
}%
\providecommand \natexlab [1]{#1}%
\providecommand \enquote  [1]{``#1''}%
\providecommand \bibnamefont  [1]{#1}%
\providecommand \bibfnamefont [1]{#1}%
\providecommand \citenamefont [1]{#1}%
\providecommand \href@noop [0]{\@secondoftwo}%
\providecommand \href [0]{\begingroup \@sanitize@url \@href}%
\providecommand \@href[1]{\@@startlink{#1}\@@href}%
\providecommand \@@href[1]{\endgroup#1\@@endlink}%
\providecommand \@sanitize@url [0]{\catcode `\\12\catcode `\$12\catcode `\&12\catcode `\#12\catcode `\^12\catcode `\_12\catcode `\%12\relax}%
\providecommand \@@startlink[1]{}%
\providecommand \@@endlink[0]{}%
\providecommand \url  [0]{\begingroup\@sanitize@url \@url }%
\providecommand \@url [1]{\endgroup\@href {#1}{\urlprefix }}%
\providecommand \urlprefix  [0]{URL }%
\providecommand \Eprint [0]{\href }%
\providecommand \doibase [0]{https://doi.org/}%
\providecommand \selectlanguage [0]{\@gobble}%
\providecommand \bibinfo  [0]{\@secondoftwo}%
\providecommand \bibfield  [0]{\@secondoftwo}%
\providecommand \translation [1]{[#1]}%
\providecommand \BibitemOpen [0]{}%
\providecommand \bibitemStop [0]{}%
\providecommand \bibitemNoStop [0]{.\EOS\space}%
\providecommand \EOS [0]{\spacefactor3000\relax}%
\providecommand \BibitemShut  [1]{\csname bibitem#1\endcsname}%
\let\auto@bib@innerbib\@empty
\bibitem [{\citenamefont {Boys}(1960)}]{boys1960}%
  \BibitemOpen
  \bibfield  {author} {\bibinfo {author} {\bibfnamefont {S.~F.}\ \bibnamefont {Boys}},\ }\href {https://doi.org/10.1103/RevModPhys.32.296} {\bibfield  {journal} {\bibinfo  {journal} {Rev. Mod. Phys.}\ }\textbf {\bibinfo {volume} {32}},\ \bibinfo {pages} {296} (\bibinfo {year} {1960})}\BibitemShut {NoStop}%
\bibitem [{\citenamefont {Foster}\ and\ \citenamefont {Boys}(1960)}]{foster1960}%
  \BibitemOpen
  \bibfield  {author} {\bibinfo {author} {\bibfnamefont {J.~M.}\ \bibnamefont {Foster}}\ and\ \bibinfo {author} {\bibfnamefont {S.~F.}\ \bibnamefont {Boys}},\ }\href {https://doi.org/10.1103/RevModPhys.32.300} {\bibfield  {journal} {\bibinfo  {journal} {Rev. Mod. Phys.}\ }\textbf {\bibinfo {volume} {32}},\ \bibinfo {pages} {300} (\bibinfo {year} {1960})}\BibitemShut {NoStop}%
\bibitem [{\citenamefont {Edmiston}\ and\ \citenamefont {Ruedenberg}(1963)}]{edmiston1963}%
  \BibitemOpen
  \bibfield  {author} {\bibinfo {author} {\bibfnamefont {C.}~\bibnamefont {Edmiston}}\ and\ \bibinfo {author} {\bibfnamefont {K.}~\bibnamefont {Ruedenberg}},\ }\href {https://doi.org/10.1103/RevModPhys.35.457} {\bibfield  {journal} {\bibinfo  {journal} {Rev. Mod. Phys.}\ }\textbf {\bibinfo {volume} {35}},\ \bibinfo {pages} {457} (\bibinfo {year} {1963})}\BibitemShut {NoStop}%
\bibitem [{\citenamefont {Zak}(1989)}]{zak1989}%
  \BibitemOpen
  \bibfield  {author} {\bibinfo {author} {\bibfnamefont {J.}~\bibnamefont {Zak}},\ }\href {https://doi.org/10.1103/PhysRevLett.62.2747} {\bibfield  {journal} {\bibinfo  {journal} {Phys. Rev. Lett.}\ }\textbf {\bibinfo {volume} {62}},\ \bibinfo {pages} {2747} (\bibinfo {year} {1989})}\BibitemShut {NoStop}%
\bibitem [{\citenamefont {King-Smith}\ and\ \citenamefont {Vanderbilt}(1993)}]{king-smith_1993}%
  \BibitemOpen
  \bibfield  {author} {\bibinfo {author} {\bibfnamefont {R.~D.}\ \bibnamefont {King-Smith}}\ and\ \bibinfo {author} {\bibfnamefont {D.}~\bibnamefont {Vanderbilt}},\ }\href {https://doi.org/10.1103/PhysRevB.47.1651} {\bibfield  {journal} {\bibinfo  {journal} {Phys. Rev. B}\ }\textbf {\bibinfo {volume} {47}},\ \bibinfo {pages} {1651} (\bibinfo {year} {1993})}\BibitemShut {NoStop}%
\bibitem [{\citenamefont {Qi}\ \emph {et~al.}(2008)\citenamefont {Qi}, \citenamefont {Hughes},\ and\ \citenamefont {Zhang}}]{qi2008}%
  \BibitemOpen
  \bibfield  {author} {\bibinfo {author} {\bibfnamefont {X.-L.}\ \bibnamefont {Qi}}, \bibinfo {author} {\bibfnamefont {T.~L.}\ \bibnamefont {Hughes}},\ and\ \bibinfo {author} {\bibfnamefont {S.-C.}\ \bibnamefont {Zhang}},\ }\href {https://doi.org/10.1103/PhysRevB.78.195424} {\bibfield  {journal} {\bibinfo  {journal} {Phys. Rev. B}\ }\textbf {\bibinfo {volume} {78}},\ \bibinfo {pages} {195424} (\bibinfo {year} {2008})}\BibitemShut {NoStop}%
\bibitem [{\citenamefont {Kitaev}(2009)}]{kitaev2009}%
  \BibitemOpen
  \bibfield  {author} {\bibinfo {author} {\bibfnamefont {A.}~\bibnamefont {Kitaev}},\ }\href {https://doi.org/10.1063/1.3149495} {\bibfield  {journal} {\bibinfo  {journal} {AIP Conference Proceedings}\ }\textbf {\bibinfo {volume} {1134}},\ \bibinfo {pages} {22} (\bibinfo {year} {2009})}\BibitemShut {NoStop}%
\bibitem [{\citenamefont {Ryu}\ \emph {et~al.}(2010)\citenamefont {Ryu}, \citenamefont {Schnyder}, \citenamefont {Furusaki},\ and\ \citenamefont {Ludwig}}]{ryu2010}%
  \BibitemOpen
  \bibfield  {author} {\bibinfo {author} {\bibfnamefont {S.}~\bibnamefont {Ryu}}, \bibinfo {author} {\bibfnamefont {A.~P.}\ \bibnamefont {Schnyder}}, \bibinfo {author} {\bibfnamefont {A.}~\bibnamefont {Furusaki}},\ and\ \bibinfo {author} {\bibfnamefont {A.~W.~W.}\ \bibnamefont {Ludwig}},\ }\href {https://doi.org/10.1088/1367-2630/12/6/065010} {\bibfield  {journal} {\bibinfo  {journal} {New Journal of Physics}\ }\textbf {\bibinfo {volume} {12}},\ \bibinfo {pages} {065010} (\bibinfo {year} {2010})}\BibitemShut {NoStop}%
\bibitem [{\citenamefont {Hasan}\ and\ \citenamefont {Kane}(2010)}]{hasan2010}%
  \BibitemOpen
  \bibfield  {author} {\bibinfo {author} {\bibfnamefont {M.~Z.}\ \bibnamefont {Hasan}}\ and\ \bibinfo {author} {\bibfnamefont {C.~L.}\ \bibnamefont {Kane}},\ }\href {https://doi.org/10.1103/RevModPhys.82.3045} {\bibfield  {journal} {\bibinfo  {journal} {Rev. Mod. Phys.}\ }\textbf {\bibinfo {volume} {82}},\ \bibinfo {pages} {3045} (\bibinfo {year} {2010})}\BibitemShut {NoStop}%
\bibitem [{\citenamefont {Qi}\ and\ \citenamefont {Zhang}(2011)}]{qi2011}%
  \BibitemOpen
  \bibfield  {author} {\bibinfo {author} {\bibfnamefont {X.-L.}\ \bibnamefont {Qi}}\ and\ \bibinfo {author} {\bibfnamefont {S.-C.}\ \bibnamefont {Zhang}},\ }\href {https://doi.org/10.1103/RevModPhys.83.1057} {\bibfield  {journal} {\bibinfo  {journal} {Rev. Mod. Phys.}\ }\textbf {\bibinfo {volume} {83}},\ \bibinfo {pages} {1057} (\bibinfo {year} {2011})}\BibitemShut {NoStop}%
\bibitem [{\citenamefont {Marzari}\ \emph {et~al.}(2012)\citenamefont {Marzari}, \citenamefont {Mostofi}, \citenamefont {Yates}, \citenamefont {Souza},\ and\ \citenamefont {Vanderbilt}}]{marzari2012}%
  \BibitemOpen
  \bibfield  {author} {\bibinfo {author} {\bibfnamefont {N.}~\bibnamefont {Marzari}}, \bibinfo {author} {\bibfnamefont {A.~A.}\ \bibnamefont {Mostofi}}, \bibinfo {author} {\bibfnamefont {J.~R.}\ \bibnamefont {Yates}}, \bibinfo {author} {\bibfnamefont {I.}~\bibnamefont {Souza}},\ and\ \bibinfo {author} {\bibfnamefont {D.}~\bibnamefont {Vanderbilt}},\ }\href {https://doi.org/10.1103/RevModPhys.84.1419} {\bibfield  {journal} {\bibinfo  {journal} {Rev. Mod. Phys.}\ }\textbf {\bibinfo {volume} {84}},\ \bibinfo {pages} {1419} (\bibinfo {year} {2012})}\BibitemShut {NoStop}%
\bibitem [{\citenamefont {R\"osner}\ \emph {et~al.}(2015)\citenamefont {R\"osner}, \citenamefont {\ifmmode \mbox{\c{S}}\else \c{S}\fi{}a\ifmmode \mbox{\c{s}}\else \c{s}\fi{}\ifmmode \imath \else \i \fi{}o\ifmmode~\breve{g}\else \u{g}\fi{}lu}, \citenamefont {Friedrich}, \citenamefont {Bl\"ugel},\ and\ \citenamefont {Wehling}}]{rosner2015}%
  \BibitemOpen
  \bibfield  {author} {\bibinfo {author} {\bibfnamefont {M.}~\bibnamefont {R\"osner}}, \bibinfo {author} {\bibfnamefont {E.}~\bibnamefont {\ifmmode \mbox{\c{S}}\else \c{S}\fi{}a\ifmmode \mbox{\c{s}}\else \c{s}\fi{}\ifmmode \imath \else \i \fi{}o\ifmmode~\breve{g}\else \u{g}\fi{}lu}}, \bibinfo {author} {\bibfnamefont {C.}~\bibnamefont {Friedrich}}, \bibinfo {author} {\bibfnamefont {S.}~\bibnamefont {Bl\"ugel}},\ and\ \bibinfo {author} {\bibfnamefont {T.~O.}\ \bibnamefont {Wehling}},\ }\href {https://doi.org/10.1103/PhysRevB.92.085102} {\bibfield  {journal} {\bibinfo  {journal} {Phys. Rev. B}\ }\textbf {\bibinfo {volume} {92}},\ \bibinfo {pages} {085102} (\bibinfo {year} {2015})}\BibitemShut {NoStop}%
\bibitem [{\citenamefont {Koshino}\ \emph {et~al.}(2018)\citenamefont {Koshino}, \citenamefont {Yuan}, \citenamefont {Koretsune}, \citenamefont {Ochi}, \citenamefont {Kuroki},\ and\ \citenamefont {Fu}}]{koshino2018}%
  \BibitemOpen
  \bibfield  {author} {\bibinfo {author} {\bibfnamefont {M.}~\bibnamefont {Koshino}}, \bibinfo {author} {\bibfnamefont {N.~F.~Q.}\ \bibnamefont {Yuan}}, \bibinfo {author} {\bibfnamefont {T.}~\bibnamefont {Koretsune}}, \bibinfo {author} {\bibfnamefont {M.}~\bibnamefont {Ochi}}, \bibinfo {author} {\bibfnamefont {K.}~\bibnamefont {Kuroki}},\ and\ \bibinfo {author} {\bibfnamefont {L.}~\bibnamefont {Fu}},\ }\href {https://doi.org/10.1103/PhysRevX.8.031087} {\bibfield  {journal} {\bibinfo  {journal} {Phys. Rev. X}\ }\textbf {\bibinfo {volume} {8}},\ \bibinfo {pages} {031087} (\bibinfo {year} {2018})}\BibitemShut {NoStop}%
\bibitem [{\citenamefont {Xie}\ \emph {et~al.}(2024)\citenamefont {Xie}, \citenamefont {Fang}, \citenamefont {Chen}, \citenamefont {Cano},\ and\ \citenamefont {Si}}]{xie2024}%
  \BibitemOpen
  \bibfield  {author} {\bibinfo {author} {\bibfnamefont {F.}~\bibnamefont {Xie}}, \bibinfo {author} {\bibfnamefont {Y.}~\bibnamefont {Fang}}, \bibinfo {author} {\bibfnamefont {L.}~\bibnamefont {Chen}}, \bibinfo {author} {\bibfnamefont {J.}~\bibnamefont {Cano}},\ and\ \bibinfo {author} {\bibfnamefont {Q.}~\bibnamefont {Si}},\ }\href {https://arxiv.org/abs/2407.08920} {\bibinfo {title} {Chern bands' optimally localized wannier functions and fractional chern insulators}} (\bibinfo {year} {2024}),\ \Eprint {https://arxiv.org/abs/2407.08920} {arXiv:2407.08920 [cond-mat.mes-hall]} \BibitemShut {NoStop}%
\bibitem [{\citenamefont {Resta}(1998)}]{resta1998}%
  \BibitemOpen
  \bibfield  {author} {\bibinfo {author} {\bibfnamefont {R.}~\bibnamefont {Resta}},\ }\href {https://doi.org/10.1103/PhysRevLett.80.1800} {\bibfield  {journal} {\bibinfo  {journal} {Phys. Rev. Lett.}\ }\textbf {\bibinfo {volume} {80}},\ \bibinfo {pages} {1800} (\bibinfo {year} {1998})}\BibitemShut {NoStop}%
\bibitem [{\citenamefont {Kivelson}(1982)}]{kielson1982}%
  \BibitemOpen
  \bibfield  {author} {\bibinfo {author} {\bibfnamefont {S.}~\bibnamefont {Kivelson}},\ }\href {https://doi.org/10.1103/PhysRevB.26.4269} {\bibfield  {journal} {\bibinfo  {journal} {Phys. Rev. B}\ }\textbf {\bibinfo {volume} {26}},\ \bibinfo {pages} {4269} (\bibinfo {year} {1982})}\BibitemShut {NoStop}%
\bibitem [{\citenamefont {Resta}(1994)}]{resta1994}%
  \BibitemOpen
  \bibfield  {author} {\bibinfo {author} {\bibfnamefont {R.}~\bibnamefont {Resta}},\ }\href {https://doi.org/10.1103/RevModPhys.66.899} {\bibfield  {journal} {\bibinfo  {journal} {Rev. Mod. Phys.}\ }\textbf {\bibinfo {volume} {66}},\ \bibinfo {pages} {899} (\bibinfo {year} {1994})}\BibitemShut {NoStop}%
\bibitem [{\citenamefont {Vanderbilt}(2018)}]{vanderbilt2018book}%
  \BibitemOpen
  \bibfield  {author} {\bibinfo {author} {\bibfnamefont {D.}~\bibnamefont {Vanderbilt}},\ }\href {https://books.google.com/books?id=485FtgEACAAJ} {\emph {\bibinfo {title} {Berry Phases in Electronic Structure Theory: Electric Polarization, Orbital Magnetization and Topological Insulators}}},\ Titolo collana\ (\bibinfo  {publisher} {Cambridge University Press},\ \bibinfo {year} {2018})\BibitemShut {NoStop}%
\bibitem [{\citenamefont {Hughes}\ \emph {et~al.}(2011)\citenamefont {Hughes}, \citenamefont {Prodan},\ and\ \citenamefont {Bernevig}}]{hughes2011}%
  \BibitemOpen
  \bibfield  {author} {\bibinfo {author} {\bibfnamefont {T.~L.}\ \bibnamefont {Hughes}}, \bibinfo {author} {\bibfnamefont {E.}~\bibnamefont {Prodan}},\ and\ \bibinfo {author} {\bibfnamefont {B.~A.}\ \bibnamefont {Bernevig}},\ }\href {https://doi.org/10.1103/PhysRevB.83.245132} {\bibfield  {journal} {\bibinfo  {journal} {Phys. Rev. B}\ }\textbf {\bibinfo {volume} {83}},\ \bibinfo {pages} {245132} (\bibinfo {year} {2011})}\BibitemShut {NoStop}%
\bibitem [{\citenamefont {Bradlyn}\ \emph {et~al.}(2017)\citenamefont {Bradlyn}, \citenamefont {Elcoro}, \citenamefont {Cano}, \citenamefont {Vergniory}, \citenamefont {Wang}, \citenamefont {Felser}, \citenamefont {Aroyo},\ and\ \citenamefont {Bernevig}}]{bradlyn2017topological}%
  \BibitemOpen
  \bibfield  {author} {\bibinfo {author} {\bibfnamefont {B.}~\bibnamefont {Bradlyn}}, \bibinfo {author} {\bibfnamefont {L.}~\bibnamefont {Elcoro}}, \bibinfo {author} {\bibfnamefont {J.}~\bibnamefont {Cano}}, \bibinfo {author} {\bibfnamefont {M.~G.}\ \bibnamefont {Vergniory}}, \bibinfo {author} {\bibfnamefont {Z.}~\bibnamefont {Wang}}, \bibinfo {author} {\bibfnamefont {C.}~\bibnamefont {Felser}}, \bibinfo {author} {\bibfnamefont {M.~I.}\ \bibnamefont {Aroyo}},\ and\ \bibinfo {author} {\bibfnamefont {B.~A.}\ \bibnamefont {Bernevig}},\ }\href {https://doi.org/10.1038/nature23268} {\bibfield  {journal} {\bibinfo  {journal} {Nature}\ }\textbf {\bibinfo {volume} {547}},\ \bibinfo {pages} {298} (\bibinfo {year} {2017})}\BibitemShut {NoStop}%
\bibitem [{\citenamefont {Cano}\ \emph {et~al.}(2018)\citenamefont {Cano}, \citenamefont {Bradlyn}, \citenamefont {Wang}, \citenamefont {Elcoro}, \citenamefont {Vergniory}, \citenamefont {Felser}, \citenamefont {Aroyo},\ and\ \citenamefont {Bernevig}}]{cano2018}%
  \BibitemOpen
  \bibfield  {author} {\bibinfo {author} {\bibfnamefont {J.}~\bibnamefont {Cano}}, \bibinfo {author} {\bibfnamefont {B.}~\bibnamefont {Bradlyn}}, \bibinfo {author} {\bibfnamefont {Z.}~\bibnamefont {Wang}}, \bibinfo {author} {\bibfnamefont {L.}~\bibnamefont {Elcoro}}, \bibinfo {author} {\bibfnamefont {M.~G.}\ \bibnamefont {Vergniory}}, \bibinfo {author} {\bibfnamefont {C.}~\bibnamefont {Felser}}, \bibinfo {author} {\bibfnamefont {M.~I.}\ \bibnamefont {Aroyo}},\ and\ \bibinfo {author} {\bibfnamefont {B.~A.}\ \bibnamefont {Bernevig}},\ }\href {https://doi.org/10.1103/PhysRevB.97.035139} {\bibfield  {journal} {\bibinfo  {journal} {Physical Review B}\ }\textbf {\bibinfo {volume} {97}},\ \bibinfo {pages} {035139} (\bibinfo {year} {2018})}\BibitemShut {NoStop}%
\bibitem [{\citenamefont {Po}\ \emph {et~al.}(2017)\citenamefont {Po}, \citenamefont {Vishwanath},\ and\ \citenamefont {Watanabe}}]{po2017symmetry-based}%
  \BibitemOpen
  \bibfield  {author} {\bibinfo {author} {\bibfnamefont {H.~C.}\ \bibnamefont {Po}}, \bibinfo {author} {\bibfnamefont {A.}~\bibnamefont {Vishwanath}},\ and\ \bibinfo {author} {\bibfnamefont {H.}~\bibnamefont {Watanabe}},\ }\href {https://doi.org/10.1038/s41467-017-00133-2} {\bibfield  {journal} {\bibinfo  {journal} {Nature Communications}\ }\textbf {\bibinfo {volume} {8}},\ \bibinfo {pages} {50} (\bibinfo {year} {2017})}\BibitemShut {NoStop}%
\bibitem [{\citenamefont {Kruthoff}\ \emph {et~al.}(2017)\citenamefont {Kruthoff}, \citenamefont {de~Boer}, \citenamefont {van Wezel}, \citenamefont {Kane},\ and\ \citenamefont {Slager}}]{kruthoff2017}%
  \BibitemOpen
  \bibfield  {author} {\bibinfo {author} {\bibfnamefont {J.}~\bibnamefont {Kruthoff}}, \bibinfo {author} {\bibfnamefont {J.}~\bibnamefont {de~Boer}}, \bibinfo {author} {\bibfnamefont {J.}~\bibnamefont {van Wezel}}, \bibinfo {author} {\bibfnamefont {C.~L.}\ \bibnamefont {Kane}},\ and\ \bibinfo {author} {\bibfnamefont {R.-J.}\ \bibnamefont {Slager}},\ }\href {https://doi.org/10.1103/PhysRevX.7.041069} {\bibfield  {journal} {\bibinfo  {journal} {Phys. Rev. X}\ }\textbf {\bibinfo {volume} {7}},\ \bibinfo {pages} {041069} (\bibinfo {year} {2017})}\BibitemShut {NoStop}%
\bibitem [{\citenamefont {Song}\ \emph {et~al.}(2018)\citenamefont {Song}, \citenamefont {Zhang}, \citenamefont {Fang},\ and\ \citenamefont {Fang}}]{song2018}%
  \BibitemOpen
  \bibfield  {author} {\bibinfo {author} {\bibfnamefont {Z.}~\bibnamefont {Song}}, \bibinfo {author} {\bibfnamefont {T.}~\bibnamefont {Zhang}}, \bibinfo {author} {\bibfnamefont {Z.}~\bibnamefont {Fang}},\ and\ \bibinfo {author} {\bibfnamefont {C.}~\bibnamefont {Fang}},\ }\href {https://doi.org/10.1038/s41467-018-06010-w} {\bibfield  {journal} {\bibinfo  {journal} {Nature Communications}\ }\textbf {\bibinfo {volume} {9}},\ \bibinfo {pages} {3530} (\bibinfo {year} {2018})}\BibitemShut {NoStop}%
\bibitem [{\citenamefont {Benalcazar}\ \emph {et~al.}(2019)\citenamefont {Benalcazar}, \citenamefont {Li},\ and\ \citenamefont {Hughes}}]{benalcazar2019}%
  \BibitemOpen
  \bibfield  {author} {\bibinfo {author} {\bibfnamefont {W.~A.}\ \bibnamefont {Benalcazar}}, \bibinfo {author} {\bibfnamefont {T.}~\bibnamefont {Li}},\ and\ \bibinfo {author} {\bibfnamefont {T.~L.}\ \bibnamefont {Hughes}},\ }\href {https://doi.org/10.1103/PhysRevB.99.245151} {\bibfield  {journal} {\bibinfo  {journal} {Phys. Rev. B}\ }\textbf {\bibinfo {volume} {99}},\ \bibinfo {pages} {245151} (\bibinfo {year} {2019})}\BibitemShut {NoStop}%
\bibitem [{\citenamefont {Vergniory}\ \emph {et~al.}(2019)\citenamefont {Vergniory}, \citenamefont {Elcoro}, \citenamefont {Felser}, \citenamefont {Regnault}, \citenamefont {Bernevig},\ and\ \citenamefont {Wang}}]{catalogue1}%
  \BibitemOpen
  \bibfield  {author} {\bibinfo {author} {\bibfnamefont {M.~G.}\ \bibnamefont {Vergniory}}, \bibinfo {author} {\bibfnamefont {L.}~\bibnamefont {Elcoro}}, \bibinfo {author} {\bibfnamefont {C.}~\bibnamefont {Felser}}, \bibinfo {author} {\bibfnamefont {N.}~\bibnamefont {Regnault}}, \bibinfo {author} {\bibfnamefont {B.~A.}\ \bibnamefont {Bernevig}},\ and\ \bibinfo {author} {\bibfnamefont {Z.}~\bibnamefont {Wang}},\ }\href {https://doi.org/10.1038/s41586-019-0954-4} {\bibfield  {journal} {\bibinfo  {journal} {Nature}\ }\textbf {\bibinfo {volume} {566}},\ \bibinfo {pages} {480} (\bibinfo {year} {2019})}\BibitemShut {NoStop}%
\bibitem [{\citenamefont {Zhang}\ \emph {et~al.}(2019)\citenamefont {Zhang}, \citenamefont {Jiang}, \citenamefont {Song}, \citenamefont {Huang}, \citenamefont {He}, \citenamefont {Fang}, \citenamefont {Weng},\ and\ \citenamefont {Fang}}]{catalogue2}%
  \BibitemOpen
  \bibfield  {author} {\bibinfo {author} {\bibfnamefont {T.}~\bibnamefont {Zhang}}, \bibinfo {author} {\bibfnamefont {Y.}~\bibnamefont {Jiang}}, \bibinfo {author} {\bibfnamefont {Z.}~\bibnamefont {Song}}, \bibinfo {author} {\bibfnamefont {H.}~\bibnamefont {Huang}}, \bibinfo {author} {\bibfnamefont {Y.}~\bibnamefont {He}}, \bibinfo {author} {\bibfnamefont {Z.}~\bibnamefont {Fang}}, \bibinfo {author} {\bibfnamefont {H.}~\bibnamefont {Weng}},\ and\ \bibinfo {author} {\bibfnamefont {C.}~\bibnamefont {Fang}},\ }\href {https://doi.org/10.1038/s41586-019-0944-6} {\bibfield  {journal} {\bibinfo  {journal} {Nature}\ }\textbf {\bibinfo {volume} {566}},\ \bibinfo {pages} {475} (\bibinfo {year} {2019})}\BibitemShut {NoStop}%
\bibitem [{\citenamefont {Cano}\ \emph {et~al.}(2022)\citenamefont {Cano}, \citenamefont {Elcoro}, \citenamefont {Aroyo}, \citenamefont {Bernevig},\ and\ \citenamefont {Bradlyn}}]{cano2022}%
  \BibitemOpen
  \bibfield  {author} {\bibinfo {author} {\bibfnamefont {J.}~\bibnamefont {Cano}}, \bibinfo {author} {\bibfnamefont {L.}~\bibnamefont {Elcoro}}, \bibinfo {author} {\bibfnamefont {M.~I.}\ \bibnamefont {Aroyo}}, \bibinfo {author} {\bibfnamefont {B.~A.}\ \bibnamefont {Bernevig}},\ and\ \bibinfo {author} {\bibfnamefont {B.}~\bibnamefont {Bradlyn}},\ }\href {https://doi.org/10.1103/PhysRevB.105.125115} {\bibfield  {journal} {\bibinfo  {journal} {Phys. Rev. B}\ }\textbf {\bibinfo {volume} {105}},\ \bibinfo {pages} {125115} (\bibinfo {year} {2022})}\BibitemShut {NoStop}%
\bibitem [{\citenamefont {Fang}\ and\ \citenamefont {Fu}(2019)}]{rotationAnomaly}%
  \BibitemOpen
  \bibfield  {author} {\bibinfo {author} {\bibfnamefont {C.}~\bibnamefont {Fang}}\ and\ \bibinfo {author} {\bibfnamefont {L.}~\bibnamefont {Fu}},\ }\href {https://doi.org/10.1126/sciadv.aat2374} {\bibfield  {journal} {\bibinfo  {journal} {Science Advances}\ }\textbf {\bibinfo {volume} {5}},\ \bibinfo {pages} {eaat2374} (\bibinfo {year} {2019})},\ \Eprint {https://arxiv.org/abs/https://www.science.org/doi/pdf/10.1126/sciadv.aat2374} {https://www.science.org/doi/pdf/10.1126/sciadv.aat2374} \BibitemShut {NoStop}%
\bibitem [{\citenamefont {Benalcazar}\ \emph {et~al.}(2017{\natexlab{a}})\citenamefont {Benalcazar}, \citenamefont {Bernevig},\ and\ \citenamefont {Hughes}}]{benalcazar2017_science}%
  \BibitemOpen
  \bibfield  {author} {\bibinfo {author} {\bibfnamefont {W.~A.}\ \bibnamefont {Benalcazar}}, \bibinfo {author} {\bibfnamefont {B.~A.}\ \bibnamefont {Bernevig}},\ and\ \bibinfo {author} {\bibfnamefont {T.~L.}\ \bibnamefont {Hughes}},\ }\href {https://doi.org/10.1126/science.aah6442} {\bibfield  {journal} {\bibinfo  {journal} {Science}\ }\textbf {\bibinfo {volume} {357}},\ \bibinfo {pages} {61} (\bibinfo {year} {2017}{\natexlab{a}})},\ \Eprint {https://arxiv.org/abs/https://www.science.org/doi/pdf/10.1126/science.aah6442} {https://www.science.org/doi/pdf/10.1126/science.aah6442} \BibitemShut {NoStop}%
\bibitem [{\citenamefont {Benalcazar}\ \emph {et~al.}(2017{\natexlab{b}})\citenamefont {Benalcazar}, \citenamefont {Bernevig},\ and\ \citenamefont {Hughes}}]{benalcazar2017}%
  \BibitemOpen
  \bibfield  {author} {\bibinfo {author} {\bibfnamefont {W.~A.}\ \bibnamefont {Benalcazar}}, \bibinfo {author} {\bibfnamefont {B.~A.}\ \bibnamefont {Bernevig}},\ and\ \bibinfo {author} {\bibfnamefont {T.~L.}\ \bibnamefont {Hughes}},\ }\href {https://doi.org/10.1103/PhysRevB.96.245115} {\bibfield  {journal} {\bibinfo  {journal} {Phys. Rev. B}\ }\textbf {\bibinfo {volume} {96}},\ \bibinfo {pages} {245115} (\bibinfo {year} {2017}{\natexlab{b}})}\BibitemShut {NoStop}%
\bibitem [{\citenamefont {Khalaf}\ \emph {et~al.}(2021)\citenamefont {Khalaf}, \citenamefont {Benalcazar}, \citenamefont {Hughes},\ and\ \citenamefont {Queiroz}}]{botp}%
  \BibitemOpen
  \bibfield  {author} {\bibinfo {author} {\bibfnamefont {E.}~\bibnamefont {Khalaf}}, \bibinfo {author} {\bibfnamefont {W.~A.}\ \bibnamefont {Benalcazar}}, \bibinfo {author} {\bibfnamefont {T.~L.}\ \bibnamefont {Hughes}},\ and\ \bibinfo {author} {\bibfnamefont {R.}~\bibnamefont {Queiroz}},\ }\href {https://doi.org/10.1103/PhysRevResearch.3.013239} {\bibfield  {journal} {\bibinfo  {journal} {Phys. Rev. Res.}\ }\textbf {\bibinfo {volume} {3}},\ \bibinfo {pages} {013239} (\bibinfo {year} {2021})}\BibitemShut {NoStop}%
\bibitem [{\citenamefont {Schindler}\ \emph {et~al.}(2019)\citenamefont {Schindler}, \citenamefont {Brzezi\ifmmode~\acute{n}\else \'{n}\fi{}ska}, \citenamefont {Benalcazar}, \citenamefont {Iraola}, \citenamefont {Bouhon}, \citenamefont {Tsirkin}, \citenamefont {Vergniory},\ and\ \citenamefont {Neupert}}]{schindler2019}%
  \BibitemOpen
  \bibfield  {author} {\bibinfo {author} {\bibfnamefont {F.}~\bibnamefont {Schindler}}, \bibinfo {author} {\bibfnamefont {M.}~\bibnamefont {Brzezi\ifmmode~\acute{n}\else \'{n}\fi{}ska}}, \bibinfo {author} {\bibfnamefont {W.~A.}\ \bibnamefont {Benalcazar}}, \bibinfo {author} {\bibfnamefont {M.}~\bibnamefont {Iraola}}, \bibinfo {author} {\bibfnamefont {A.}~\bibnamefont {Bouhon}}, \bibinfo {author} {\bibfnamefont {S.~S.}\ \bibnamefont {Tsirkin}}, \bibinfo {author} {\bibfnamefont {M.~G.}\ \bibnamefont {Vergniory}},\ and\ \bibinfo {author} {\bibfnamefont {T.}~\bibnamefont {Neupert}},\ }\href {https://doi.org/10.1103/PhysRevResearch.1.033074} {\bibfield  {journal} {\bibinfo  {journal} {Phys. Rev. Res.}\ }\textbf {\bibinfo {volume} {1}},\ \bibinfo {pages} {033074} (\bibinfo {year} {2019})}\BibitemShut {NoStop}%
\bibitem [{\citenamefont {Sgiarovello}\ \emph {et~al.}(2001)\citenamefont {Sgiarovello}, \citenamefont {Peressi},\ and\ \citenamefont {Resta}}]{sgiarovello2001}%
  \BibitemOpen
  \bibfield  {author} {\bibinfo {author} {\bibfnamefont {C.}~\bibnamefont {Sgiarovello}}, \bibinfo {author} {\bibfnamefont {M.}~\bibnamefont {Peressi}},\ and\ \bibinfo {author} {\bibfnamefont {R.}~\bibnamefont {Resta}},\ }\href {https://doi.org/10.1103/PhysRevB.64.115202} {\bibfield  {journal} {\bibinfo  {journal} {Phys. Rev. B}\ }\textbf {\bibinfo {volume} {64}},\ \bibinfo {pages} {115202} (\bibinfo {year} {2001})}\BibitemShut {NoStop}%
\bibitem [{\citenamefont {Alexandradinata}\ \emph {et~al.}(2014)\citenamefont {Alexandradinata}, \citenamefont {Dai},\ and\ \citenamefont {Bernevig}}]{alexandradinata2014}%
  \BibitemOpen
  \bibfield  {author} {\bibinfo {author} {\bibfnamefont {A.}~\bibnamefont {Alexandradinata}}, \bibinfo {author} {\bibfnamefont {X.}~\bibnamefont {Dai}},\ and\ \bibinfo {author} {\bibfnamefont {B.~A.}\ \bibnamefont {Bernevig}},\ }\href {https://doi.org/10.1103/PhysRevB.89.155114} {\bibfield  {journal} {\bibinfo  {journal} {Phys. Rev. B}\ }\textbf {\bibinfo {volume} {89}},\ \bibinfo {pages} {155114} (\bibinfo {year} {2014})}\BibitemShut {NoStop}%
\bibitem [{\citenamefont {Zeng}\ \emph {et~al.}(2023)\citenamefont {Zeng}, \citenamefont {Duan},\ and\ \citenamefont {Huang}}]{zeng2023}%
  \BibitemOpen
  \bibfield  {author} {\bibinfo {author} {\bibfnamefont {H.}~\bibnamefont {Zeng}}, \bibinfo {author} {\bibfnamefont {W.}~\bibnamefont {Duan}},\ and\ \bibinfo {author} {\bibfnamefont {H.}~\bibnamefont {Huang}},\ }\href {https://doi.org/10.1103/PhysRevResearch.5.L042003} {\bibfield  {journal} {\bibinfo  {journal} {Phys. Rev. Res.}\ }\textbf {\bibinfo {volume} {5}},\ \bibinfo {pages} {L042003} (\bibinfo {year} {2023})}\BibitemShut {NoStop}%
\bibitem [{\citenamefont {Marzari}\ and\ \citenamefont {Vanderbilt}(1997)}]{marzari1997}%
  \BibitemOpen
  \bibfield  {author} {\bibinfo {author} {\bibfnamefont {N.}~\bibnamefont {Marzari}}\ and\ \bibinfo {author} {\bibfnamefont {D.}~\bibnamefont {Vanderbilt}},\ }\href {https://doi.org/10.1103/PhysRevB.56.12847} {\bibfield  {journal} {\bibinfo  {journal} {Phys. Rev. B}\ }\textbf {\bibinfo {volume} {56}},\ \bibinfo {pages} {12847} (\bibinfo {year} {1997})}\BibitemShut {NoStop}%
\bibitem [{\citenamefont {Souza}\ \emph {et~al.}(2001)\citenamefont {Souza}, \citenamefont {Marzari},\ and\ \citenamefont {Vanderbilt}}]{souza2001}%
  \BibitemOpen
  \bibfield  {author} {\bibinfo {author} {\bibfnamefont {I.}~\bibnamefont {Souza}}, \bibinfo {author} {\bibfnamefont {N.}~\bibnamefont {Marzari}},\ and\ \bibinfo {author} {\bibfnamefont {D.}~\bibnamefont {Vanderbilt}},\ }\href {https://doi.org/10.1103/PhysRevB.65.035109} {\bibfield  {journal} {\bibinfo  {journal} {Phys. Rev. B}\ }\textbf {\bibinfo {volume} {65}},\ \bibinfo {pages} {035109} (\bibinfo {year} {2001})}\BibitemShut {NoStop}%
\bibitem [{\citenamefont {Mostofi}\ \emph {et~al.}(2008)\citenamefont {Mostofi}, \citenamefont {Yates}, \citenamefont {Lee}, \citenamefont {Souza}, \citenamefont {Vanderbilt},\ and\ \citenamefont {Marzari}}]{wannier90}%
  \BibitemOpen
  \bibfield  {author} {\bibinfo {author} {\bibfnamefont {A.~A.}\ \bibnamefont {Mostofi}}, \bibinfo {author} {\bibfnamefont {J.~R.}\ \bibnamefont {Yates}}, \bibinfo {author} {\bibfnamefont {Y.-S.}\ \bibnamefont {Lee}}, \bibinfo {author} {\bibfnamefont {I.}~\bibnamefont {Souza}}, \bibinfo {author} {\bibfnamefont {D.}~\bibnamefont {Vanderbilt}},\ and\ \bibinfo {author} {\bibfnamefont {N.}~\bibnamefont {Marzari}},\ }\href {https://doi.org/https://doi.org/10.1016/j.cpc.2007.11.016} {\bibfield  {journal} {\bibinfo  {journal} {Computer Physics Communications}\ }\textbf {\bibinfo {volume} {178}},\ \bibinfo {pages} {685} (\bibinfo {year} {2008})}\BibitemShut {NoStop}%
\bibitem [{\citenamefont {Pizzi}\ \emph {et~al.}(2020)\citenamefont {Pizzi}, \citenamefont {Vitale}, \citenamefont {Arita}, \citenamefont {Blügel}, \citenamefont {Freimuth}, \citenamefont {Géranton}, \citenamefont {Gibertini}, \citenamefont {Gresch}, \citenamefont {Johnson}, \citenamefont {Koretsune}, \citenamefont {Ibañez-Azpiroz}, \citenamefont {Lee}, \citenamefont {Lihm}, \citenamefont {Marchand}, \citenamefont {Marrazzo}, \citenamefont {Mokrousov}, \citenamefont {Mustafa}, \citenamefont {Nohara}, \citenamefont {Nomura}, \citenamefont {Paulatto}, \citenamefont {Poncé}, \citenamefont {Ponweiser}, \citenamefont {Qiao}, \citenamefont {Thöle}, \citenamefont {Tsirkin}, \citenamefont {Wierzbowska}, \citenamefont {Marzari}, \citenamefont {Vanderbilt}, \citenamefont {Souza}, \citenamefont {Mostofi},\ and\ \citenamefont {Yates}}]{wannier90_v2}%
  \BibitemOpen
  \bibfield  {author} {\bibinfo {author} {\bibfnamefont {G.}~\bibnamefont {Pizzi}}, \bibinfo {author} {\bibfnamefont {V.}~\bibnamefont {Vitale}}, \bibinfo {author} {\bibfnamefont {R.}~\bibnamefont {Arita}}, \bibinfo {author} {\bibfnamefont {S.}~\bibnamefont {Blügel}}, \bibinfo {author} {\bibfnamefont {F.}~\bibnamefont {Freimuth}}, \bibinfo {author} {\bibfnamefont {G.}~\bibnamefont {Géranton}}, \bibinfo {author} {\bibfnamefont {M.}~\bibnamefont {Gibertini}}, \bibinfo {author} {\bibfnamefont {D.}~\bibnamefont {Gresch}}, \bibinfo {author} {\bibfnamefont {C.}~\bibnamefont {Johnson}}, \bibinfo {author} {\bibfnamefont {T.}~\bibnamefont {Koretsune}}, \bibinfo {author} {\bibfnamefont {J.}~\bibnamefont {Ibañez-Azpiroz}}, \bibinfo {author} {\bibfnamefont {H.}~\bibnamefont {Lee}}, \bibinfo {author} {\bibfnamefont {J.-M.}\ \bibnamefont {Lihm}}, \bibinfo {author} {\bibfnamefont {D.}~\bibnamefont {Marchand}}, \bibinfo {author} {\bibfnamefont {A.}~\bibnamefont {Marrazzo}}, \bibinfo {author} {\bibfnamefont
  {Y.}~\bibnamefont {Mokrousov}}, \bibinfo {author} {\bibfnamefont {J.~I.}\ \bibnamefont {Mustafa}}, \bibinfo {author} {\bibfnamefont {Y.}~\bibnamefont {Nohara}}, \bibinfo {author} {\bibfnamefont {Y.}~\bibnamefont {Nomura}}, \bibinfo {author} {\bibfnamefont {L.}~\bibnamefont {Paulatto}}, \bibinfo {author} {\bibfnamefont {S.}~\bibnamefont {Poncé}}, \bibinfo {author} {\bibfnamefont {T.}~\bibnamefont {Ponweiser}}, \bibinfo {author} {\bibfnamefont {J.}~\bibnamefont {Qiao}}, \bibinfo {author} {\bibfnamefont {F.}~\bibnamefont {Thöle}}, \bibinfo {author} {\bibfnamefont {S.~S.}\ \bibnamefont {Tsirkin}}, \bibinfo {author} {\bibfnamefont {M.}~\bibnamefont {Wierzbowska}}, \bibinfo {author} {\bibfnamefont {N.}~\bibnamefont {Marzari}}, \bibinfo {author} {\bibfnamefont {D.}~\bibnamefont {Vanderbilt}}, \bibinfo {author} {\bibfnamefont {I.}~\bibnamefont {Souza}}, \bibinfo {author} {\bibfnamefont {A.~A.}\ \bibnamefont {Mostofi}},\ and\ \bibinfo {author} {\bibfnamefont {J.~R.}\ \bibnamefont {Yates}},\ }\href
  {https://doi.org/10.1088/1361-648X/ab51ff} {\bibfield  {journal} {\bibinfo  {journal} {Journal of Physics: Condensed Matter}\ }\textbf {\bibinfo {volume} {32}},\ \bibinfo {pages} {165902} (\bibinfo {year} {2020})}\BibitemShut {NoStop}%
\bibitem [{\citenamefont {Bianco}\ and\ \citenamefont {Resta}(2011)}]{bianco2011localchern}%
  \BibitemOpen
  \bibfield  {author} {\bibinfo {author} {\bibfnamefont {R.}~\bibnamefont {Bianco}}\ and\ \bibinfo {author} {\bibfnamefont {R.}~\bibnamefont {Resta}},\ }\href {https://doi.org/10.1103/PhysRevB.84.241106} {\bibfield  {journal} {\bibinfo  {journal} {Phys. Rev. B}\ }\textbf {\bibinfo {volume} {84}},\ \bibinfo {pages} {241106} (\bibinfo {year} {2011})}\BibitemShut {NoStop}%
\bibitem [{\citenamefont {Marrazzo}\ and\ \citenamefont {Resta}(2019)}]{marrazzo2019}%
  \BibitemOpen
  \bibfield  {author} {\bibinfo {author} {\bibfnamefont {A.}~\bibnamefont {Marrazzo}}\ and\ \bibinfo {author} {\bibfnamefont {R.}~\bibnamefont {Resta}},\ }\href {https://doi.org/10.1103/PhysRevLett.122.166602} {\bibfield  {journal} {\bibinfo  {journal} {Phys. Rev. Lett.}\ }\textbf {\bibinfo {volume} {122}},\ \bibinfo {pages} {166602} (\bibinfo {year} {2019})}\BibitemShut {NoStop}%
\bibitem [{\citenamefont {Neupert}\ \emph {et~al.}(2012)\citenamefont {Neupert}, \citenamefont {Santos}, \citenamefont {Ryu}, \citenamefont {Chamon},\ and\ \citenamefont {Mudry}}]{neupert2012}%
  \BibitemOpen
  \bibfield  {author} {\bibinfo {author} {\bibfnamefont {T.}~\bibnamefont {Neupert}}, \bibinfo {author} {\bibfnamefont {L.}~\bibnamefont {Santos}}, \bibinfo {author} {\bibfnamefont {S.}~\bibnamefont {Ryu}}, \bibinfo {author} {\bibfnamefont {C.}~\bibnamefont {Chamon}},\ and\ \bibinfo {author} {\bibfnamefont {C.}~\bibnamefont {Mudry}},\ }\href {https://doi.org/10.1103/PhysRevB.86.035125} {\bibfield  {journal} {\bibinfo  {journal} {Phys. Rev. B}\ }\textbf {\bibinfo {volume} {86}},\ \bibinfo {pages} {035125} (\bibinfo {year} {2012})}\BibitemShut {NoStop}%
\bibitem [{\citenamefont {Brouder}\ \emph {et~al.}(2007)\citenamefont {Brouder}, \citenamefont {Panati}, \citenamefont {Calandra}, \citenamefont {Mourougane},\ and\ \citenamefont {Marzari}}]{brouder2007exponential}%
  \BibitemOpen
  \bibfield  {author} {\bibinfo {author} {\bibfnamefont {C.}~\bibnamefont {Brouder}}, \bibinfo {author} {\bibfnamefont {G.}~\bibnamefont {Panati}}, \bibinfo {author} {\bibfnamefont {M.}~\bibnamefont {Calandra}}, \bibinfo {author} {\bibfnamefont {C.}~\bibnamefont {Mourougane}},\ and\ \bibinfo {author} {\bibfnamefont {N.}~\bibnamefont {Marzari}},\ }\href {https://doi.org/10.1103/PhysRevLett.98.046402} {\bibfield  {journal} {\bibinfo  {journal} {Phys. Rev. Lett.}\ }\textbf {\bibinfo {volume} {98}},\ \bibinfo {pages} {046402} (\bibinfo {year} {2007})}\BibitemShut {NoStop}%
\bibitem [{\citenamefont {Cao}\ \emph {et~al.}(2018)\citenamefont {Cao}, \citenamefont {Fatemi}, \citenamefont {Fang}, \citenamefont {Watanabe}, \citenamefont {Taniguchi}, \citenamefont {Kaxiras},\ and\ \citenamefont {Jarillo-Herrero}}]{cao2018}%
  \BibitemOpen
  \bibfield  {author} {\bibinfo {author} {\bibfnamefont {Y.}~\bibnamefont {Cao}}, \bibinfo {author} {\bibfnamefont {V.}~\bibnamefont {Fatemi}}, \bibinfo {author} {\bibfnamefont {S.}~\bibnamefont {Fang}}, \bibinfo {author} {\bibfnamefont {K.}~\bibnamefont {Watanabe}}, \bibinfo {author} {\bibfnamefont {T.}~\bibnamefont {Taniguchi}}, \bibinfo {author} {\bibfnamefont {E.}~\bibnamefont {Kaxiras}},\ and\ \bibinfo {author} {\bibfnamefont {P.}~\bibnamefont {Jarillo-Herrero}},\ }\href {https://doi.org/10.1038/nature26160} {\bibfield  {journal} {\bibinfo  {journal} {Nature}\ }\textbf {\bibinfo {volume} {556}},\ \bibinfo {pages} {43} (\bibinfo {year} {2018})}\BibitemShut {NoStop}%
\bibitem [{\citenamefont {Sharpe}\ \emph {et~al.}(2019)\citenamefont {Sharpe}, \citenamefont {Fox}, \citenamefont {Barnard}, \citenamefont {Finney}, \citenamefont {Watanabe}, \citenamefont {Taniguchi}, \citenamefont {Kastner},\ and\ \citenamefont {Goldhaber-Gordon}}]{sharpe2019}%
  \BibitemOpen
  \bibfield  {author} {\bibinfo {author} {\bibfnamefont {A.~L.}\ \bibnamefont {Sharpe}}, \bibinfo {author} {\bibfnamefont {E.~J.}\ \bibnamefont {Fox}}, \bibinfo {author} {\bibfnamefont {A.~W.}\ \bibnamefont {Barnard}}, \bibinfo {author} {\bibfnamefont {J.}~\bibnamefont {Finney}}, \bibinfo {author} {\bibfnamefont {K.}~\bibnamefont {Watanabe}}, \bibinfo {author} {\bibfnamefont {T.}~\bibnamefont {Taniguchi}}, \bibinfo {author} {\bibfnamefont {M.~A.}\ \bibnamefont {Kastner}},\ and\ \bibinfo {author} {\bibfnamefont {D.}~\bibnamefont {Goldhaber-Gordon}},\ }\href {https://doi.org/10.1126/science.aaw3780} {\bibfield  {journal} {\bibinfo  {journal} {Science}\ }\textbf {\bibinfo {volume} {365}},\ \bibinfo {pages} {605} (\bibinfo {year} {2019})}\BibitemShut {NoStop}%
\bibitem [{\citenamefont {Serlin}\ \emph {et~al.}(2020)\citenamefont {Serlin}, \citenamefont {Tschirhart}, \citenamefont {Polshyn}, \citenamefont {Zhang}, \citenamefont {Zhu}, \citenamefont {Watanabe}, \citenamefont {Taniguchi}, \citenamefont {Balents},\ and\ \citenamefont {Young}}]{serlin2020}%
  \BibitemOpen
  \bibfield  {author} {\bibinfo {author} {\bibfnamefont {M.}~\bibnamefont {Serlin}}, \bibinfo {author} {\bibfnamefont {C.~L.}\ \bibnamefont {Tschirhart}}, \bibinfo {author} {\bibfnamefont {H.}~\bibnamefont {Polshyn}}, \bibinfo {author} {\bibfnamefont {Y.}~\bibnamefont {Zhang}}, \bibinfo {author} {\bibfnamefont {J.}~\bibnamefont {Zhu}}, \bibinfo {author} {\bibfnamefont {K.}~\bibnamefont {Watanabe}}, \bibinfo {author} {\bibfnamefont {T.}~\bibnamefont {Taniguchi}}, \bibinfo {author} {\bibfnamefont {L.}~\bibnamefont {Balents}},\ and\ \bibinfo {author} {\bibfnamefont {A.~F.}\ \bibnamefont {Young}},\ }\href {https://doi.org/10.1126/science.aay5533} {\bibfield  {journal} {\bibinfo  {journal} {Science}\ }\textbf {\bibinfo {volume} {367}},\ \bibinfo {pages} {900} (\bibinfo {year} {2020})}\BibitemShut {NoStop}%
\bibitem [{\citenamefont {Xie}\ \emph {et~al.}(2021)\citenamefont {Xie}, \citenamefont {Pierce}, \citenamefont {Park}, \citenamefont {Parker}, \citenamefont {Khalaf}, \citenamefont {Ledwith}, \citenamefont {Cao}, \citenamefont {Lee}, \citenamefont {Chen}, \citenamefont {Forrester}, \citenamefont {Watanabe}, \citenamefont {Taniguchi}, \citenamefont {Vishwanath}, \citenamefont {Jarillo-Herrero},\ and\ \citenamefont {Yacoby}}]{xie2021}%
  \BibitemOpen
  \bibfield  {author} {\bibinfo {author} {\bibfnamefont {Y.}~\bibnamefont {Xie}}, \bibinfo {author} {\bibfnamefont {A.~T.}\ \bibnamefont {Pierce}}, \bibinfo {author} {\bibfnamefont {J.~M.}\ \bibnamefont {Park}}, \bibinfo {author} {\bibfnamefont {D.~E.}\ \bibnamefont {Parker}}, \bibinfo {author} {\bibfnamefont {E.}~\bibnamefont {Khalaf}}, \bibinfo {author} {\bibfnamefont {P.}~\bibnamefont {Ledwith}}, \bibinfo {author} {\bibfnamefont {Y.}~\bibnamefont {Cao}}, \bibinfo {author} {\bibfnamefont {S.~H.}\ \bibnamefont {Lee}}, \bibinfo {author} {\bibfnamefont {S.}~\bibnamefont {Chen}}, \bibinfo {author} {\bibfnamefont {P.~R.}\ \bibnamefont {Forrester}}, \bibinfo {author} {\bibfnamefont {K.}~\bibnamefont {Watanabe}}, \bibinfo {author} {\bibfnamefont {T.}~\bibnamefont {Taniguchi}}, \bibinfo {author} {\bibfnamefont {A.}~\bibnamefont {Vishwanath}}, \bibinfo {author} {\bibfnamefont {P.}~\bibnamefont {Jarillo-Herrero}},\ and\ \bibinfo {author} {\bibfnamefont {A.}~\bibnamefont {Yacoby}},\ }\href
  {https://doi.org/10.1038/s41586-021-04002-3} {\bibfield  {journal} {\bibinfo  {journal} {Nature}\ }\textbf {\bibinfo {volume} {600}},\ \bibinfo {pages} {439} (\bibinfo {year} {2021})}\BibitemShut {NoStop}%
\bibitem [{\citenamefont {Zhou}\ \emph {et~al.}(2021)\citenamefont {Zhou}, \citenamefont {Xie}, \citenamefont {Taniguchi}, \citenamefont {Watanabe},\ and\ \citenamefont {Young}}]{zhou2021}%
  \BibitemOpen
  \bibfield  {author} {\bibinfo {author} {\bibfnamefont {H.}~\bibnamefont {Zhou}}, \bibinfo {author} {\bibfnamefont {T.}~\bibnamefont {Xie}}, \bibinfo {author} {\bibfnamefont {T.}~\bibnamefont {Taniguchi}}, \bibinfo {author} {\bibfnamefont {K.}~\bibnamefont {Watanabe}},\ and\ \bibinfo {author} {\bibfnamefont {A.~F.}\ \bibnamefont {Young}},\ }\href {https://doi.org/10.1038/s41586-021-03926-0} {\bibfield  {journal} {\bibinfo  {journal} {Nature}\ }\textbf {\bibinfo {volume} {598}},\ \bibinfo {pages} {434} (\bibinfo {year} {2021})}\BibitemShut {NoStop}%
\bibitem [{\citenamefont {Cai}\ \emph {et~al.}(2023)\citenamefont {Cai}, \citenamefont {Anderson}, \citenamefont {Wang}, \citenamefont {Zhang}, \citenamefont {Liu}, \citenamefont {Holtzmann}, \citenamefont {Zhang}, \citenamefont {Fan}, \citenamefont {Taniguchi}, \citenamefont {Watanabe}, \citenamefont {Ran}, \citenamefont {Cao}, \citenamefont {Fu}, \citenamefont {Xiao}, \citenamefont {Yao},\ and\ \citenamefont {Xu}}]{cai2023}%
  \BibitemOpen
  \bibfield  {author} {\bibinfo {author} {\bibfnamefont {J.}~\bibnamefont {Cai}}, \bibinfo {author} {\bibfnamefont {E.}~\bibnamefont {Anderson}}, \bibinfo {author} {\bibfnamefont {C.}~\bibnamefont {Wang}}, \bibinfo {author} {\bibfnamefont {X.}~\bibnamefont {Zhang}}, \bibinfo {author} {\bibfnamefont {X.}~\bibnamefont {Liu}}, \bibinfo {author} {\bibfnamefont {W.}~\bibnamefont {Holtzmann}}, \bibinfo {author} {\bibfnamefont {Y.}~\bibnamefont {Zhang}}, \bibinfo {author} {\bibfnamefont {F.}~\bibnamefont {Fan}}, \bibinfo {author} {\bibfnamefont {T.}~\bibnamefont {Taniguchi}}, \bibinfo {author} {\bibfnamefont {K.}~\bibnamefont {Watanabe}}, \bibinfo {author} {\bibfnamefont {Y.}~\bibnamefont {Ran}}, \bibinfo {author} {\bibfnamefont {T.}~\bibnamefont {Cao}}, \bibinfo {author} {\bibfnamefont {L.}~\bibnamefont {Fu}}, \bibinfo {author} {\bibfnamefont {D.}~\bibnamefont {Xiao}}, \bibinfo {author} {\bibfnamefont {W.}~\bibnamefont {Yao}},\ and\ \bibinfo {author} {\bibfnamefont {X.}~\bibnamefont {Xu}},\ }\href
  {https://doi.org/10.1038/s41586-023-06289-w} {\bibfield  {journal} {\bibinfo  {journal} {Nature}\ }\textbf {\bibinfo {volume} {622}},\ \bibinfo {pages} {63} (\bibinfo {year} {2023})}\BibitemShut {NoStop}%
\bibitem [{\citenamefont {Lu}\ \emph {et~al.}(2024)\citenamefont {Lu}, \citenamefont {Han}, \citenamefont {Yao}, \citenamefont {Reddy}, \citenamefont {Yang}, \citenamefont {Seo}, \citenamefont {Watanabe}, \citenamefont {Taniguchi}, \citenamefont {Fu},\ and\ \citenamefont {Ju}}]{lu2024}%
  \BibitemOpen
  \bibfield  {author} {\bibinfo {author} {\bibfnamefont {Z.}~\bibnamefont {Lu}}, \bibinfo {author} {\bibfnamefont {T.}~\bibnamefont {Han}}, \bibinfo {author} {\bibfnamefont {Y.}~\bibnamefont {Yao}}, \bibinfo {author} {\bibfnamefont {A.~P.}\ \bibnamefont {Reddy}}, \bibinfo {author} {\bibfnamefont {J.}~\bibnamefont {Yang}}, \bibinfo {author} {\bibfnamefont {J.}~\bibnamefont {Seo}}, \bibinfo {author} {\bibfnamefont {K.}~\bibnamefont {Watanabe}}, \bibinfo {author} {\bibfnamefont {T.}~\bibnamefont {Taniguchi}}, \bibinfo {author} {\bibfnamefont {L.}~\bibnamefont {Fu}},\ and\ \bibinfo {author} {\bibfnamefont {L.}~\bibnamefont {Ju}},\ }\href {https://doi.org/10.1038/s41586-023-07010-7} {\bibfield  {journal} {\bibinfo  {journal} {Nature}\ }\textbf {\bibinfo {volume} {626}},\ \bibinfo {pages} {759} (\bibinfo {year} {2024})}\BibitemShut {NoStop}%
\bibitem [{\citenamefont {Xia}\ \emph {et~al.}(2025)\citenamefont {Xia}, \citenamefont {Han}, \citenamefont {Watanabe}, \citenamefont {Taniguchi}, \citenamefont {Shan},\ and\ \citenamefont {Mak}}]{xia2025}%
  \BibitemOpen
  \bibfield  {author} {\bibinfo {author} {\bibfnamefont {Y.}~\bibnamefont {Xia}}, \bibinfo {author} {\bibfnamefont {Z.}~\bibnamefont {Han}}, \bibinfo {author} {\bibfnamefont {K.}~\bibnamefont {Watanabe}}, \bibinfo {author} {\bibfnamefont {T.}~\bibnamefont {Taniguchi}}, \bibinfo {author} {\bibfnamefont {J.}~\bibnamefont {Shan}},\ and\ \bibinfo {author} {\bibfnamefont {K.~F.}\ \bibnamefont {Mak}},\ }\href {https://doi.org/10.1038/s41586-024-08116-2} {\bibfield  {journal} {\bibinfo  {journal} {Nature}\ }\textbf {\bibinfo {volume} {637}},\ \bibinfo {pages} {833} (\bibinfo {year} {2025})}\BibitemShut {NoStop}%
\bibitem [{\citenamefont {Guo}\ \emph {et~al.}(2025)\citenamefont {Guo}, \citenamefont {Pack}, \citenamefont {Swann}, \citenamefont {Holtzman}, \citenamefont {Cothrine}, \citenamefont {Watanabe}, \citenamefont {Taniguchi}, \citenamefont {Mandrus}, \citenamefont {Barmak}, \citenamefont {Hone}, \citenamefont {Millis}, \citenamefont {Pasupathy},\ and\ \citenamefont {Dean}}]{guo2025}%
  \BibitemOpen
  \bibfield  {author} {\bibinfo {author} {\bibfnamefont {Y.}~\bibnamefont {Guo}}, \bibinfo {author} {\bibfnamefont {J.}~\bibnamefont {Pack}}, \bibinfo {author} {\bibfnamefont {J.}~\bibnamefont {Swann}}, \bibinfo {author} {\bibfnamefont {L.}~\bibnamefont {Holtzman}}, \bibinfo {author} {\bibfnamefont {M.}~\bibnamefont {Cothrine}}, \bibinfo {author} {\bibfnamefont {K.}~\bibnamefont {Watanabe}}, \bibinfo {author} {\bibfnamefont {T.}~\bibnamefont {Taniguchi}}, \bibinfo {author} {\bibfnamefont {D.~G.}\ \bibnamefont {Mandrus}}, \bibinfo {author} {\bibfnamefont {K.}~\bibnamefont {Barmak}}, \bibinfo {author} {\bibfnamefont {J.}~\bibnamefont {Hone}}, \bibinfo {author} {\bibfnamefont {A.~J.}\ \bibnamefont {Millis}}, \bibinfo {author} {\bibfnamefont {A.}~\bibnamefont {Pasupathy}},\ and\ \bibinfo {author} {\bibfnamefont {C.~R.}\ \bibnamefont {Dean}},\ }\href {https://doi.org/10.1038/s41586-024-08381-1} {\bibfield  {journal} {\bibinfo  {journal} {Nature}\ }\textbf {\bibinfo {volume} {637}},\ \bibinfo {pages} {839}
  (\bibinfo {year} {2025})}\BibitemShut {NoStop}%
\bibitem [{\citenamefont {Kang}\ and\ \citenamefont {Vafek}(2018)}]{kang2018}%
  \BibitemOpen
  \bibfield  {author} {\bibinfo {author} {\bibfnamefont {J.}~\bibnamefont {Kang}}\ and\ \bibinfo {author} {\bibfnamefont {O.}~\bibnamefont {Vafek}},\ }\href {https://doi.org/10.1103/PhysRevX.8.031088} {\bibfield  {journal} {\bibinfo  {journal} {Phys. Rev. X}\ }\textbf {\bibinfo {volume} {8}},\ \bibinfo {pages} {031088} (\bibinfo {year} {2018})}\BibitemShut {NoStop}%
\bibitem [{\citenamefont {Seo}\ \emph {et~al.}(2019)\citenamefont {Seo}, \citenamefont {Kotov},\ and\ \citenamefont {Uchoa}}]{seo2019}%
  \BibitemOpen
  \bibfield  {author} {\bibinfo {author} {\bibfnamefont {K.}~\bibnamefont {Seo}}, \bibinfo {author} {\bibfnamefont {V.~N.}\ \bibnamefont {Kotov}},\ and\ \bibinfo {author} {\bibfnamefont {B.}~\bibnamefont {Uchoa}},\ }\href {https://doi.org/10.1103/PhysRevLett.122.246402} {\bibfield  {journal} {\bibinfo  {journal} {Phys. Rev. Lett.}\ }\textbf {\bibinfo {volume} {122}},\ \bibinfo {pages} {246402} (\bibinfo {year} {2019})}\BibitemShut {NoStop}%
\bibitem [{\citenamefont {Hejazi}\ \emph {et~al.}(2021)\citenamefont {Hejazi}, \citenamefont {Chen},\ and\ \citenamefont {Balents}}]{hejazi2021}%
  \BibitemOpen
  \bibfield  {author} {\bibinfo {author} {\bibfnamefont {K.}~\bibnamefont {Hejazi}}, \bibinfo {author} {\bibfnamefont {X.}~\bibnamefont {Chen}},\ and\ \bibinfo {author} {\bibfnamefont {L.}~\bibnamefont {Balents}},\ }\href {https://doi.org/10.1103/PhysRevResearch.3.013242} {\bibfield  {journal} {\bibinfo  {journal} {Phys. Rev. Res.}\ }\textbf {\bibinfo {volume} {3}},\ \bibinfo {pages} {013242} (\bibinfo {year} {2021})}\BibitemShut {NoStop}%
\bibitem [{\citenamefont {Herzog-Arbeitman}\ \emph {et~al.}(2024)\citenamefont {Herzog-Arbeitman}, \citenamefont {Wang}, \citenamefont {Liu}, \citenamefont {Tam}, \citenamefont {Qi}, \citenamefont {Jia}, \citenamefont {Efetov}, \citenamefont {Vafek}, \citenamefont {Regnault}, \citenamefont {Weng}, \citenamefont {Wu}, \citenamefont {Bernevig},\ and\ \citenamefont {Yu}}]{herzog2024}%
  \BibitemOpen
  \bibfield  {author} {\bibinfo {author} {\bibfnamefont {J.}~\bibnamefont {Herzog-Arbeitman}}, \bibinfo {author} {\bibfnamefont {Y.}~\bibnamefont {Wang}}, \bibinfo {author} {\bibfnamefont {J.}~\bibnamefont {Liu}}, \bibinfo {author} {\bibfnamefont {P.~M.}\ \bibnamefont {Tam}}, \bibinfo {author} {\bibfnamefont {Z.}~\bibnamefont {Qi}}, \bibinfo {author} {\bibfnamefont {Y.}~\bibnamefont {Jia}}, \bibinfo {author} {\bibfnamefont {D.~K.}\ \bibnamefont {Efetov}}, \bibinfo {author} {\bibfnamefont {O.}~\bibnamefont {Vafek}}, \bibinfo {author} {\bibfnamefont {N.}~\bibnamefont {Regnault}}, \bibinfo {author} {\bibfnamefont {H.}~\bibnamefont {Weng}}, \bibinfo {author} {\bibfnamefont {Q.}~\bibnamefont {Wu}}, \bibinfo {author} {\bibfnamefont {B.~A.}\ \bibnamefont {Bernevig}},\ and\ \bibinfo {author} {\bibfnamefont {J.}~\bibnamefont {Yu}},\ }\href {https://doi.org/10.1103/PhysRevB.109.205122} {\bibfield  {journal} {\bibinfo  {journal} {Phys. Rev. B}\ }\textbf {\bibinfo {volume} {109}},\ \bibinfo {pages} {205122} (\bibinfo
  {year} {2024})}\BibitemShut {NoStop}%
\bibitem [{\citenamefont {Okuma}(2026)}]{okuma2026}%
  \BibitemOpen
  \bibfield  {author} {\bibinfo {author} {\bibfnamefont {N.}~\bibnamefont {Okuma}},\ }\href {https://arxiv.org/abs/2602.13698} {\bibinfo {title} {Localized-basis formulation of interacting hamiltonians in flat topological bands: coherent states and coherent-like states for fractional physics}} (\bibinfo {year} {2026}),\ \Eprint {https://arxiv.org/abs/2602.13698} {arXiv:2602.13698 [cond-mat.str-el]} \BibitemShut {NoStop}%
\bibitem [{\citenamefont {Fradkin}(2013)}]{fradkin2013field}%
  \BibitemOpen
  \bibfield  {author} {\bibinfo {author} {\bibfnamefont {E.}~\bibnamefont {Fradkin}},\ }\href {https://books.google.com/books?id=x7_6MX4ye_wC} {\emph {\bibinfo {title} {Field Theories of Condensed Matter Physics}}},\ Field Theories of Condensed Matter Physics\ (\bibinfo  {publisher} {Cambridge University Press},\ \bibinfo {year} {2013})\BibitemShut {NoStop}%
\bibitem [{\citenamefont {Thouless}(1984)}]{thouless1984_WF_OC}%
  \BibitemOpen
  \bibfield  {author} {\bibinfo {author} {\bibfnamefont {D.~J.}\ \bibnamefont {Thouless}},\ }\href {https://doi.org/10.1088/0022-3719/17/12/003} {\bibfield  {journal} {\bibinfo  {journal} {Journal of Physics C: Solid State Physics}\ }\textbf {\bibinfo {volume} {17}},\ \bibinfo {pages} {L325} (\bibinfo {year} {1984})}\BibitemShut {NoStop}%
\bibitem [{\citenamefont {Girvin}\ and\ \citenamefont {Jach}(1984)}]{girvin1984}%
  \BibitemOpen
  \bibfield  {author} {\bibinfo {author} {\bibfnamefont {S.~M.}\ \bibnamefont {Girvin}}\ and\ \bibinfo {author} {\bibfnamefont {T.}~\bibnamefont {Jach}},\ }\href {https://doi.org/10.1103/PhysRevB.29.5617} {\bibfield  {journal} {\bibinfo  {journal} {Phys. Rev. B}\ }\textbf {\bibinfo {volume} {29}},\ \bibinfo {pages} {5617} (\bibinfo {year} {1984})}\BibitemShut {NoStop}%
\bibitem [{\citenamefont {Wannier}(1937)}]{Wannier1937}%
  \BibitemOpen
  \bibfield  {author} {\bibinfo {author} {\bibfnamefont {G.~H.}\ \bibnamefont {Wannier}},\ }\href {https://doi.org/10.1103/PhysRev.52.191} {\bibfield  {journal} {\bibinfo  {journal} {Physical Review}\ }\textbf {\bibinfo {volume} {52}},\ \bibinfo {pages} {191} (\bibinfo {year} {1937})}\BibitemShut {NoStop}%
\bibitem [{\citenamefont {Yu}\ \emph {et~al.}(2011)\citenamefont {Yu}, \citenamefont {Qi}, \citenamefont {Bernevig}, \citenamefont {Fang},\ and\ \citenamefont {Dai}}]{yu2011}%
  \BibitemOpen
  \bibfield  {author} {\bibinfo {author} {\bibfnamefont {R.}~\bibnamefont {Yu}}, \bibinfo {author} {\bibfnamefont {X.~L.}\ \bibnamefont {Qi}}, \bibinfo {author} {\bibfnamefont {A.}~\bibnamefont {Bernevig}}, \bibinfo {author} {\bibfnamefont {Z.}~\bibnamefont {Fang}},\ and\ \bibinfo {author} {\bibfnamefont {X.}~\bibnamefont {Dai}},\ }\href {https://doi.org/10.1103/PhysRevB.84.075119} {\bibfield  {journal} {\bibinfo  {journal} {Phys. Rev. B}\ }\textbf {\bibinfo {volume} {84}},\ \bibinfo {pages} {075119} (\bibinfo {year} {2011})}\BibitemShut {NoStop}%
\bibitem [{Note1()}]{Note1}%
  \BibitemOpen
  \bibinfo {note} {Note that the Resta Spatial Localizer in~\protect \eqref {eq:RestaLocalizer} is equivalent to the standard operator used in eigenvalue equations, $O - \lambda I$, scaled by $\lambda ^{-1}$.}\BibitemShut {Stop}%
\bibitem [{Note2()}]{Note2}%
  \BibitemOpen
  \bibinfo {note} {There are other embeddings for localizers under PBC. For example, see Ref.~\cite {doll2024}, where the embedding is into $S^2$ using Pauli matrices. We have chosen an embedding into the Clifford torus instead because it is locally flat everywhere, resembling both the periodicity and flat nature of the physical space for a translationally invariant system.}\BibitemShut {Stop}%
\bibitem [{Note3()}]{Note3}%
  \BibitemOpen
  \bibinfo {note} {Similar considerations of symmetry are also used in optimization methods such as those in Ref.~\cite {marzari1997}}\BibitemShut {NoStop}%
\bibitem [{Note4()}]{Note4}%
  \BibitemOpen
  \bibinfo {note} {The model itself is an effective model for the monolayer of the W sublattice, which takes into account both (i) direct W-to-W hoppings and (ii) W-to-W hoppings mediated by the Se sublattice, and which predominantly accounts for the conduction and valence bands of $\protect \mathrm {WSe_2}$~\cite {wse2_model}.}\BibitemShut {Stop}%
\bibitem [{\citenamefont {Holbrook}\ \emph {et~al.}(2024)\citenamefont {Holbrook}, \citenamefont {Ingham}, \citenamefont {Kaplan}, \citenamefont {Holtzman}, \citenamefont {Bierman}, \citenamefont {Olson}, \citenamefont {Nashabeh}, \citenamefont {Liu}, \citenamefont {Zhu}, \citenamefont {Rhodes}, \citenamefont {Barmak}, \citenamefont {Hone}, \citenamefont {Queiroz},\ and\ \citenamefont {Pasupathy}}]{holbrook2024realspaceimagingbandtopology}%
  \BibitemOpen
  \bibfield  {author} {\bibinfo {author} {\bibfnamefont {M.}~\bibnamefont {Holbrook}}, \bibinfo {author} {\bibfnamefont {J.}~\bibnamefont {Ingham}}, \bibinfo {author} {\bibfnamefont {D.}~\bibnamefont {Kaplan}}, \bibinfo {author} {\bibfnamefont {L.}~\bibnamefont {Holtzman}}, \bibinfo {author} {\bibfnamefont {B.}~\bibnamefont {Bierman}}, \bibinfo {author} {\bibfnamefont {N.}~\bibnamefont {Olson}}, \bibinfo {author} {\bibfnamefont {L.}~\bibnamefont {Nashabeh}}, \bibinfo {author} {\bibfnamefont {S.}~\bibnamefont {Liu}}, \bibinfo {author} {\bibfnamefont {X.}~\bibnamefont {Zhu}}, \bibinfo {author} {\bibfnamefont {D.}~\bibnamefont {Rhodes}}, \bibinfo {author} {\bibfnamefont {K.}~\bibnamefont {Barmak}}, \bibinfo {author} {\bibfnamefont {J.}~\bibnamefont {Hone}}, \bibinfo {author} {\bibfnamefont {R.}~\bibnamefont {Queiroz}},\ and\ \bibinfo {author} {\bibfnamefont {A.}~\bibnamefont {Pasupathy}},\ }\href {https://arxiv.org/abs/2412.02813} {\bibinfo {title} {Real-space imaging of the band topology of transition metal
  dichalcogenides}} (\bibinfo {year} {2024}),\ \Eprint {https://arxiv.org/abs/2412.02813} {arXiv:2412.02813 [cond-mat.mtrl-sci]} \BibitemShut {NoStop}%
\bibitem [{\citenamefont {Li}\ \emph {et~al.}(2020)\citenamefont {Li}, \citenamefont {Zhu}, \citenamefont {Benalcazar},\ and\ \citenamefont {Hughes}}]{li2020fractional}%
  \BibitemOpen
  \bibfield  {author} {\bibinfo {author} {\bibfnamefont {T.}~\bibnamefont {Li}}, \bibinfo {author} {\bibfnamefont {P.}~\bibnamefont {Zhu}}, \bibinfo {author} {\bibfnamefont {W.~A.}\ \bibnamefont {Benalcazar}},\ and\ \bibinfo {author} {\bibfnamefont {T.~L.}\ \bibnamefont {Hughes}},\ }\href {https://doi.org/10.1103/PhysRevB.101.115115} {\bibfield  {journal} {\bibinfo  {journal} {Phys. Rev. B}\ }\textbf {\bibinfo {volume} {101}},\ \bibinfo {pages} {115115} (\bibinfo {year} {2020})}\BibitemShut {NoStop}%
\bibitem [{\citenamefont {Teo}\ and\ \citenamefont {Kane}(2010)}]{teo2010}%
  \BibitemOpen
  \bibfield  {author} {\bibinfo {author} {\bibfnamefont {J.~C.~Y.}\ \bibnamefont {Teo}}\ and\ \bibinfo {author} {\bibfnamefont {C.~L.}\ \bibnamefont {Kane}},\ }\href {https://doi.org/10.1103/PhysRevB.82.115120} {\bibfield  {journal} {\bibinfo  {journal} {Phys. Rev. B}\ }\textbf {\bibinfo {volume} {82}},\ \bibinfo {pages} {115120} (\bibinfo {year} {2010})}\BibitemShut {NoStop}%
\bibitem [{\citenamefont {Teo}\ and\ \citenamefont {Hughes}(2013)}]{teo2013}%
  \BibitemOpen
  \bibfield  {author} {\bibinfo {author} {\bibfnamefont {J.~C.~Y.}\ \bibnamefont {Teo}}\ and\ \bibinfo {author} {\bibfnamefont {T.~L.}\ \bibnamefont {Hughes}},\ }\href {https://doi.org/10.1103/PhysRevLett.111.047006} {\bibfield  {journal} {\bibinfo  {journal} {Phys. Rev. Lett.}\ }\textbf {\bibinfo {volume} {111}},\ \bibinfo {pages} {047006} (\bibinfo {year} {2013})}\BibitemShut {NoStop}%
\bibitem [{\citenamefont {Benalcazar}\ \emph {et~al.}(2014)\citenamefont {Benalcazar}, \citenamefont {Teo},\ and\ \citenamefont {Hughes}}]{benalcazar2014classification}%
  \BibitemOpen
  \bibfield  {author} {\bibinfo {author} {\bibfnamefont {W.~A.}\ \bibnamefont {Benalcazar}}, \bibinfo {author} {\bibfnamefont {J.}~\bibnamefont {Teo}},\ and\ \bibinfo {author} {\bibfnamefont {T.~L.}\ \bibnamefont {Hughes}},\ }\href {https://doi.org/10.1103/PhysRevB.89.224503} {\bibfield  {journal} {\bibinfo  {journal} {Phys. Rev. B}\ }\textbf {\bibinfo {volume} {89}},\ \bibinfo {pages} {224503} (\bibinfo {year} {2014})}\BibitemShut {NoStop}%
\bibitem [{\citenamefont {Deng}\ \emph {et~al.}(2022)\citenamefont {Deng}, \citenamefont {Benalcazar}, \citenamefont {Chen}, \citenamefont {Oudich}, \citenamefont {Ma},\ and\ \citenamefont {Jing}}]{deng2022observation}%
  \BibitemOpen
  \bibfield  {author} {\bibinfo {author} {\bibfnamefont {Y.}~\bibnamefont {Deng}}, \bibinfo {author} {\bibfnamefont {W.~A.}\ \bibnamefont {Benalcazar}}, \bibinfo {author} {\bibfnamefont {Z.-G.}\ \bibnamefont {Chen}}, \bibinfo {author} {\bibfnamefont {M.}~\bibnamefont {Oudich}}, \bibinfo {author} {\bibfnamefont {G.}~\bibnamefont {Ma}},\ and\ \bibinfo {author} {\bibfnamefont {Y.}~\bibnamefont {Jing}},\ }\href {https://doi.org/10.1103/PhysRevLett.128.174301} {\bibfield  {journal} {\bibinfo  {journal} {Phys. Rev. Lett.}\ }\textbf {\bibinfo {volume} {128}},\ \bibinfo {pages} {174301} (\bibinfo {year} {2022})}\BibitemShut {NoStop}%
\bibitem [{Note5()}]{Note5}%
  \BibitemOpen
  \bibinfo {note} {Based on studies of operators with the same mathematical structure of our localizers, we conclude that the existence of $N_{\protect \text {occ}}$ WCs, where $\mu ({\protect \bf r}^W)\rightarrow 0$, is guaranteed to exist for localizers consisting of two position operators (e.g., our 1D PBC or 2D OBC Spatial Localizers). The general existence of zero spectra for localizers with three or more operators (spatial dimensions) has been conjectured, but not proven~\cite {kisil1996mobius,jefferies1998weyl}. However, we provide an argument for the existence of $N_\protect \text {occ}$ points for the 2D PBC Spatial Localizer that converge to zero as $\mu (r^W)\propto N^{-2}$, while generic points scale as $N^{-1}$ [Appendix C]}\BibitemShut {NoStop}%
\bibitem [{\citenamefont {Robertson}(1929)}]{robertson_1929}%
  \BibitemOpen
  \bibfield  {author} {\bibinfo {author} {\bibfnamefont {H.~P.}\ \bibnamefont {Robertson}},\ }\href {https://doi.org/10.1103/PhysRev.34.163} {\bibfield  {journal} {\bibinfo  {journal} {Phys. Rev.}\ }\textbf {\bibinfo {volume} {34}},\ \bibinfo {pages} {163} (\bibinfo {year} {1929})}\BibitemShut {NoStop}%
\bibitem [{\citenamefont {Löwdin}(1950)}]{lowdinortho1950}%
  \BibitemOpen
  \bibfield  {author} {\bibinfo {author} {\bibfnamefont {P.}~\bibnamefont {Löwdin}},\ }\href {https://doi.org/10.1063/1.1747632} {\bibfield  {journal} {\bibinfo  {journal} {The Journal of Chemical Physics}\ }\textbf {\bibinfo {volume} {18}},\ \bibinfo {pages} {365} (\bibinfo {year} {1950})}\BibitemShut {NoStop}%
\bibitem [{Note6()}]{Note6}%
  \BibitemOpen
  \bibinfo {note} {This is a standard orthogonalization method, widely used in numerical approaches that find Wannier bases~\cite {marzari1997}. For one band, this procedure amounts to simply setting equal magnitude to the coefficients in the expansion of $\mathinner {|{p_1({\protect \bf r})}\rangle }$ in terms of occupied Bloch states}\BibitemShut {NoStop}%
\bibitem [{\citenamefont {Panati}(2007)}]{panati2007}%
  \BibitemOpen
  \bibfield  {author} {\bibinfo {author} {\bibfnamefont {G.}~\bibnamefont {Panati}},\ }\href {https://doi.org/10.1007/s00023-007-0326-8} {\bibfield  {journal} {\bibinfo  {journal} {Annales Henri Poincar{\'e}}\ }\textbf {\bibinfo {volume} {8}},\ \bibinfo {pages} {995} (\bibinfo {year} {2007})}\BibitemShut {NoStop}%
\bibitem [{\citenamefont {Monaco}\ \emph {et~al.}(2018)\citenamefont {Monaco}, \citenamefont {Panati}, \citenamefont {Pisante},\ and\ \citenamefont {Teufel}}]{monaco2018}%
  \BibitemOpen
  \bibfield  {author} {\bibinfo {author} {\bibfnamefont {D.}~\bibnamefont {Monaco}}, \bibinfo {author} {\bibfnamefont {G.}~\bibnamefont {Panati}}, \bibinfo {author} {\bibfnamefont {A.}~\bibnamefont {Pisante}},\ and\ \bibinfo {author} {\bibfnamefont {S.}~\bibnamefont {Teufel}},\ }\href {https://doi.org/10.1007/s00220-017-3067-7} {\bibfield  {journal} {\bibinfo  {journal} {Communications in Mathematical Physics}\ }\textbf {\bibinfo {volume} {359}},\ \bibinfo {pages} {61} (\bibinfo {year} {2018})}\BibitemShut {NoStop}%
\bibitem [{\citenamefont {Qi}\ \emph {et~al.}(2006)\citenamefont {Qi}, \citenamefont {Wu},\ and\ \citenamefont {Zhang}}]{qi2006}%
  \BibitemOpen
  \bibfield  {author} {\bibinfo {author} {\bibfnamefont {X.-L.}\ \bibnamefont {Qi}}, \bibinfo {author} {\bibfnamefont {Y.-S.}\ \bibnamefont {Wu}},\ and\ \bibinfo {author} {\bibfnamefont {S.-C.}\ \bibnamefont {Zhang}},\ }\href {https://doi.org/10.1103/PhysRevB.74.085308} {\bibfield  {journal} {\bibinfo  {journal} {Phys. Rev. B}\ }\textbf {\bibinfo {volume} {74}},\ \bibinfo {pages} {085308} (\bibinfo {year} {2006})}\BibitemShut {NoStop}%
\bibitem [{\citenamefont {Okuma}(2024)}]{okuma2024}%
  \BibitemOpen
  \bibfield  {author} {\bibinfo {author} {\bibfnamefont {N.}~\bibnamefont {Okuma}},\ }\href {https://doi.org/10.1103/PhysRevB.110.245112} {\bibfield  {journal} {\bibinfo  {journal} {Phys. Rev. B}\ }\textbf {\bibinfo {volume} {110}},\ \bibinfo {pages} {245112} (\bibinfo {year} {2024})}\BibitemShut {NoStop}%
\bibitem [{\citenamefont {Tao}(1986)}]{tao1986_WF_OC}%
  \BibitemOpen
  \bibfield  {author} {\bibinfo {author} {\bibfnamefont {R.}~\bibnamefont {Tao}},\ }\href {https://doi.org/10.1088/0022-3719/19/27/003} {\bibfield  {journal} {\bibinfo  {journal} {Journal of Physics C: Solid State Physics}\ }\textbf {\bibinfo {volume} {19}},\ \bibinfo {pages} {L619} (\bibinfo {year} {1986})}\BibitemShut {NoStop}%
\bibitem [{\citenamefont {Perelomov}(1971)}]{Perelomov1971}%
  \BibitemOpen
  \bibfield  {author} {\bibinfo {author} {\bibfnamefont {A.~M.}\ \bibnamefont {Perelomov}},\ }\href {https://doi.org/10.1007/BF01036577} {\bibfield  {journal} {\bibinfo  {journal} {Theoretical and Mathematical Physics}\ }\textbf {\bibinfo {volume} {6}},\ \bibinfo {pages} {156} (\bibinfo {year} {1971})}\BibitemShut {NoStop}%
\bibitem [{\citenamefont {Gunawardana}\ \emph {et~al.}(2024)\citenamefont {Gunawardana}, \citenamefont {Turner},\ and\ \citenamefont {Barnett}}]{gunawardana2024}%
  \BibitemOpen
  \bibfield  {author} {\bibinfo {author} {\bibfnamefont {T.~M.}\ \bibnamefont {Gunawardana}}, \bibinfo {author} {\bibfnamefont {A.~M.}\ \bibnamefont {Turner}},\ and\ \bibinfo {author} {\bibfnamefont {R.}~\bibnamefont {Barnett}},\ }\href {https://doi.org/10.1103/PhysRevResearch.6.023046} {\bibfield  {journal} {\bibinfo  {journal} {Phys. Rev. Res.}\ }\textbf {\bibinfo {volume} {6}},\ \bibinfo {pages} {023046} (\bibinfo {year} {2024})}\BibitemShut {NoStop}%
\bibitem [{\citenamefont {Li}\ \emph {et~al.}(2024)\citenamefont {Li}, \citenamefont {Dong}, \citenamefont {Ledwith},\ and\ \citenamefont {Khalaf}}]{li2024}%
  \BibitemOpen
  \bibfield  {author} {\bibinfo {author} {\bibfnamefont {Q.}~\bibnamefont {Li}}, \bibinfo {author} {\bibfnamefont {J.}~\bibnamefont {Dong}}, \bibinfo {author} {\bibfnamefont {P.~J.}\ \bibnamefont {Ledwith}},\ and\ \bibinfo {author} {\bibfnamefont {E.}~\bibnamefont {Khalaf}},\ }\href {https://arxiv.org/abs/2407.02561} {\bibinfo {title} {Constraints on real space representations of chern bands}} (\bibinfo {year} {2024}),\ \Eprint {https://arxiv.org/abs/2407.02561} {arXiv:2407.02561 [cond-mat.str-el]} \BibitemShut {NoStop}%
\bibitem [{\citenamefont {Gunawardana}\ \emph {et~al.}(2025)\citenamefont {Gunawardana}, \citenamefont {Schindler}, \citenamefont {Turner},\ and\ \citenamefont {Barnett}}]{gunawardana2025}%
  \BibitemOpen
  \bibfield  {author} {\bibinfo {author} {\bibfnamefont {T.~M.}\ \bibnamefont {Gunawardana}}, \bibinfo {author} {\bibfnamefont {F.}~\bibnamefont {Schindler}}, \bibinfo {author} {\bibfnamefont {A.~M.}\ \bibnamefont {Turner}},\ and\ \bibinfo {author} {\bibfnamefont {R.}~\bibnamefont {Barnett}},\ }\href {https://arxiv.org/abs/2502.17735} {\bibinfo {title} {Microscopic theory of chern polarization}} (\bibinfo {year} {2025}),\ \Eprint {https://arxiv.org/abs/2502.17735} {arXiv:2502.17735 [cond-mat.str-el]} \BibitemShut {NoStop}%
\bibitem [{\citenamefont {Coh}\ and\ \citenamefont {Vanderbilt}(2009)}]{coh2009}%
  \BibitemOpen
  \bibfield  {author} {\bibinfo {author} {\bibfnamefont {S.}~\bibnamefont {Coh}}\ and\ \bibinfo {author} {\bibfnamefont {D.}~\bibnamefont {Vanderbilt}},\ }\href {https://doi.org/10.1103/PhysRevLett.102.107603} {\bibfield  {journal} {\bibinfo  {journal} {Phys. Rev. Lett.}\ }\textbf {\bibinfo {volume} {102}},\ \bibinfo {pages} {107603} (\bibinfo {year} {2009})}\BibitemShut {NoStop}%
\bibitem [{\citenamefont {Vaidya}\ \emph {et~al.}(2024)\citenamefont {Vaidya}, \citenamefont {Rechtsman},\ and\ \citenamefont {Benalcazar}}]{vaidya2024}%
  \BibitemOpen
  \bibfield  {author} {\bibinfo {author} {\bibfnamefont {S.}~\bibnamefont {Vaidya}}, \bibinfo {author} {\bibfnamefont {M.~C.}\ \bibnamefont {Rechtsman}},\ and\ \bibinfo {author} {\bibfnamefont {W.~A.}\ \bibnamefont {Benalcazar}},\ }\href {https://doi.org/10.1103/PhysRevLett.132.116602} {\bibfield  {journal} {\bibinfo  {journal} {Phys. Rev. Lett.}\ }\textbf {\bibinfo {volume} {132}},\ \bibinfo {pages} {116602} (\bibinfo {year} {2024})}\BibitemShut {NoStop}%
\bibitem [{\citenamefont {Zhang}\ and\ \citenamefont {Barkeshli}(2025)}]{zhang2025}%
  \BibitemOpen
  \bibfield  {author} {\bibinfo {author} {\bibfnamefont {Y.}~\bibnamefont {Zhang}}\ and\ \bibinfo {author} {\bibfnamefont {M.}~\bibnamefont {Barkeshli}},\ }\href {https://doi.org/10.1103/frws-n1l7} {\bibfield  {journal} {\bibinfo  {journal} {Phys. Rev. B}\ }\textbf {\bibinfo {volume} {112}},\ \bibinfo {pages} {115124} (\bibinfo {year} {2025})}\BibitemShut {NoStop}%
\bibitem [{\citenamefont {Doll}\ \emph {et~al.}(2025)\citenamefont {Doll}, \citenamefont {Loring},\ and\ \citenamefont {Schulz-Baldes}}]{doll2024}%
  \BibitemOpen
  \bibfield  {author} {\bibinfo {author} {\bibfnamefont {N.}~\bibnamefont {Doll}}, \bibinfo {author} {\bibfnamefont {T.}~\bibnamefont {Loring}},\ and\ \bibinfo {author} {\bibfnamefont {H.}~\bibnamefont {Schulz-Baldes}},\ }\href {https://doi.org/10.1007/s11040-025-09508-0} {\bibfield  {journal} {\bibinfo  {journal} {Mathematical Physics, Analysis and Geometry}\ }\textbf {\bibinfo {volume} {28}},\ \bibinfo {pages} {13} (\bibinfo {year} {2025})}\BibitemShut {NoStop}%
\bibitem [{\citenamefont {Liu}\ \emph {et~al.}(2013)\citenamefont {Liu}, \citenamefont {Shan}, \citenamefont {Yao}, \citenamefont {Yao},\ and\ \citenamefont {Xiao}}]{wse2_model}%
  \BibitemOpen
  \bibfield  {author} {\bibinfo {author} {\bibfnamefont {G.-B.}\ \bibnamefont {Liu}}, \bibinfo {author} {\bibfnamefont {W.-Y.}\ \bibnamefont {Shan}}, \bibinfo {author} {\bibfnamefont {Y.}~\bibnamefont {Yao}}, \bibinfo {author} {\bibfnamefont {W.}~\bibnamefont {Yao}},\ and\ \bibinfo {author} {\bibfnamefont {D.}~\bibnamefont {Xiao}},\ }\href {https://doi.org/10.1103/PhysRevB.88.085433} {\bibfield  {journal} {\bibinfo  {journal} {Phys. Rev. B}\ }\textbf {\bibinfo {volume} {88}},\ \bibinfo {pages} {085433} (\bibinfo {year} {2013})}\BibitemShut {NoStop}%
\bibitem [{\citenamefont {Kisil}(1996)}]{kisil1996mobius}%
  \BibitemOpen
  \bibfield  {author} {\bibinfo {author} {\bibfnamefont {V.}~\bibnamefont {Kisil}},\ }\href@noop {} {\bibfield  {journal} {\bibinfo  {journal} {Electronic Research Announcements of the American Mathematical Society}\ }\textbf {\bibinfo {volume} {2}},\ \bibinfo {pages} {26} (\bibinfo {year} {1996})}\BibitemShut {NoStop}%
\bibitem [{\citenamefont {Jefferies}\ and\ \citenamefont {McIntosh}(1998)}]{jefferies1998weyl}%
  \BibitemOpen
  \bibfield  {author} {\bibinfo {author} {\bibfnamefont {B.}~\bibnamefont {Jefferies}}\ and\ \bibinfo {author} {\bibfnamefont {A.}~\bibnamefont {McIntosh}},\ }\href@noop {} {\bibfield  {journal} {\bibinfo  {journal} {Bulletin of the Australian Mathematical Society}\ }\textbf {\bibinfo {volume} {57}},\ \bibinfo {pages} {329} (\bibinfo {year} {1998})}\BibitemShut {NoStop}%
\bibitem [{\citenamefont {Brauer}\ and\ \citenamefont {Weyl}(1935)}]{brauer1935}%
  \BibitemOpen
  \bibfield  {author} {\bibinfo {author} {\bibfnamefont {R.}~\bibnamefont {Brauer}}\ and\ \bibinfo {author} {\bibfnamefont {H.}~\bibnamefont {Weyl}},\ }\href {http://www.jstor.org/stable/2371218} {\bibfield  {journal} {\bibinfo  {journal} {American Journal of Mathematics}\ }\textbf {\bibinfo {volume} {57}},\ \bibinfo {pages} {425} (\bibinfo {year} {1935})}\BibitemShut {NoStop}%
\bibitem [{\citenamefont {Loring}(2015)}]{loring2015}%
  \BibitemOpen
  \bibfield  {author} {\bibinfo {author} {\bibfnamefont {T.~A.}\ \bibnamefont {Loring}},\ }\href {https://doi.org/https://doi.org/10.1016/j.aop.2015.02.031} {\bibfield  {journal} {\bibinfo  {journal} {Annals of Physics}\ }\textbf {\bibinfo {volume} {356}},\ \bibinfo {pages} {383} (\bibinfo {year} {2015})}\BibitemShut {NoStop}%
\bibitem [{\citenamefont {Cerjan}\ and\ \citenamefont {Loring}(2022)}]{cerjan2022}%
  \BibitemOpen
  \bibfield  {author} {\bibinfo {author} {\bibfnamefont {A.}~\bibnamefont {Cerjan}}\ and\ \bibinfo {author} {\bibfnamefont {T.~A.}\ \bibnamefont {Loring}},\ }\href {https://doi.org/10.1103/PhysRevB.106.064109} {\bibfield  {journal} {\bibinfo  {journal} {Phys. Rev. B}\ }\textbf {\bibinfo {volume} {106}},\ \bibinfo {pages} {064109} (\bibinfo {year} {2022})}\BibitemShut {NoStop}%
\bibitem [{\citenamefont {Monkhorst}\ and\ \citenamefont {Pack}(1976)}]{monkhorst1976}%
  \BibitemOpen
  \bibfield  {author} {\bibinfo {author} {\bibfnamefont {H.~J.}\ \bibnamefont {Monkhorst}}\ and\ \bibinfo {author} {\bibfnamefont {J.~D.}\ \bibnamefont {Pack}},\ }\href {https://doi.org/10.1103/PhysRevB.13.5188} {\bibfield  {journal} {\bibinfo  {journal} {Phys. Rev. B}\ }\textbf {\bibinfo {volume} {13}},\ \bibinfo {pages} {5188} (\bibinfo {year} {1976})}\BibitemShut {NoStop}%
\bibitem [{\citenamefont {Trifonov}(1994)}]{trifonov1994}%
  \BibitemOpen
  \bibfield  {author} {\bibinfo {author} {\bibfnamefont {D.~A.}\ \bibnamefont {Trifonov}},\ }\href {https://doi.org/10.1063/1.530553} {\bibfield  {journal} {\bibinfo  {journal} {Journal of Mathematical Physics}\ }\textbf {\bibinfo {volume} {35}},\ \bibinfo {pages} {2297} (\bibinfo {year} {1994})}\BibitemShut {NoStop}%
\bibitem [{Note7()}]{Note7}%
  \BibitemOpen
  \bibinfo {note} {Note that the choice of embedding structure for PBC localizer will slightly alter the structure of the bounds, but the general method used here is applicable to other embedding structures, such as that used in ~\cite {doll2024}.}\BibitemShut {Stop}%
\bibitem [{Note8()}]{Note8}%
  \BibitemOpen
  \bibinfo {note} {Note that this component is gauge-invariant with respect to a Wannier basis, where one fixes the distribution of weights in the BZ to be flat when constructing (Wannier) states. For a more general wave packet, the smallest singular value of the operator $X(I-P)X$ acts as a lower bound on the spread.}\BibitemShut {Stop}%
\bibitem [{\citenamefont {Maccone}\ and\ \citenamefont {Pati}(2014)}]{maccone_uncertainty}%
  \BibitemOpen
  \bibfield  {author} {\bibinfo {author} {\bibfnamefont {L.}~\bibnamefont {Maccone}}\ and\ \bibinfo {author} {\bibfnamefont {A.~K.}\ \bibnamefont {Pati}},\ }\href {https://doi.org/10.1103/PhysRevLett.113.260401} {\bibfield  {journal} {\bibinfo  {journal} {Phys. Rev. Lett.}\ }\textbf {\bibinfo {volume} {113}},\ \bibinfo {pages} {260401} (\bibinfo {year} {2014})}\BibitemShut {NoStop}%
\bibitem [{\citenamefont {Ledwith}\ \emph {et~al.}(2023)\citenamefont {Ledwith}, \citenamefont {Vishwanath},\ and\ \citenamefont {Parker}}]{ledwith2023}%
  \BibitemOpen
  \bibfield  {author} {\bibinfo {author} {\bibfnamefont {P.~J.}\ \bibnamefont {Ledwith}}, \bibinfo {author} {\bibfnamefont {A.}~\bibnamefont {Vishwanath}},\ and\ \bibinfo {author} {\bibfnamefont {D.~E.}\ \bibnamefont {Parker}},\ }\href {https://doi.org/10.1103/PhysRevB.108.205144} {\bibfield  {journal} {\bibinfo  {journal} {Phys. Rev. B}\ }\textbf {\bibinfo {volume} {108}},\ \bibinfo {pages} {205144} (\bibinfo {year} {2023})}\BibitemShut {NoStop}%
\bibitem [{Note9()}]{Note9}%
  \BibitemOpen
  \bibinfo {note} {There is an underlying assumption that the OBC position operators are formulated such that there is no issue of the WF in question wrapping around the (periodic) boundary, which would artificially increase the spread.}\BibitemShut {Stop}%
\bibitem [{\citenamefont {Provost}\ and\ \citenamefont {Vallee}(1980)}]{provost1980}%
  \BibitemOpen
  \bibfield  {author} {\bibinfo {author} {\bibfnamefont {J.~P.}\ \bibnamefont {Provost}}\ and\ \bibinfo {author} {\bibfnamefont {G.}~\bibnamefont {Vallee}},\ }\href {https://doi.org/10.1007/BF02193559} {\bibfield  {journal} {\bibinfo  {journal} {Communications in Mathematical Physics}\ }\textbf {\bibinfo {volume} {76}},\ \bibinfo {pages} {289} (\bibinfo {year} {1980})}\BibitemShut {NoStop}%
\bibitem [{\citenamefont {Kohn}(1964)}]{kohn1964}%
  \BibitemOpen
  \bibfield  {author} {\bibinfo {author} {\bibfnamefont {W.}~\bibnamefont {Kohn}},\ }\href {https://doi.org/10.1103/PhysRev.133.A171} {\bibfield  {journal} {\bibinfo  {journal} {Phys. Rev.}\ }\textbf {\bibinfo {volume} {133}},\ \bibinfo {pages} {A171} (\bibinfo {year} {1964})}\BibitemShut {NoStop}%
\bibitem [{\citenamefont {Souza}\ \emph {et~al.}(2000)\citenamefont {Souza}, \citenamefont {Wilkens},\ and\ \citenamefont {Martin}}]{souza2000}%
  \BibitemOpen
  \bibfield  {author} {\bibinfo {author} {\bibfnamefont {I.}~\bibnamefont {Souza}}, \bibinfo {author} {\bibfnamefont {T.}~\bibnamefont {Wilkens}},\ and\ \bibinfo {author} {\bibfnamefont {R.~M.}\ \bibnamefont {Martin}},\ }\href {https://doi.org/10.1103/PhysRevB.62.1666} {\bibfield  {journal} {\bibinfo  {journal} {Phys. Rev. B}\ }\textbf {\bibinfo {volume} {62}},\ \bibinfo {pages} {1666} (\bibinfo {year} {2000})}\BibitemShut {NoStop}%
\bibitem [{\citenamefont {Resta}\ and\ \citenamefont {Sorella}(1999)}]{resta1999}%
  \BibitemOpen
  \bibfield  {author} {\bibinfo {author} {\bibfnamefont {R.}~\bibnamefont {Resta}}\ and\ \bibinfo {author} {\bibfnamefont {S.}~\bibnamefont {Sorella}},\ }\href {https://doi.org/10.1103/PhysRevLett.82.370} {\bibfield  {journal} {\bibinfo  {journal} {Phys. Rev. Lett.}\ }\textbf {\bibinfo {volume} {82}},\ \bibinfo {pages} {370} (\bibinfo {year} {1999})}\BibitemShut {NoStop}%
\bibitem [{\citenamefont {Resta}(2017)}]{resta2017_talk}%
  \BibitemOpen
  \bibfield  {author} {\bibinfo {author} {\bibfnamefont {R.}~\bibnamefont {Resta}}} (\bibinfo {year} {2017}),\ \bibinfo {note} {lecture at the Autumn School on Correlated Electrons: \emph{The Physics of Correlated Insulators, Metals, and Superconductors}, Forschungszentrum J\"ulich, Germany}\BibitemShut {NoStop}%
\bibitem [{\citenamefont {Bloch}(1929)}]{Bloch1929}%
  \BibitemOpen
  \bibfield  {author} {\bibinfo {author} {\bibfnamefont {F.}~\bibnamefont {Bloch}},\ }\href {https://doi.org/10.1007/BF01339455} {\bibfield  {journal} {\bibinfo  {journal} {Zeitschrift f{\"u}r Physik}\ }\textbf {\bibinfo {volume} {52}},\ \bibinfo {pages} {555} (\bibinfo {year} {1929})}\BibitemShut {NoStop}%
\bibitem [{\citenamefont {Cerjan}\ \emph {et~al.}(2023)\citenamefont {Cerjan}, \citenamefont {Loring},\ and\ \citenamefont {Vides}}]{cerjan2023}%
  \BibitemOpen
  \bibfield  {author} {\bibinfo {author} {\bibfnamefont {A.}~\bibnamefont {Cerjan}}, \bibinfo {author} {\bibfnamefont {T.~A.}\ \bibnamefont {Loring}},\ and\ \bibinfo {author} {\bibfnamefont {F.}~\bibnamefont {Vides}},\ }\href {https://doi.org/10.1063/5.0098336} {\bibfield  {journal} {\bibinfo  {journal} {Journal of Mathematical Physics}\ }\textbf {\bibinfo {volume} {64}},\ \bibinfo {pages} {023501} (\bibinfo {year} {2023})}\BibitemShut {NoStop}%
\end{thebibliography}%

\newpage

\appendix

\section{Wannier Centers and Wilson Loops}\label{app:wilson_loops}

In this appendix, we discuss the relation of Wannier centers (WCs) as eigenvalues of a Wilson loop and the connection to the projected Resta position operator \cite{alexandradinata2014,benalcazar2017}. This connection is crucial, as the eigenvalues of a Wilson loop, which correspond directly to WCs, are powers of the eigenvalues of the projected position operator. We can then use a Spatial Localizer, constructed from projected Resta position operators, to search for approximate eigenstates (approximate maximally localized Wannier functions) of those operators.

We begin with a discussion of Bloch states and the Resta position operator. We then project the Resta position operator onto a space of occupied bands and discuss how WCs manifest in the spectrum of a Wilson loop (a Wilson loop in this context is understood as the repeated application of the projected Resta position operator).
For a more detailed discussion, see Section III.B of~\cite{benalcazar2017}. 

Our starting point is the set of Bloch states
\begin{align}
    \ket{\psi_{n,\kv}} &= \frac{1}{N}\sum_\transvec e^{\ii \kv \cdot \transvec} \ket{u_{n,\kv}} \otimes \ket{\transvec} \nonumber \\
    &= \ket{u_{n,\kv}} \otimes \ket{\kv}  ,
\end{align}
which diagonalize the Hamiltonian,
\begin{align}
    H \ket{\psi_{n,\kv}}= E_{n,\kv} \ket{\psi_{n,\kv}},
\end{align}
where $n$ labels the band index, $\transvec$ is a direct lattice vector, $\kv$ is the crystal momentum defined over the Brillouin zone (BZ) of the crystal, and $\ket{u_{n,\kv}}$ is defined over the unit cell degrees of freedom, and obey $\braket{u_{n,\kv}|u_{m,\kv}}=\delta_{m,n}$. The Hamiltonian eigenstates are orthonormal, $\braket{\psi_{n,\kv}|\psi_{m,{\bf q}}}=\delta_{\kv,{\bf q}}\delta_{m,n}$. Under periodic boundary conditions (PBC), we use the Resta position operator~\cite{resta1998} so that the position operator is well-defined on the periodic Hilbert space that $H$ acts on. 

We now move to a 1D analysis for simplicity (i.e., $\transvec\rightarrow x$ and $\kv\rightarrow k$). In position space, the 1D Resta position operator is formulated as 
\begin{equation}
    X_R = \sum_x e^{\ii \Delta k x} I \otimes \ketbra{x} ,
\end{equation}
where I is the $N_\textrm{b} \times N_\textrm{b}$ identity matrix with $N_b$ as the number of bands. 
Furthermore, we have $\Delta k = 2\pi/N$ and $N$ as the number of unit cells in the system. 

Using the Fourier transform $\ket{x} = \frac{1}{\sqrt{N}} \sum_k e^{-\ii kx}\ket{k}$, we have

\begin{equation}
    \begin{split}
        X_R &=\sum_x e^{\ii \Delta k x} \frac{1}{N}\sum_{k,k'} e^{-\ii kx} e^{\ii k' x} I \otimes \ketbra{k}{k'}  \\
        & = \sum_{k,k'} \frac{1}{N} \sum_x  e^{\ii (k' + \Delta k - k)x} I \otimes\ketbra{k}{k'}  \\
        &= \sum_{k} I \otimes \ketbra{k + \Delta k}{k},
    \end{split}
\end{equation}
where we used $\delta_{k, k'+\Delta k} = \frac{1}{N} \sum_x  e^{\ii (k' + \Delta k - k)x}$ in the last line. 

To project $X_R$ into occupied energy bands, we can define the projector as 

\begin{equation}
    P = \sum_k P_k \ketbra{k}{k} , \quad P_k=\sum_n \ketbra{u_{n,k}}{u_{n,k}},
\end{equation}
where $n$ runs over occupied bands and $\ket{u_{n,k}}$ is the periodic component of the Bloch state $\ket{\psi_{n,k}}$. Furthermore, we use the notation $P_k \ketbra{k} \equiv P_k \otimes \ketbra{k}$ for brevity. Projecting $X_R$ into the occupied subspace, we have 
\begin{align}
PX_RP = \sum_k P_{k+ \Delta k}P_{k} \ketbra{k+ \Delta k}{k}.
\end{align}
To simplify our analysis, we proceed on the basis of occupied states, labeled only by the occupied bands and $k$. We move to this basis by using the ``half projector,'' $\halfproj=\sum_{n,k} \ketbra{\psi_{n,k}}{n,k}$, which is a matrix with the occupied states as column vectors; the usual ``full projector'' is $P=\halfproj\halfproj^\dagger$. Here, the sum over $n$ runs over the occupied bands. Note that this is equivalent to the construction 

\begin{equation}
    \halfproj = \sum_k V_k \ketbra{k}{k}, \ \ V_k = \sum_n \ketbra{u_{n,k}}{n}.
\end{equation}
Furthermore, the definition of $\halfproj$ above is equivalent to the definition in the main text, $\halfproj=\sum_{E_i<E_F} \ketbra{\psi_i}{i}$, given some Fermi energy $E_F$ (see Appendix~\ref{app:sec:general_spatial_localizer}).
The use of $\halfproj$ in place of $P$ yields

\begin{equation}
    \halfproj^\dagger X_R\halfproj = \begin{pmatrix}
0 & 0 & 0 & \cdots & G_{k_N} \\
G_{k_1} & 0 & 0 & \cdots & 0 \\
0 & G_{k_2} & 0 & \cdots & 0 \\
\vdots & \vdots & \vdots & \ddots & \vdots \\
0 & 0 & 0 & \cdots & 0
\end{pmatrix},
\end{equation}
where $[G_k]_{mn} = \bra{u_{m,k+\Delta k}}u_{n,k}\rangle$ and both $m$ and $n$ run over occupied bands, such that $[\halfproj^\dagger X_R\halfproj]_{k+\Delta k,k} = [G_k]_{mn}$. Note that $\halfproj^\dagger X_R\halfproj$ is unitary only in the thermodynamic limit ($N\rightarrow \infty$). To obtain a unitary $\halfproj^\dagger X_R\halfproj$ for finite $N$, one can set the singular values of $[G_k]_{mn}$ to 1~\cite{souza2001}, i.e., we take the singular value decomposition $[G_k]_{mn} = USV^\dagger$ and replace $S$ with the identity matrix. Let $[F_{k_i}]^{mn}=UV^\dagger$. 

We now look to solve for the eigenvalues of $\halfproj^\dagger X_R\halfproj$, for which the WCs
manifest as phase angles of the eigenvalues of $\halfproj^\dagger X_R\halfproj$.
Consider the eigenvalue equation $\halfproj^\dagger X_R\halfproj \ket{\psi^j} = E^j \ket{\psi^j}$. This is equivalent to 

\begin{equation}\label{app:eq:wilson_loop_mat_eq}
    \begin{pmatrix}
0 & 0 & 0 & \cdots & F_{k_N} \\
F_{k_1} & 0 & 0 & \cdots & 0 \\
0 & F_{k_2} & 0 & \cdots & 0 \\
\vdots & \vdots & \vdots & \ddots & \vdots \\
0 & 0 & 0 & \cdots & 0
\end{pmatrix}
\begin{pmatrix}
\nu_{k_1} \\
\nu_{k_2} \\
\nu_{k_3} \\
\vdots \\
\nu_{k_N}
\end{pmatrix}^{j}
=
E^j
\begin{pmatrix}
\nu_{k_1} \\
\nu_{k_2} \\
\nu_{k_3} \\
\vdots \\
\nu_{k_N}
\end{pmatrix}^{j}.
\end{equation}
By repeated application of $\halfproj^\dagger X_R\halfproj$ on $\ket{\psi^j}$, we have the Wilson loop $\mathcal{W}_{k} = \Pi _{i=1} ^N F_{k_i}$ where $[F_{k_i}]^{mn} = \bra{u_{m,k_i+\Delta k}}{u_{n,k_i}}\rangle$.
The eigenvalue equation $\mathcal{W}_k \ket{\nu^j_k} = e^{i2\pi x^{Wj}} \ket{\nu^j_k}$ then yields WCs as the phase angles $x^{Wj}$ of the Wilson loop spectrum.

\section{General Spatial Localizer and notation} \label{app:sec:general_spatial_localizer}

In this appendix, we define a general construction for Spatial Localizers in $d$ dimensions and the relevant notation used in this work. This allows one to construct and utilize a Spatial Localizer in higher spatial dimensions/on non-cubic lattices.

We begin by revisiting the half projector $\halfproj$ defined in Appendix~\ref{app:wilson_loops} and the main text. We then discuss the general form of the Clifford elements $\Gamma_j$ used to construct Spatial Localizers. Finally, we provide the general form of Spatial Localizers under open boundary conditions (OBC) and PBC. During the discussion of periodic Spatial Localizers, we also provide a general construction of $\tX_{j,C}$ and $\tX_{j,S}$. This general construction is useful for the case of a periodic Spatial Localizer formulated in momentum-space, where the Resta position operator is no longer diagonal. We end this appendix with a discussion on Spatial Localizers with chiral symmetry (arising from the embedding structure) and we specify the scheme used in this work to discretize the Brillouin zone.

We consider a Hamiltonian $H$ with a band gap (e.g., $H$ is an insulator). 
We denote occupied energy eigenstates by $\ket{\Psi_i}$ such that $H\ket{\Psi_i} = E_i\ket{\Psi_i}$ and $E_i<E_F$ for some Fermi energy $E_F$ within a band gap. 
One may also define the set of occupied states by any isolated (by a finite band gap) set of bands or optimal subspace of ``entangled'' bands~\cite{souza2001}. From these occupied energy eigenstates, we have the half-projector $\halfproj=\sum_{E_i<E_F} \ketbra{\Psi_i}{i}$ as in both the main text and Appendix~\ref{app:wilson_loops}.

We denote the Clifford elements used in the construction of a Spatial Localizer by $\Gamma_j$ such that $\acomm{\Gamma_j}{\Gamma_k}=2\delta_{jk}$ and $\Gamma_j=\Gamma_j^\dagger$. 
One can use the generators of any Euclidean Clifford algebra $\mathrm{Cl}_{n,0}$ to construct $\Gamma_j$. 
In this work, we use the Weyl-Brauer matrices~\cite{brauer1935} in the case of $n\geq6$ generators. We note that, in the case of even $n$, one can use a modified version of the pseudoscalar $\Gamma_1\Gamma_2...\Gamma_n$ as $\Gamma_{n+1}$, where a scalar may be needed to satisfy $\Gamma_{n+1}^2=I$. 
We define $g$ to be the dimension of the matrix representation of $\Gamma_j$.

We denote the Hilbert spaces of occupied states and Clifford elements respectively by $\hs_{\text{occ}}$ and $\mathcal{H}_\Gamma$. 
A Spatial Localizer, $L(\rv)$, then exists in the space of bounded linear operators, $L(\rv) \in \mathcal{B}(\hs_{\text{occ}} \otimes \mathcal{H}_\Gamma)$, that act on the Hilbert space $\hs_{\text{occ}} \otimes \mathcal{H}_\Gamma$.

Under PBC, we assume the system to be described by a Bravais lattice of unit cells with primitive and reciprocal lattice vectors respectively denoted as $\mathbf{a}_j$ and $\mathbf{b}_j$ for $j=1,2,...,d$ such that $\mathbf{a}_j \cdot\mathbf{b}_k = 2\pi \delta_{jk}$. 
We assume $\norm{\mathbf{a}_i}=1 \  \forall \  i$. 
In position space, $N_j$ corresponds to the number of unit cells considered in the $\mathbf{a}_j$ direction for a periodicity over the length $N_j\norm{\mathbf{a}_j}$ in the $\mathbf{a}_j$ direction. 
In momentum space, $N_j$ corresponds to the number of unique k points along $\mathbf{b}_j$ direction. 
We define the position operator $\hat{\Rv} = \sum_j X_j \hat{x}_j$ which acts on the position basis as $\hat{\Rv} \ket{\rv} = \rv \ket{\rv}$ where $\hat{x}_j$ are basis vectors for $\mathbb{R}^d$.
We then have the Resta position operators

\begin{align}
X_{j,R}(\rv) = \exp[\frac{\ii}{N_j}(\hat{\Rv} - \rv I) \cdot \bv_j],
\end{align}
and the projected Resta position operators 
\begin{align}
\tX_{j,R}(\rv) = \halfproj^\dagger X_{j,R}(\rv) \halfproj.    
\end{align}
Note that one may need to consider $d' > d$ Resta position operators in the case of non-cubic Bravais lattices (see Appendix B of \cite{marzari1997}).
We then define the real and imaginary components of $\tX_{j,R}(\rv)-\halfproj^\dagger \halfproj$ as 

\begin{equation} \label{app:eq_resta_real_imag}
    \begin{split}
        &\tX_{j,C}(\mathbf{r}) = \frac{1}{2}\left(\tilde{X}_{j,R}(\mathbf{r}) + \tilde{X}_{j,R}^\dagger(\mathbf{r})\right) - \halfproj^\dagger \halfproj, \\
        &\tX_{j,S}(\mathbf{r}) = \frac{1}{2\mathrm{i}}\left(\tilde{X}_{j,R}(\mathbf{r}) - \tilde{X}_{j,R}^\dagger(\mathbf{r})\right), \\
    \end{split}
\end{equation}
which correspond to the terms used in the explicit construction of the $d=1,2$ dimensional periodic Spatial Localizers in the main text. 
While equivalent to the definition in the main text, the relations in \eqref{app:eq_resta_real_imag} are useful in momentum space where $X_{j,R}$ are no longer diagonal operators (see Appendix \ref{app:wilson_loops}). We then construct a PBC Spatial Localizer in $d$ spatial dimensions as 

\begin{equation}
\label{app:eq:general_PBC_loc}
    L^\textrm{PBC}_{d\textrm{D}}(\rv) = \sum_{j} \tX_{j,C}(\rv) \otimes \Gamma_{2j-1} + \tX_{j,S}(\rv) \otimes \Gamma_{2j}.
\end{equation}

Under OBC, we denote the $d$ position operators by $X_j$ for $j=1,2,...,d$ and the projected position operators as $\tX_j(r_j) = \halfproj^\dagger (X_j - r_jI) \halfproj$.
We then construct an OBC Spatial Localizer in $d$ spatial dimensions as

\begin{equation}
\label{app:eq:general_OBC_loc}
    L^\textrm{OBC}_{d\textrm{D}}(\rv) = \sum_j \tX_j(r_j)\otimes \Gamma_j.
\end{equation}

Sometimes, we omit the explicit $\rv$ dependence, i.e., $\tX_j (\rv)=\tX_j$, when the discussion is independent of $\rv$ or for brevity. 
When considering a general localizer structure, we may use $\tX_j$ to refer to either the OBC elements of Eq. \eqref{app:eq:general_OBC_loc} or a real/imaginary component of Eq. \eqref{app:eq:general_PBC_loc}. This comes from the general structure of a localizer~\cite{loring2015,cerjan2022}, where the only restriction is that the $\tX_j$ are Hermitian.

By the chiral symmetry of our OBC (in $d=1,2$ spatial dimensions) and PBC (in all spatial dimensions) Spatial Localizers, the localizer indicator function (LIF), defined as

\begin{equation}
    \mu(\rv;L) = \min[|\sigma(L(\rv))|],
\end{equation}
will correspond to at least two distinct localizer spectra, manifesting from a spectral band above and below zero. However, the difference between the localizer eigenstates related by chiral symmetry is a transformation on the embedding degrees of freedom (DOF), and thus will have no effect on the physical Schmidt vectors we use to construct local state representations. As such, we employ the convention of only considering nonnegative localizer spectra in the presence of chiral symmetry. In the considered PBC models, we use a Monkhorst-Pack~\cite{monkhorst1976} mesh to discretize the BZ.

\section{Localizer indicator function, Schmidt values, and uncertainty relations}\label{app:sec:alt_bound}

In this appendix, we derive the general form of the bound on the variance of physical Schmidt vectors $\ket{\psch_1(\rv)}$ presented in Eq. (9) of the main text. This bound strengthens the validity of a Spatial Localizer as a tool for constructing localized bases, as we find the upper component of the bound to outperform previous bounds, and both the upper and lower bounds provide a squeeze theorem viewpoint on the degree to which the first physical Schmidt vectors, $\ket{\psch_1}$, are ``intelligent states''~\cite{trifonov1994} from the perspective of saturating the Heisenberg-Robertson uncertainty relation (HRUR)~\cite{robertson_1929}. 

After presenting the results and context, we derive the general lower and upper bounds in Sections~\ref{app:sec:sum_HRUR} and~\ref{app:sec:upper_bound_derivation}, respectively. Section~\ref{app:sec:alt_bound_2D} discusses some properties of the LIF and the first Schmidt value that we have derived in the case of two spatial dimensions. We then discuss conditions for equivalence between OBC and PBC localizer bounds in Section~\ref{app:sec:OBC_PBC_equivalence}. Furthermore, we quantify the strength (tightness/degree of saturation) of our derived bounds in Section~\ref{app:sec:bounds_metrics}. We note that, while this section is discussed in relation to the Spatial Localizer, the bounds derived apply for any localizer, e.g., the \emph{spectral localizer}~\cite{loring2015,cerjan2022}, constructed as in Eq.~\eqref{app:eq:general_OBC_loc} using Hermitian operators $\tX_j(r_j)$ and Clifford components $\Gamma_j$. The discussion of OBC vs. PBC Spatial Localizers is also applicable to general OBC~\cite{loring2015,cerjan2022} and PBC~\footnote{Note that the choice of embedding structure for PBC localizer will slightly alter the structure of the bounds, but the general method used here is applicable to other embedding structures, such as that used in ~\cite{doll2024}.} localizers.
    
We are interested in characterizing how tightly localized are the \emph{localized states} derived from the Spatial Localizer eigenstates $\ket{\psi^L(\rv)} \in \hs_{\text{occ}}\otimes \hs_\Gamma$ such that $L(\rv)\ket{\psi^L(\rv)}=\mu(\rv)\ket{\psi^L(\rv)}$. Spatial localization is commonly characterized by the spread of position operators, i.e., 

\begin{align}
    \Delta X_j^2 = \expval{X_j^2}-\expval{X_j}^2.
    \label{eq:variance_global}
\end{align}
However, Spatial Localizers are constructed purely in the occupied physical Hilbert space, $\hs_{\text{occ}}$, while $X_j$ act on the global physical Hilbert space. As such, we can naturally use properties of a Spatial Localizer, such as the LIF and Schmidt values, to establish bounds on the spread of the localized states within the occupied subspace, characterized by their variances $\Delta \tilde{X}_j^2$ within the occupied subspace along each physical direction $j$, 
\begin{align}
    \Delta \tX_j^2 = \expval{\tX_j^2}-\expval{\tX_j}^2,
    \label{eq:variance_occ}
\end{align}
where $\tX_j$ are the projected position operators used in the construction of the Spatial Localizer. We note that the difference between the spreads in~\eqref{eq:variance_global} and~\eqref{eq:variance_occ} differ by the additive factor 

\begin{equation}
    \Delta X_j^2 - \Delta \tX_j^2 = \expval{X(1-P)X},
\end{equation}
which is commonly referred to as the gauge-invariant~\footnote{Note that this component is gauge-invariant with respect to a Wannier basis, where one fixes the distribution of weights in the BZ to be flat when constructing (Wannier) states. For a more general wave packet, the smallest singular value of the operator $X(I-P)X$ acts as a lower bound on the spread.} spread where $P$ is the projector onto occupied states. 

In the case of an OBC Spatial Localizer, the $\tX_j$ operators take the form
\begin{align}
\tX_j(r_j) = \halfproj^\dagger(X_j- r_jI)\halfproj,
\label{eq:tXoperator}
\end{align}
where $r_j$ is the coordinate in position space from which we ``extract'' the Spatial Localizer eigenstates (i.e., the coordinate $\rv$ in $L(\rv)$). In what follows, we assume the expectation values are always with respect to states within $\hs_{\text{occ}}$. First note that 
\begin{align}
    \expval{\tX_j(r_j)}=\expval{\tX_j(0)}-r_j
\end{align}
In particular, if the state is a Wannier function (WF), we have that
\begin{align}
\expval{\tX_j(r_j)}=r^W_j-r_j    
\end{align}
is the distance along direction $j$ from where we extract the state to the state's WC $\rv^W$.

The structure of localizers naturally yield bounds on the quantity $\texpval{\tX_j^2(r_j)}$, which can be related to the variance $\Delta\tX_j^2$ by

\begin{equation}
    \begin{split}
        \expval{\tX_j^2(r_j)}&= \expval{\tX_j^2(r_j)} - \expval{\tX_j(r_j)}^2 + \expval{\tX_j(r_j)}^2 \\
        &= \Delta\tX_j^2 + \expval{\tX_j(r_j)}^2,
    \end{split}
\end{equation}
i.e., $\expval{\tX_j^2(r_j)}$ and the variance are related by the squared deviation of the mean from the point $r_j$. Note that we omit the $r_j$ dependence of $\Delta\tX_j^2$ as the variance calculated from the operators \eqref{eq:tXoperator} is independent of $r_j$, i.e., 
\begin{align}
    \Delta \tX_j^2 &= \expval{\tX_j^2(r_j)}-\expval{\tX_j(r_j)}^2 \nonumber\\
    &= \expval{\tX_j^2(0)}-\expval{\tX_j(0)}^2
\end{align}
 The most general form of a Spatial Localizer is
\begin{equation}
\label{app:eq:general_OBC_loc_2}
    L(\rv) = \sum_j \tX_j(r_j)\otimes \Gamma_j.
\end{equation}
We consider this localizer $L(\rv)$ constructed from $d$ Hermitian operators $\tX_j(r_j)$ and Clifford elements $\Gamma_j$. We omit an OBC/PBC label on $L(\rv)$ as the entire lower bound and the derivation of the upper bound up to~\eqref{app:eq:final_bound_before_variance_form} is independent of whether or not the localizer of interest is periodic. That is, the only requirement is the Hermiticity of $\tX_j$, which holds for both the OBC components and the PBC components $\tX_{j,C}$ and $\tX_{j,S}$. (See Equations~\eqref{app:eq:general_OBC_loc} and~\eqref{app:eq:general_PBC_loc} respectively for the general formulations of OBC and PBC Spatial Localizers). 
Furthermore, for a PBC Spatial Localizer as in Eq. (4) of the main text, the derivation below bounds the quantity

\begin{equation}
    \sum_{j} \expval{\tX^2_{j,C}} + \expval{\tX^2_{j,S}},
\end{equation}
which we argue to be a relevant quantity for localization in PBC systems in Appendix~\ref{app:sec:MV_section}. We further note that we expect the OBC bound to hold for PBC systems in the thermodynamic limit as $\texpval{\tX^2_{j,C}} + \texpval{\tX^2_{j,S}} \approx \Delta k ^2 \texpval{\tX_j^2}$, as discussed in Section~\ref{app:sec:OBC_PBC_equivalence}.

A previously derived bound (see Lemma 1.2 of~\cite{loring2015}) on the localization of states extracted from a localizer is
    \begin{equation}
    \label{eq:loring_bound}
        \sum_j \expval{\tX_j^2} \leq g \bigg (\mu(\mathbf{r})^2 + \sum_{j < k} \norm{\comm{\tilde{X}_j}{\tilde{X}_k}} \bigg),
    \end{equation}
where $g$ is the dimension of the Clifford elements $\Gamma_j$ and $\norm{\cdot}$ is the spectral operator norm (i.e., largest singular value), can always be satisfied. If the expectation value is with respect to the first physical Schmidt vector, $\ket{\psch_1}$, we can replace $g$ with $\frac{1}{s_1}$. 
    
While this bound holds for a Spatial Localizer, we often observe the commutator norms between OBC projected position operators to diverge in the thermodynamic limit, diluting the information that the bound can give us.
Furthermore, using the method of choosing a state in the physical Hilbert space in~\cite{loring2015}, the lower component in~\eqref{eq:loring_bound} is highly dependent on the choice of basis for $\Gamma_j$, while the physical Schmidt vectors $\ket{\psch_i}$ that we utilize are independent of the choice of basis for $\Gamma_j$.

The use of the Schmidt decomposition of the localizer's eigenstate $\ket{\psi^L}=\sum_i s_i \ket{\psch_i} \otimes \ket{\esch_i}$, as considered in the main text, allows us to establish tighter bounds on the variance of the physical vectors $\ket{\psch_1}$. Crucially, these bounds (and the states $\ket{\psch_i}$) are independent of the basis choice for the embedding degrees of freedom (Clifford elements). In what follows, any implicit expectation values presented here are either with respect to $\ket{\psch_1}$ or $\ket{s_1}$, depending on whether the operator we are taking the expectation value of acts on $\hs_{\text{occ}}$ or $\hs_{\text{occ}} \otimes \hs_\Gamma$; e.g., 

    \begin{equation}\nonumber
        \begin{split}
            &\expval{\tX^2_j} \equiv \bra{\psch_1} \tX^2_j \ket{\psch_1}, \\
            &\expval{\tX^2_j \otimes I} \equiv \bra{s_1} \tX^2_j \otimes I \ket{s_1} . \\
        \end{split}
    \end{equation}

In light of the vantage point provided by the Schmidt decomposition of localizer eigenstates, we construct new upper bounds on the quantity $\sum_j {\texpval{\tX^2_j(\rv)}}$. 
    First, we define the relevant decomposition of $\ket{\psi^L}$, 
    \begin{equation}
    \begin{split}
        &\ket{\psi^L} = s_1 \ket{s_1} + \sqrt{1-s_1^2} \ket{\phi}, \\
        &\ket{\phi} = \sum_{i\neq 1} \frac{s_i}{\sqrt{1-s_1^2}}\ket{s_i}, \\
        &\ket{s_i} = \ket{\psch_i} \otimes \ket{\esch_i}, \\
    \end{split}
    \end{equation}
where $\ket{\phi}$ holds the remainder of the Schmidt decomposition. 
    
We can now present the upper and lower bounds on the variance of $\ket{\psch_1}$ for OBC Spatial Localizers, with 
    \begin{equation}
        \begin{split}\label{app:eq:general_bound}
    &\frac{\Delta\tilde{R}_H^2}{d-1} \leq \sum_j \Delta\tX_j^2  \leq  \Delta\tilde{R}_H^2 + \satresidual, \\ 
    &\Delta\tilde{R}_H^2 = \sum_{j < k} \abs{\expval{\comm{\tilde{X}_j}{\tilde{X}_k}}},\\
    &\satresidual = \mu^2 - \sum_j \left(\expval{\tX_j(r_j)}\right)^2 \\& \ \ \ \ \ \ \ + \sum_{j<k} \sum_{i\neq 1} \frac{s_i}{s_1} \abs{\bra{\psch_1} \comm{\tilde{X}_j}{\tilde{X}_k} \ket{\psch_i}},
    \end{split}
    \end{equation}
where it follows that, in the case of $d=2$, $\mu(\rv)=0$ and $s_1=1$, the HRUR is saturated ($\leq \ \rightarrow \ =$) via a squeeze theorem discussed in the next section. In the case of $d>2$, $\mu(\rv)=0$ and $s_1=1$, the bound is not quite saturated. This prompts an investigation into stronger uncertainty relations ~\cite{maccone_uncertainty}, to be discussed in future work. 

For completeness, we include the bound that is relevant for periodic localizers when not taking the thermodynamic limit approximation. We have the bounds 

\begin{equation}\label{app:eq:pbc_general_bound}
        \begin{split}
    &\frac{\Delta\tilde{R}_H^2}{2d'-1} \leq \sum_j \Delta\tX^2_{j,C} + \Delta\tX^2_{j,S}  \leq  \Delta\tilde{R}_H^2 + \satresidual, \\ 
    &\Delta\tilde{R}_H^2 = \sum_{j< k} \sum_{\substack{J=S,C \\ K = S,C}} \abs{\expval{\comm{\tilde{X}_{j,J}}{\tilde{X}_{k,K}}}},\\
    &\satresidual = \mu^2 - \left(\sum_j \expval{\tX_{j,C}}^2 + \expval{\tX_{j,S}}^2\right) \\ & \ \ \ \ \  \quad+ \sum_{j < k} \sum_{\substack{J=S,C \\ K = S,C}} \sum_{i\neq 1} \frac{s_i}{s_1} \abs{\bra{\psch_1} \comm{\tilde{X}_{j,J}}{\tilde{X}_{k,K}} \ket{\psch_i}}, 
    \end{split}
    \end{equation}
where $d'\geq d$ is the number of Resta position operators used to construct a periodic Spatial Localizer, thus requiring $2d'$ Hermitian operators (one real and one imaginary component per Resta position operator) to be included in the construction of a periodic Spatial Localizer. We have omitted the commutator terms of the form $\comm{\tX_{j,C}}{\tX_{j,S}}$ as those terms vanish if $\tX_{j,R}$ are unitary, which can approximated in the approach of the thermodynamic limit or manually enforced via a singular value decomposition of $\tX_{j,R}$ (See Appendix~\ref{app:wilson_loops}).

\subsection{Derivation of lower bound on a sum of variances} \label{app:sec:sum_HRUR}

Our starting point is the Heisenberg-Robertson uncertainty relation (HRUR)~\cite{robertson_1929} for two observables (Hermitian operators) $\tX$ and $\tilde{Y}$, 
\begin{equation}
    \frac12\abs{\expval{\comm{\tX}{\tY}}}\leq(\Delta\tX)(\Delta\tilde{Y}),
\end{equation}
which holds for \emph{any} state in the Hilbert space that $\tX$ and $\tilde{Y}$ act on. While this bound can be informative, the localizer structure naturally provides bounds on a sum of variances, rather than a product. Therefore, we note the relation
\begin{equation} \nonumber
    2\Delta\tX \Delta\tilde{Y} \leq (\Delta\tX)^2 + (\Delta\tilde{Y})^2,
\end{equation}
which follows from $(\Delta\tX - \Delta\tilde{Y})^2 \geq 0$. We then have 
\begin{equation}
\label{eq:uncertainty_relation}
    \abs{\expval{\comm{\tX}{\tilde{Y}}}} \leq (\Delta\tX)^2 + (\Delta\tilde{Y})^2.
\end{equation}
The bound~\eqref{eq:uncertainty_relation} was also derived in~\cite{maccone_uncertainty}. While ~\eqref{eq:uncertainty_relation} is useful in the case of two operators, we wish to generalize it to $d \geq 2$ operators since the structure of a localizer generalizes to arbitrarily many (Hermitian) operators. To do this, we note that
\begin{align} 
    \sum_{j=1}^d \Delta\tX_j^2 =  \sum_{j<k,k=2}^d \left(\frac{\Delta\tX_j^2 + \Delta\tX_k^2}{d-1} \right), \\
     \sum_{j<k,k=2}^d \left(\frac{\Delta\tX_j^2 + \Delta\tX_k^2}{d-1} \right) \geq \sum_{j<k} \frac{\abs{\expval{\comm{\tX_j}{\tX_k}}}}{d-1}.
\end{align}
We now have a lower bound on the sum of the variances of arbitrarily many operators that is in the spirit of the HRUR. Namely, we have
\begin{equation}
    \sum_{j<k} \frac{\abs{\expval{\comm{\tX_j}{\tX_k}}}}{d-1} \leq \sum_{j=1}^d \Delta\tX_j^2 \leq \sum_{j=1}^d \expval{\tX^2_j}.
\end{equation}

\subsection{Derivation of upper bound on a sum of variances} \label{app:sec:upper_bound_derivation}

We note that, in this derivation, we omit the explicit $\rv$ dependence on operators and states for brevity until the $\rv$ dependence is relevant in Eq.~\eqref{app:eq:variance_extraction}. To find an upper bound for the sum of variances, consider a localizer eigenstate $\ket{\psi^L}\in\hs_{\text{occ}}\otimes\hs_\Gamma$ such that $L\ket{\psi^L}=\mu\ket{\psi^L}$ and $L^2\ket{\psi^L}=\mu^2\ket{\psi^L}$.
$\ket{\psi^L}$ admits a Schmidt decomposition 
\begin{equation} \nonumber
    \ket{\psi^L}=\sum_i s_i \ket{s_i} = \sum_i s_i \ket{\psch_i} \otimes \ket{\esch_i}.
\end{equation}
The bound discussed in Lemma 1.2 of~\cite{loring2015} and reproduced in \eqref{eq:loring_bound} informs us that $\ket{\psch_1}$ may be optimally localized (i.e, with minimal variance) with respect to the operators used to construct a localizer. As such, we consider $\ket{\psi^L}$ decomposed as
\begin{equation}
    \ket{\psi^L} = s_1 \ket{s_1} + \sqrt{1-s_1^2} \ket{\phi},
\end{equation}
where $\ket{\phi}$ contains the rest of the Schmidt decomposition and is orthogonal to $\ket{s_1}$, i.e., $\bra{s_1}{\phi}\rangle=0$.

In a similar spirit, we aim to provide an upper bound on the variance of $\ket{\psch_1}$ that we can directly relate to the Heisenberg-Robertson uncertainty relation. We consider the expectation value of  the scalar component of $L^2$ [see eq.~\eqref{eq:app_loc_squared}] with respect to $\ket{s_1}$
\begin{equation} \nonumber
        {\sum_j \expval{\tilde{X}^2_j \otimes I}} =  \abs{\expval{L^2} - \sum_{j < k} \expval{\comm{\tilde{X}_j}{\tilde{X}_k} \otimes \Gamma_j\Gamma_k}}.
\end{equation}
We first note that 
\begin{equation}\nonumber
    \begin{split}
        \bra{s_1} {\tilde{X}^2_j} \otimes I \ket{s_1} &= \bra{\psch_1} {\tilde{X}^2_j} \ket{\psch_1} \bra{\esch_1} I \ket{\esch_1} \\ 
        &= \bra{\psch_1} {\tilde{X}^2_j} \ket{\psch_1},
    \end{split}
\end{equation}
yielding
\begin{equation} \nonumber
        {\sum_j \expval{\tilde{X}^2_j}} =  \abs{\expval{L^2} - \sum_{j < k} \expval{\comm{\tilde{X}_j}{\tilde{X}_k} \otimes \Gamma_j\Gamma_k}}.
\end{equation}
By the triangle inequality, $\abs{a+b} \leq \abs{a} + \abs{b}$, we have
\begin{equation} \nonumber
        \sum_j \expval{\tilde{X}^2_j} \leq  \abs{\expval{L^2}} + \sum_{j < k} \abs{\expval{\comm{\tilde{X}_j}{\tilde{X}_k} \otimes \Gamma_j\Gamma_k}}.
\end{equation}

Now we estimate the commutator term. Since the expectation value is w.r.t. a product state $\ket{s_1} = \ket{\psch_1} \otimes \ket{\esch_1}$, we have
\begin{equation}\nonumber
    \begin{split}
        \abs{\expval {\comm{\tilde{X}_j}{\tilde{X}_k} \otimes \Gamma_j\Gamma_k} } = \abs{\bra{\psch_1} \comm{\tilde{X}_j}{\tilde{X}_k} \ket{\psch_1} \bra{\esch_1} \Gamma_j\Gamma_k \ket{\esch_1}}.
    \end{split}
\end{equation}
We then use the fact that $\norm{\Gamma_j\Gamma_k} \leq \norm{\Gamma_j}\norm{\Gamma_k}$ and $\norm{\Gamma_j}=1$ to remove the Clifford element component, yielding 
\begin{equation}\nonumber
    \begin{split}
        \abs{\bra{s_1} \comm{\tilde{X}_j}{\tilde{X}_k} \otimes \Gamma_j\Gamma_k \ket{s_1}} \leq \abs{\bra{\psch_1} \comm{\tilde{X}_j}{\tilde{X}_k} \ket{\psch_1} }.
    \end{split}
\end{equation}
We can now substitute back into the main bound, yielding 

\begin{equation} \nonumber
        \sum_j \expval{\tilde{X}^2_j} \leq  \abs{\expval{L^2}} + \sum_{j < k} \abs{\expval{\comm{\tilde{X}_j}{\tilde{X}_k}}}.
\end{equation}

Note that $L^2$ is a positive-semidefinite operator, i.e., $\bra{s_1} L^2 \ket{s_1} \geq 0$. This makes the absolute value around $\expval{L^2}$ optional, although we retain it for this derivation so that terms will collect into a simpler expression. We now aim to make estimates of $\expval{L^2}$. We consider

\begin{equation}\nonumber
    \begin{split}
        &\bra{s_1}L^2\ket{\psi^L} = \bra{s_1} L^2 \left(s_1 \ket{s_1} + \sqrt{1-s_1^2} \ket{\phi}\right), \\
        \implies &\mu^2 s_1 = s_1\bra{s_1} L^2 \ket{s_1} + \sqrt{1-s_1^2}\bra{s_1} L^2 \ket{\phi}, \\
        \implies &\bra{s_1} L^2 \ket{s_1} = \mu^2 - \frac{\sqrt{1-s_1^2}}{s_1} \bra{s_1} L^2 \ket{\phi}.
    \end{split} 
\end{equation}

 We then have the updated bound,

\begin{equation}
    \begin{split}
\sum_j \expval{\tilde{X}^2_j} \leq \ & \mu^2 + \frac{\sqrt{1-s_1^2}}{s_1} \abs{\bra{s_1} L^2 \ket{\phi}} \\ + &\sum_{j < k} \abs{\expval{\comm{\tilde{X}_j}{\tilde{X}_k}}},
    \end{split}
\end{equation}
where again we utilized the triangle inequality.
We now estimate the $\abs{\bra{s_1} L^2 \ket{\phi}}$ term by,

\begin{equation}\label{app:eq:expanded_cross_term}
    \begin{split}
    \abs{\bra{s_1} L^2 \ket{\phi}} \leq &\sum_j \abs{\bra{s_1} \tX_j^2 \otimes I \ket{\phi}} \\
    &+ \sum_{j<k}\abs{\bra{s_1}  \comm{\tilde{X}_j}{\tilde{X}_k} \otimes \Gamma_j \Gamma_k \ket{\phi}}.
    \end{split}
\end{equation}
Looking at the first term, we have 

\begin{equation} \nonumber
    \bra{s_1} \tX_j^2 \otimes I \ket{\phi} = \sum_{i\neq 1} \frac{s_i}{\sqrt{1-s_1^2}} \bra{\psch_1}\tX_j^2\ket{p_i}\bra{\esch_1} I\ket{\esch_i},
\end{equation}
which vanishes due to $\bra{\esch_1} I\ket{\esch_i}=\bra{\esch_1}{\esch_i}\rangle=0$ for $i\neq 1$ by the properties of the Schmidt decomposition. We now make estimates of the commutator terms in Eq. \eqref{app:eq:expanded_cross_term},

\begin{equation}
    \begin{split}
    &\abs{\bra{s_1}  \comm{\tilde{X}_j}{\tilde{X}_k} \otimes \Gamma_j \Gamma_k \ket{\phi}} \\ &= \abs{\sum_{i\neq 1}\frac{s_i}{\sqrt{1-s_1^2}} \bra{\psch_1}\comm{\tX_j}{\tX_k}\ket{\psch_i}\bra{\esch_1}\Gamma_j\Gamma_k\ket{\esch_i}}, \\
    &\leq \sum_{i\neq 1}\frac{s_i}{\sqrt{1-s_1^2}}\abs{\bra{\psch_1}\comm{\tX_j}{\tX_k}\ket{\psch_i}},
    \end{split}
\end{equation}
yielding the bound 

\begin{equation}\label{app:eq:final_bound_before_variance_form}
    \begin{split}
\sum_j \expval{\tilde{X}^2_j} \leq \ & \mu^2 + \sum_{j<k} \sum_{i\neq 1} \frac{s_i}{s_1} \abs{\bra{\psch_1} \comm{\tilde{X}_j}{\tilde{X}_k} \ket{\psch_i}} \\ + &\sum_{j < k} \abs{\expval{\comm{\tilde{X}_j}{\tilde{X}_k}}}.
    \end{split}
\end{equation}
Lastly, using 

\begin{equation}\label{app:eq:variance_extraction}
    \sum_j  {\expval{\tX^2_j(r_j)}} = \sum_j \left( \Delta\tX_j^2 + \expval{(\tX_j(0) - r_j)}^2\right ),
\end{equation}
we have the final upper bound

\begin{equation}
    \begin{split}\label{eq:final_bound}
\sum_j \Delta\tX_j^2 \leq \ & \mu^2(\rv) - \sum_j \left(\expval{\tX_j(0)} - r_j\right)^2 + \sum_{j < k} \abs{\expval{\comm{\tilde{X}_j}{\tilde{X}_k}}} \\ + & \sum_{j<k} \sum_{i\neq 1} \frac{s_i}{s_1} \abs{\bra{\psch_1(\rv)} \comm{\tilde{X}_j}{\tilde{X}_k} \ket{\psch_i(\rv)}}.
    \end{split}
\end{equation}

Note that the step in~\eqref{app:eq:variance_extraction} is only meaningfully valid for PBC localizers in the thermodynamic limit. See Appendices~\ref{app:sec:OBC_PBC_equivalence} and ~\ref{app:sec:MV_section} for discussion on the bounds for PBC Spatial Localizers. From Eq.~\eqref{eq:final_bound}, we see that the LIF of $L(\mathbf{r})$, the Schmidt values of $\ket{\psi^L(\rv)}$ and the first off diagonal of the commutators $\comm{\tilde{X}_j}{\tilde{X}_k}$ projected into the basis of physical Schmidt vectors play a role in determining an upper bound on the variance of the physical Schmidt vector $\ket{\psch_1(\rv)}$. We note that the LIF can be replaced with any eigenvalue of $L(\rv)$, so long as one considers physical Schmidt vectors of the corresponding localizer eigenvector.

\subsection{Special case of 2 spatial dimensions: saturating the uncertainty relation} \label{app:sec:alt_bound_2D}

In 2 spatial dimensions, we have $\tX$ and $\tilde{Y}$, with the upper bound 

\begin{equation}\nonumber
    \begin{split}
     \Delta \tilde{X}^2 + \Delta \tilde{Y}^2 \leq &\mu^2 +  \sum_{i\neq 1} \frac{s_i}{s_1} \abs{\bra{\psch_1} \comm{\tilde{X}}{\tilde{Y}} \ket{\psch_i}}\\ + &\abs{\expval{\comm{\tilde{X}}{\tilde{Y}}}} - \norm{\expval{\tilde{\mathbf{R}}(\rv)}}^2,
     \end{split}
\end{equation}
Such a bound is always valid in the case of OBC, and is valid in the case of PBC in the thermodynamic limit, provided $\ket{p_1}=\ket{p_1(\rv)}$ only has support near the extraction point, $\rv$ (Appendix~\ref{app:sec:OBC_PBC_equivalence}). Additionally, we have the uncertainty relation of Eq. \eqref{eq:uncertainty_relation}, yielding

\begin{equation}\label{app:eq:2D_bound}
    \begin{split}
        &\Delta\tilde{R}_H^2 \leq \Delta \tilde{X}^2 + \Delta \tilde{Y}^2 \leq \Delta\tilde{R}_H^2+\satresidual, \\
        &\Delta\tilde{R}_H^2 = \abs{\expval{\comm{\tilde{X}}{\tilde{Y}}}} , \\
        &\satresidual = \mu^2 - \norm{\expval{\tilde{\mathbf{R}}(\rv)}}^2 +  \sum_{i\neq 1} \frac{s_i}{s_1} \abs{\bra{\psch_1} \comm{\tilde{X}}{\tilde{Y}} \ket{\psch_i}} ,  \\
    \end{split}
\end{equation}
where we see that the uncertainty bound is saturated if we have $\mu=0$ and $s_1=1$ (recall that $\sum_is_i^2=1$). Thus, for a 2D Spatial Localizer, the combination of $\mu$ and $s_1$ directly indicates how close we can get to saturating the HRUR with physical Schmidt vectors. 

A natural further question is ``when are we guaranteed to have $\mu=0$ and/or $s_1=1$?'' 
In 2D OBC, we will always have $N_{occ}$ unique zeros of the LIF ($\mu=0$ points), where by ``unique'' we mean that the zeros correspond to distinct $\ket{p_1}$. To see this, consider the square of a 2D OBC localizer, 
\begin{equation}
    \begin{split}
        L^2(\rv) &= (\tX^2 + \tY^2) \otimes I + \ii \comm{\tX}{\tY} \otimes \sigma_z,\\
        &= \tZ^\dagger \tZ \oplus \tZ \tZ^\dagger,
    \end{split} 
\end{equation}
where $\tZ=\tZ(\rv)=\tX(x) + \ii \tY(y)$. Thus, the LIF of $L(\rv)$ is equivalent to the square root of the LIF of $\tZ^\dagger(\rv) \tZ(\rv)$. Thus, the LIF of $\tZ^\dagger(\rv) \tZ(\rv)$ will have $N_{\mathrm{occ}}$ LIF zeros, with the locations corresponding to the spectrum of $\tZ(\mathbf{0})$. This can be seen by looking at an alternate representation of $\tZ(\rv)$,

\begin{equation}
\begin{split}
    \tZ(\rv) &= \tX(x) + \ii\tY(y), \\
    &= \tX(0) + \ii\tY(0) - (x + \ii y)\halfproj^\dagger \halfproj,\\
    &= \tZ(\mathbf{0}) - \lambda(\rv)\halfproj^\dagger \halfproj,
\end{split}
\end{equation}
where $\lambda(\rv)$ is a number in the complex plane, determined by our choice of $\rv$. From a spectral decomposition of $\tZ(\mathbf{0})$, one can see that $\tZ(\rv)$ will have a zero eigenvalue iff $\lambda(\rv)$ is equal to one of the $N_{\mathrm{occ}}$ eigenvalues of $\tZ(\mathbf{0})$. $\tZ(\rv)$ then has $N_{\mathrm{occ}}$ locations where $\mu(\rv;\tZ)=0$ (counting degeneracies). $\tZ^\dagger(\rv)\tZ(\rv)$ and $\tZ(\rv)\tZ^\dagger(\rv)$ then have zero eigenvalues at the same locations as $\tZ(\rv)$ since $\sigma(\tZ(\rv) \tZ^\dagger(\rv)) = \sigma(\tZ^\dagger(\rv) \tZ(\rv)) = \{s_i(\tZ(\rv))^2\}$ where $s_i(\tZ(\rv))$ are the singular values of $\tZ(\rv)$ (for non-normal operators, such as $\tZ$, the LIF should be defined using the singular value spectrum rather than the eigenvalue spectrum). We can then find the minima of the LIF for a 2D OBC system by simply diagonalizing $\tZ(\mathbf{0})$. Additionally, at the locations where $\lambda(\rextract)$ is an eigenvalue of $\tZ(\mathbf{0})$, the Spatial Localizer $L(\rextract)$ will have at least a two-fold degenerate spectrum at zero by the chiral symmetry of $L(\rv)$. Diagonalizing this subspace with respect to the chiral operator ($I \otimes \sigma_z$ for 2D OBC) yields distinct localizer eigenstates corresponding to right/left eigenstates of $\tZ(\mathbf{0})$ corresponding to the eigenvalue $\lambda(\rextract)$. To see this, note that the 2D OBC Spatial Localizer has the form 

\begin{equation}
    L(\rv) = \begin{pmatrix} 
        0 & \tZ^\dagger(\rv) \\
        \tZ(\rv) & 0
    \end{pmatrix}.
\end{equation}
In the case of a two-fold degenerate subspace corresponding to $\mu(\rextract)=0$, the basis of this subspace can be chosen as the eigenstates of the chiral operator, notated as $\ket{\psi_1(\rextract)}$ and $\ket{\psi_2(\rextract)}$ with the form

\begin{equation}
\begin{split}
    &\ket{\psi_1(\rextract)} = \begin{pmatrix}
        \ket{\psi_\tZ(\lambda(\rextract)} \\ 0
    \end{pmatrix},\\ 
    &\ket{\psi_2(\rextract)} = \begin{pmatrix}
        0 \\ \ket{\psi_{\tZ^\dagger}(\lambda(\rextract)}
    \end{pmatrix},
\end{split}
\end{equation}

where $\ket{\psi_\tZ(\lambda(\rextract)}$ ($\ket{\psi_{\tZ^\dagger}(\lambda(\rextract)}$) is the right (left) eigenvector of $\tZ(\mathbf{0})$ corresponding to the eigenvalue $\lambda(\rv^*)$ and saturate the HRUR~\cite{trifonov1994}. We note that operators like $\tZ(\rv)$ have been referred to as a ``vortex function'' in the context of the vortexability of Chern bands~\cite{ledwith2023,okuma2024}. However, our use case in this work is more general, as we are interested in both atomic and topological bands.

\subsection{Equivalence between OBC and PBC in the thermodynamic limit}\label{app:sec:OBC_PBC_equivalence}

In this section, we discuss the equivalence in properties of OBC and PBC Spatial Localizers in the thermodynamic limit. Specifically, we (i) make the argument that the bounds derived for OBC Spatial Localizers can apply to PBC Spatial Localizers (in the thermodynamic limit) provided $\ket{\psch_1}$ is localized and (ii) show that there exist at least $N_\text{occ}$ points for a PBC Spatial Localizer where the LIF scales as $\mu(r^*)\propto N^{-2}$, which correspond to the WCs.

\subsubsection{Small angle approximation on localized states}
Recall that the bounds for a PBC Spatial Localizer contain terms that look like $\bra{p_1(\rv)}\tX_{j,J}(\rv) \tX_{k,K}(\rv)\ket{p_i(\rv)}$ where $j,k$ denote spatial directions and $J,K\in\{S,C\}$ denote the real/imaginary components of the projected Resta position operators used to construct a PBC Spatial Localizer.
Concretely, we have the components 

\begin{equation}
\label{app:eq:Xc_Xs}
\begin{split}
    \tX_{j,C}(\rv) &= P\cos\left(\frac{(\hat{\Rv} - \rv I) \cdot \bv_j}{N_j}\right)P - P, \\
    \tX_{j,S}(\rv) &= P\sin\left(\frac{(\hat{\Rv} - \rv I) \cdot \bv_j}{N_j}\right)P,
\end{split}
\end{equation}
which are equivalent to the definition given in Equation~\eqref{app:eq_resta_real_imag} up to whether we use the full projector, $P$, or the half-projector, $\halfproj$. We emphasize that the choice of $P$ or $\halfproj$ amounts to a choice of basis so long as one is working with states in the occupied subspace corresponding to the projector $P$. Furthermore, we note that the effect of shifting $\rv$ on the operators \eqref{app:eq:Xc_Xs} amounts to a shift in the extraction point of the Spatial Localizer.

We now consider a localized wave packet, $\ket{\psi_{\rv'}}$ that lies purely in the occupied subspace and is localized around some point $\rv'$, i.e.,

\begin{equation}
\begin{split}
    &P\ket{\psi_{\rv'}} = \ket{\psi_{\rv'}}, \\
    &\bra{\rv}{\psi_{\rv'}}\rangle \approx 0 \ \  \text{for} \ \  \abs{\rv-\rv'} > \epsilon,
\end{split}
\end{equation}
where $\epsilon$ is some positive number that denotes the radius of the region in which $\ket{\psi_{\rv'}}$ has support. We represent the expansion of $\ket{\psi_{\rv'}}$ in the position basis as $\ket{\psi_{\rv'}}=\sum_\rv \alpha_\rv\ket{\rv}$ where $\alpha_\rv\equiv\bra{\rv}{\psi_{\rv'}}\rangle$. We now consider the action of $\tX_{j,C}(\rv')$ and $\tX_{j,S}(\rv')$ on $\ket{\psi_{\rv'}}$. For $\tX_{j,C}(\rv')\ket{\psi_{\rv'}}$, we have 

\begin{equation}
\begin{split}
    \tX_{j,C}&(\rv')\ket{\psi_{\rv'}} = \left[P\cos\left(\frac{(\hat{\Rv} - \rv' I) \cdot \bv_j}{N_j}\right)P - P\right]\ket{\psi_{\rv'}} \\
    &= P\cos\left(\frac{(\hat{\Rv} - \rv' I) \cdot \bv_j}{N_j}\right)\ket{\psi_{\rv'}} - P\ket{\psi_{\rv'}} \\
    &= P\cos\left(\frac{(\hat{\Rv} - \rv' I) \cdot \bv_j}{N_j}\right)\sum_\rv \alpha_\rv\ket{\rv} - P\sum_\rv \alpha_\rv\ket{\rv} \\
    &= P\sum_\rv \alpha_\rv \cos\left(\frac{(\hat{\Rv} - \rv' I) \cdot \bv_j}{N_j}\right)\ket{\rv} - P\sum_\rv \alpha_\rv\ket{\rv} \\
    &= P\sum_\rv \alpha_\rv \cos\left(\frac{(\rv - \rv') \cdot \bv_j}{N_j}\right)\ket{\rv} - P\sum_\rv \alpha_\rv\ket{\rv} \\
    &= P\sum_\rv \alpha_\rv \left[\cos\left(\frac{(\rv - \rv') \cdot \bv_j}{N_j}\right) - 1\right]\ket{\rv}. \\
\end{split}
\end{equation}
For $\tX_{j,S}(\rv')\ket{\psi_{\rv'}}$, we have 

\begin{equation}
    \begin{split}
        \tX_{j,S}(\rv')\ket{\psi_{\rv'}} &= P\sin\left(\frac{(\hat{\Rv} - \rv' I) \cdot \bv_j}{N_j}\right)P\ket{\psi_{\rv'}} \\
        &= P\sin\left(\frac{(\hat{\Rv} - \rv' I) \cdot \bv_j}{N_j}\right)\ket{\psi_{\rv'}} \\
        &= P\sin\left(\frac{(\hat{\Rv} - \rv' I) \cdot \bv_j}{N_j}\right)\sum_\rv \alpha_\rv\ket{\rv} \\
        &= P\sum_\rv \alpha_\rv\sin\left(\frac{(\hat{\Rv} - \rv' I) \cdot \bv_j}{N_j}\right)\ket{\rv} \\
        &= P\sum_\rv \alpha_\rv\sin\left(\frac{(\rv - \rv') \cdot \bv_j}{N_j}\right)\ket{\rv}. \\
    \end{split}
\end{equation}
For clarity of discussion, we have the net results of the application of $\tX_{j,J}(\rv')$ on $\ket{\psi_{\rv'}}$ as 

\begin{equation}
    \begin{split}
        \tX_{j,C}(\rv')\ket{\psi_{\rv'}} &= P\sum_\rv \alpha_\rv \left[\cos\left(\frac{(\rv - \rv') \cdot \bv_j}{N_j}\right) - 1\right]\ket{\rv} \\
        \tX_{j,S}(\rv')\ket{\psi_{\rv'}} &=P\sum_\rv \alpha_\rv\sin\left(\frac{(\rv - \rv') \cdot \bv_j}{N_j}\right)\ket{\rv}.
    \end{split}
\end{equation}
Recalling that $\alpha_\rv\approx0$ for $\abs{\rv-\rv'} > \epsilon$, we can make a small angle approximation of the cosine and sine terms for small $\frac{\epsilon}{N_j}\ll 1$. This yields 

\begin{equation}
    \begin{split}
        \tX_{j,C}(\rv')\ket{\psi_{\rv'}} &\approx P\sum_\rv \alpha_\rv \left[1 - 1\right]\ket{\rv} = 0, \\
        \tX_{j,S}(\rv')\ket{\psi_{\rv'}} &\approx P\sum_\rv \alpha_\rv\left(\frac{(\rv - \rv') \cdot \bv_j}{N_j}\right)\ket{\rv} \\
        &=\frac{2\pi}{N_j}\tX_j(r'_j) \ket{\psi_{\rv'}}.
    \end{split}
\end{equation}

We have now shown that, for a state $\ket{\psi_{\rv'}}$ localized around $\rv'$, the action of $\tX_{j,C}(\rv')$ and $\tX_{j,S}(\rv')$ on $\ket{\psi_\rv'}$ can be approximated respectively as 0 and $\frac{2\pi}{N_j}\tX_j(r'_j)$ for $2\pi r'_j = \rv' \cdot \bv_j$ due to the reciprocal lattice relation $\mathbf{a}_i \cdot \bv_j=2\pi\delta_{ij}$. Thus, if the first physical Schmidt vector, $\ket{\psch_1(\rv')}$, remains localized around the point $\rv'$ as $N_j$ increases (which we observe for all PBC systems tested), the OBC and PBC bounds on the sum of variances derived earlier in this section become equivalent in the thermodynamic limit. 
We note that, while exponential localization of $\ket{\psch_1(\rv')}$ is optimal for the equivalence to hold at smaller $N_j$, the equivalence holds as $N\rightarrow \infty$ so long as $\alpha_\rv\approx0$ for $\abs{\rv-\rv'} > \epsilon$ regardless of other qualities of the distribution of $\alpha_\rv$.

\subsubsection{$\order{N^{-2}}$ indicator scaling at $N_\textrm{occ}$ points}
In this section, we discuss the existence of $N_\textrm{occ}$ points where the LIF scales as $\mu(\rv)\propto N^{-2}$ in the thermodynamic limit. We begin by stating some assumptions required for the scaling behavior.

\begin{assumption}[Existence of localized states]
\label{ass:localized_sector}
Fix a radius $\epsilon>0$ that is independent of system size. We consider $\mathcal{S}_{\text{loc}(\rv';\epsilon)}\subset \hs_\text{occ}$ to be the subset of localized states whose position-space weight is concentrated within a distance $\epsilon$ of $\rv'$. We assume $\mathcal{S}_{\text{loc}(\rv';\epsilon)}\subset \hs_\text{occ}$ is non-empty at points of interest.
\end{assumption}
We define the operator norm on the restricted sector as 
\begin{equation}
    \norm{A}_{\mathrm{loc}(\rv')} = \max_{\ket{\psi} \in \mathcal{S}_{\text{loc}(\rv';\epsilon)}} \norm{A\ket{\psi}},
\end{equation}
where we have $\norm{\ket{\psi}}=1$. Note that $\norm{A}_{\mathrm{loc}(\rv')}$ is not the full operator norm of $A$, but only the norm of $A$ acting on localized states.

\begin{assumption}[Small norms on localized states]
\label{ass:small_norms}
    There is a constant $c_\epsilon$ (independent of system size) such that, for all $\ket{\psi} \in \mathcal{S}_{\text{loc}(\rv';\epsilon)}$, 

    \begin{equation}
        \norm{\frac{2\pi}{N_x}\tX(x') \ket{\psi}} < \frac{{c_\epsilon}}{N_X}, \quad \norm{\frac{2\pi}{N_y}\tY(y') \ket{\psi}} < \frac{{c_\epsilon}}{N_y}, 
    \end{equation}

    where $\rv'=(x',y')$. 

\end{assumption}

We now note that the 2D PBC Spatial Localizer can be written as 

\begin{equation}
    L_N(\rv) = L_\text{eff}(\rv) + \Delta_N(\rv),
\end{equation}
where
\begin{equation}
    L_\text{eff}(\rv) = \frac{2\pi}{N}\tX(x)\otimes \Gamma_2 + \frac{2\pi}{N}\tY(y)\otimes \Gamma_4,
\end{equation}
with $\Delta_N(\rv)$ holding the rest of the Spatial Localizer, and we assume $N_x=N_y=N$. We then have

\begin{equation}
    \norm{\Delta_N(\rv')}_{\mathrm{loc}(\rv')} = \order{N^{-2}}
\end{equation}
since $\epsilon$ is independent of system size.
$L_\text{eff}(\rv)$ can be seen as a 2D OBC Spatial Localizer utilizing a reducible representation of $\text{Cl}_{2,0}$, and thus there exists $N_\text{occ}$ choices of $\rv'=(x',y')$ such that $L_\text{eff}(\rv')$ has at least two zero modes where $L_\text{eff}(\rv')\ket{\psi^L(\rv')}=0$ where the $\rv'$ points are determined by the spectrum of $\tZ(\mathbf{0})=\tX(0) + \tY(0)$. These zero modes can be made to manifest as product states between $\hs_\text{occ}\otimes \hs_\Gamma$ where the physical Schmidt vectors are the right/left eigenvectors of $\tZ(\mathbf{0})$ and the embedding Schmidt vectors are eigenstates of the chiral operator (Appendix~\ref{app:sec:alt_bound_2D}). 
Let the right (left) eigenstates of $\tZ(\mathbf{0})$ notated as $\ket{\psi_\tZ(\lambda(\rv^*)}$ ($\ket{\psi_{\tZ^\dagger}(\lambda(\rv^*)}$) where $\lambda(\rv^*)=x^* + \ii y^*$ is a corresponding eigenvalue of $\tZ(\mathbf{0})$.

\begin{assumption}[Localization of $\tZ$ eigenstates]
\label{ass:psi_a_localized}
    For each $\rv^*$, we have $\ket{\psi_\tZ(\lambda(\rv^*)}\in\mathcal{S}_{\text{loc}(\rv^*;\epsilon)}$ for some $\epsilon$ that is independent of system size. 
\end{assumption}

Given assumptions \ref{ass:localized_sector}, \ref{ass:small_norms} and \ref{ass:psi_a_localized}, we have 

\begin{equation}
    \mu(\rv^*;L_N(\rv^*)) \leq \order{N^{-2}} \quad (N\rightarrow\infty).
\end{equation}

To see this, consider a (normalized) state $\ket{\Psi(\rv^*)} = \ket{\psi_\tZ(\lambda(\rv^*)} \otimes \ket{\zeta}$ where $\ket{\zeta}\in\hs_\Gamma$ is a unit vector such that $L_\text{eff}(\rv^*)\ket{\Psi(\rv^*)}=0$. The choice of $\ket{\zeta}$ can be the corresponding eigenvector of the embedding component of the chiral symmetry operator of the Spatial Localizer, as all PBC Spatial Localizers possess chiral symmetry. We then have

\begin{equation}
\begin{split}
    \norm{L_N(\rv^*) \ket{\Psi(\rv^*)}} &= \norm{\left(L_\text{eff}(\rv^*) + \Delta_N(\rv^*)\right)\ket{\Psi(\rv^*)}} \\
    &= \norm{\Delta_N(\rv^*)\ket{\Psi(\rv^*)}} \\
    &\leq \norm{\Delta_N(\rv^*)}_{\mathrm{loc}(\rv')}\norm{\ket{\Psi(\rv^*)}} \\
    \norm{L_N(\rv^*) \ket{\Psi(\rv^*)}} & \leq \order{N^{-2}}. \\
\end{split}
\end{equation}

\begin{figure}
    \centering
    \includegraphics[width=1\linewidth]{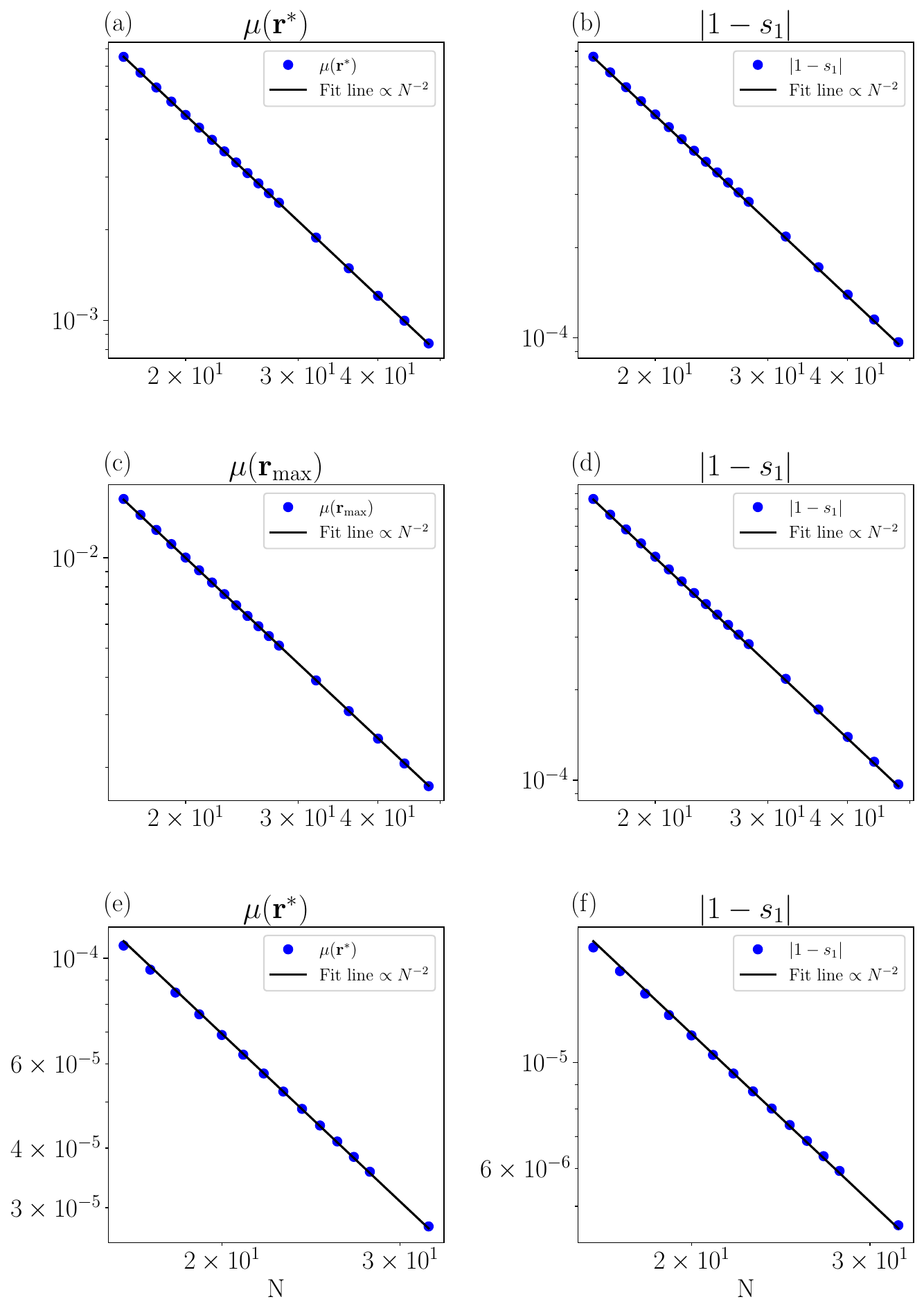}
    \caption{Scaling of the LIF (left) and first singular value (right) of the corresponding localizer eigenvector of the topological and trivial QWZ model. (a) and (b) correspond to the topological QWZ model at the minimum of the LIF, $\rv^*=(0,0)$. (c) and (d) correspond to the topological QWZ model at the maximum of the LIF, $\rv_\text{max}=(0.5,0.5)$ in units of lattice constant. (e) and (f) correspond to the trivial QWZ model at the minimum of the LIF, $\rv^*=(0,0)$. In all figures, the scatter points are numerical data and the solid lines are fit lines proportional to $N^{-2}$.}
    \label{fig:mu_s1_scaling}
\end{figure}

We have now proven the existence (given our assumption of $\ket{\psi_\tZ(\lambda(\rv^*)}$ being localized) of a state such that $\norm{L_N(\rv^*) \ket{\Psi(\rv^*)}} \leq \order{N^{-2}}$. By the variational principle, we then have that

\begin{equation}
    \mu(\rv^*;L_N(\rv^*)) \leq \order{N^{-2}} \quad (N\rightarrow\infty)
\end{equation}
since $\mu(\rv^*;L_N(\rv^*))\leq \norm{L_N(\rv^*)\ket{\Psi}} \ \forall \ \ket{\Psi}$ such that $\norm{\ket{\Psi}}=1$. As an example of this scaling, we present $\mu(\rv^*)$ for both the topological regimes of the QWZ model in Figure~\ref{fig:mu_s1_scaling} where we see $N^{-2}$ scaling in the LIF, $\mu$, for both the trivial and topological regimes of the Qi-Wu-Zhang (QWZ) model~\cite{qi2006} at the minimum of $\mu$. Furthermore, we see the same scaling behavior of $\mu$ at the maximal point in the unit cell for the topological QWZ model[Fig.~\ref{fig:mu_s1_scaling}]. This further justifies that we are obtaining a continuum of coherent-like states (zero or quasi-zero modes of $\tZ^\dagger(\rv)$) from the Spatial Localizer. Additionally, we observe the quantity $\abs{1-s_1}\propto N^{-2}$, suggesting that the localizer eigenstate corresponding to the LIF converges to the product state $\ket{\Psi(\rv^*)}$ (or its chiral counterpart) in the thermodynamic limit.

\subsection{Measuring the saturation of a bound}\label{app:sec:bounds_metrics}

\begin{figure}
    \centering
    \includegraphics[width=1\linewidth]{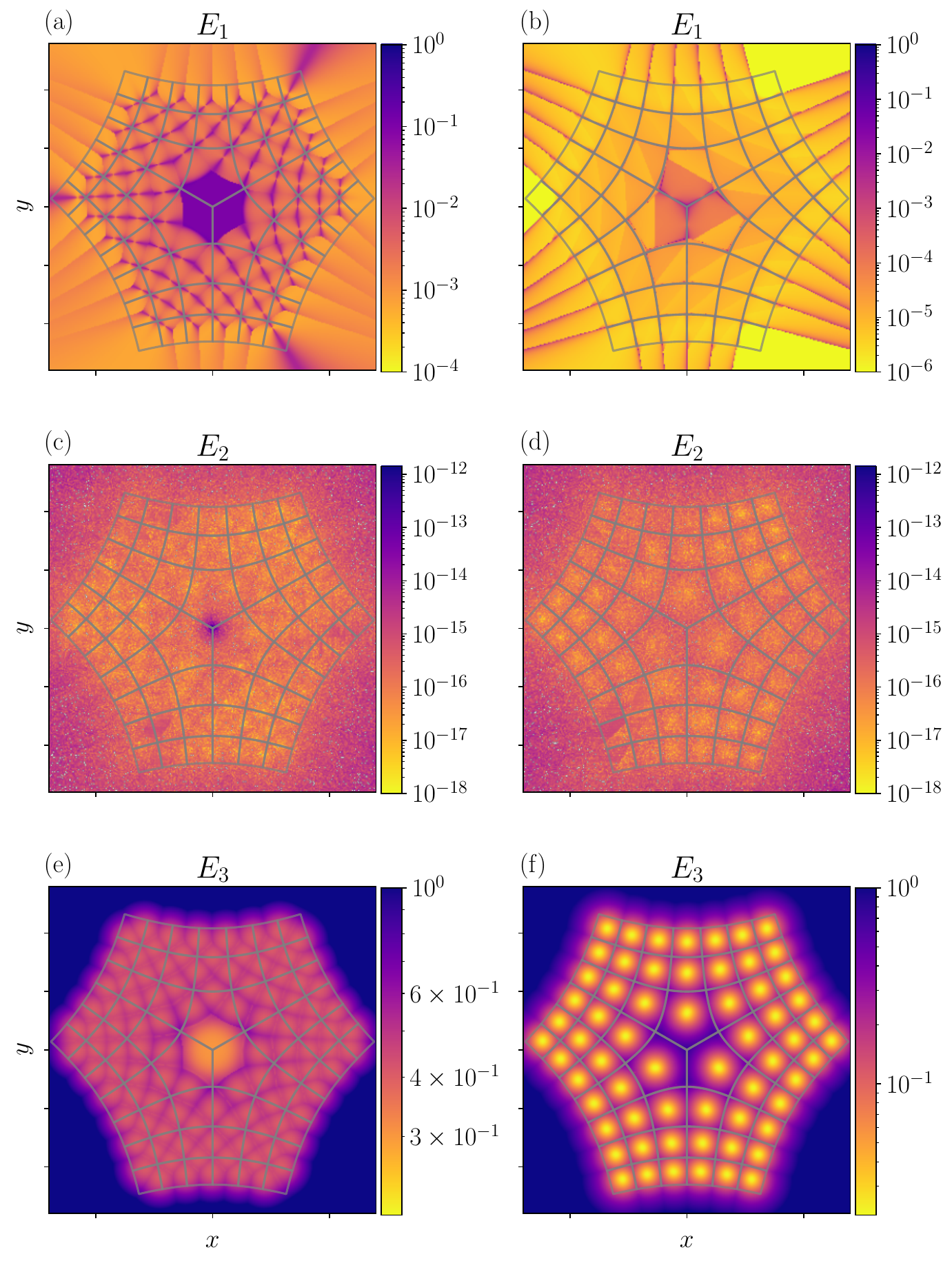}
    \caption{Measure of saturation of bounds for the topological (left) and trivial (right) disclination lattice presented in the main text. (a) and (b) correspond to the lower bound in Eq.~\eqref{app:eq:2D_bound}, with measure $E_1$. (c) and (d) correspond to the upper bound in Eq.~\eqref{app:eq:2D_bound}, with measure $E_2$. (e) and (f) correspond to the upper bound in Eq.~\eqref{eq:loring_bound}, with measure $E_3$. The general expressions used to calculate these quantities are presented in Section~\ref{app:sec:bounds_metrics}.}
    \label{fig:bound_demonstration} 
\end{figure}

Here, we write the quantities used to test the saturation of the bounds on variance discussed in previous sections. For any bound, $a \leq b$, we measure the saturation of the bound by the quantity $b-a$, which is non-negative provided $a \leq b$. We say that a bound has saturated if $b-a=0$. We use this measure for three different bounds. The first two are the lower (upper) bounds we derive in Appendix~\ref{app:sec:alt_bound}, referred to by $E_1$ ($E_2$). Furthermore, we test the previously derived~\cite{loring2015} upper bound in ~\eqref{eq:loring_bound}, which we refer to by $E_3$. The three $E_i$ are then 

\begin{equation}\label{app:eq:bound_test_quantities}
    \begin{split}
        E_1 &= \sum_j \Delta \tX_j^2 - \frac{\Delta\tilde{R}_H^2}{d-1}\\
        E_2 &= \Delta\tilde{R}_H^2 + \satresidual - \sum_j \Delta \tX_j^2 \\ 
        E_3 &= g \bigg (\mu(\mathbf{r})^2 + \sum_{j < k} \norm{\comm{\tilde{X}_j}{\tilde{X}_k}} \bigg) - \sum_j \expval{\tX_j^2},
    \end{split}
\end{equation}
where $\Delta\tilde{R}_H^2$ and $\satresidual$ are defined in Eq.~\eqref{app:eq:general_bound}. 
Note that, for PBC localizers, $d$ corresponds to twice the number of Resta position operators considered in the construction of the corresponding Spatial Localizer due to splitting each Resta position operator into real and imaginary components. We present these measures for the lattice with a disclination model in Figure~\ref{fig:bound_demonstration}.

\section{Periodic Spatial Localizers and the Marzari-Vanderbilt localization functional}\label{app:sec:MV_section}

Here, we discuss connections between the upper bound derived in Appendix \ref{app:sec:alt_bound} for PBC Spatial Localizers and the Marzari-Vanderbilt (MV) localization functional~\cite{marzari1997}. Specifically, we illustrate an alternative viewpoint on why our derived bound is relevant for periodic Spatial Localizers. That is, a viewpoint beyond the argument that $\tX_{j,C}(\rv) \approx 0$ and $\tX_{j,S}(\rv) \approx \frac{2\pi}{N}\tX_{j}(r_j)$ in the thermodynamic limit, where we then obtain the same viewpoint as in a OBC Spatial Localizer. We restrict our analysis to 1D here, but extending to multiple spatial dimensions and non-cubic lattices is straightforward and amounts to including additional projected Resta position operators in the definition of a Spatial Localizer. Concretely, we use one Resta position operator for each pair of $\pm \bv$ vectors used in the method of MV~\cite{marzari1997} to define an isotropic gradient in $k$-space. 

We note that this section considers localizers that use the full projector $P$ as opposed to the ``half-projector'' $\halfproj$ defined in the main text. Since we restrict our analysis to states that live in the occupied Hilbert space, the difference between the use of $P$ and $\halfproj$ amounts to a choice of basis. We use $\halfproj$ in the main text to remove the null space of $P$, which otherwise would pollute the LIF and the spectral bands of a Spatial Localizer.

\subsection{The Marzari-Vanderbilt localization functional}

The MV localization functional~\cite{marzari1997} can be described as the sum of variances of the WFs,

\begin{equation}
    \Omega = \sum_{n,j} \bra{\transvec n} X_j^2\ket{\transvec n} - \bra{\transvec n} X_j\ket{\transvec n}^2,
\end{equation}
where $\ket{\transvec n}$ is the Wannier state of band $n$ at unit cell $\transvec$ and the sum is over occupied bands $n$ and position directions $j$; where $X_j$ are the OBC position operators~\footnote{There is an underlying assumption that the OBC position operators are formulated such that there is no issue of the WF in question wrapping around the (periodic) boundary, which would artificially increase the spread.}. $\Omega$ can be split into

\begin{equation}
    \Omega = \Omega_I + \tilde{\Omega}, \ \ \tilde{\Omega} = \tilde{\Omega}_\text{D} + \tilde{\Omega}_\text{OD},
\end{equation}
where $\Omega_I$ is the gauge-independent component and  $\tilde{\Omega}$ the gauge-dependent component, which can be further split into the diagonal and off-diagonal components $\tilde{\Omega}_\text{D}$ and $\tilde{\Omega}_\text{OD}$. 
In position-space, the components of $\Omega$ arise from the manipulation,
\begin{equation}
    \begin{split}
        \Omega &= \Omega \\ &  + \sum_{n,j} \bra{\transvec n} PX_jPX_jP\ket{\transvec n} - \bra{\transvec n} PX_jPX_jP\ket{\transvec n} \\ 
        &= \Omega_I + \tilde{\Omega},
    \end{split}
\end{equation}
where,
\begin{equation}
    \begin{split}
        \Omega_I &= \sum_{n,j}\bra{\transvec n} PX_jQX_jP\ket{\transvec n}, \\
        \tilde{{\Omega}} &= \sum_{n,j}\bra{\transvec n} (PX_jP)^2\ket{\transvec n} - \bra{\transvec n} PX_jP\ket{\transvec n}^2,
    \end{split}
\end{equation}
with $Q=I-P$ as the projector onto the unoccupied bands, and we have utilized $P\ket{\transvec n}=\ket{\transvec n}$ to explicitly represent $\tilde{\Omega}$ as the variance of the projected position operator. We note that the Spatial Localizer is a tool that precisely minimizes the variance of the projected position operator.

In crystal momentum space, these quantities can be approximated as (see Section IV of~\cite{marzari1997}) 
\begin{equation}\label{app:eq:MV_variance_components}
    \begin{split}
        \Omega_I &= \frac{1}{N_\textrm{BZ}}\sum_{\kv,\mathbf{b}} w_b \Tr[P_\kv Q_{\kv + \mathbf{b}}], \\
        \tilde{\Omega}_\text{OD} &= \frac{1}{N_\textrm{BZ}}\sum_{\kv,\mathbf{b}} w_b \sum_{m\neq n} \abs{M_{mn}^{\kv,\mathbf{b}}}^2, \\
        \tilde{\Omega}_\text{D} &= \frac{1}{N_\textrm{BZ}}\sum_{\kv,\mathbf{b}} w_b \sum_{n} (-\text{Im}\ln M_{nn}^{\kv,\mathbf{b}} - \mathbf{b}\cdot\rv^W_n)^2,
    \end{split}
\end{equation}
where we have the quantities 
\begin{equation}
\begin{split}
    &M_{mn}^{\kv,\mathbf{b}} = \langle{u_{m\kv}}\ket{u_{n,\kv+\mathbf{b}}}, \\
    &\rv^W_n = -\frac{1}{N_\textrm{BZ}}\sum_{\kv,\mathbf{b}}w_b \mathbf{b} \text{Im} \ln M_{nn}^{\kv,\mathbf{b}}, \\
    &\sum_{n,j} \bra{\transvec n} X_j\ket{\transvec n}^2 = \sum_n \rv^W_n \cdot \rv^W_n.
\end{split}
\end{equation}

The vectors $\mathbf{b}$ and weights $w_b$ are chosen such that one has an isotropic gradient in $k$-space (see Appendix B of~\cite{marzari1997}) and $N_\textrm{BZ}=N_xN_yN_z$ is the number of states per band, equal to the number of crystal momenta in the Brillouin zone. In 1D, we have two $\mathbf{b}$ vectors, corresponding to $\{\pm \Delta k \hat{\mathbf{b}}_1\}$ where $\mathbf{b}_1$ is the reciprocal lattice vector. On a square lattice, we have four $\mathbf{b}$ vectors, corresponding to $\{\pm \Delta k \hat{\mathbf{b}}_1,\pm \Delta k \hat{\mathbf{b}}_2\}$ where $\mathbf{b}_1$ and $\mathbf{b}_2$ are the reciprocal lattice vectors. Furthermore, $P_\kv=\sum_n \ketbra{u_{n,\kv}}$ for $n$ over the occupied bands and $Q_\kv=I-P_\kv$ are the local projectors onto occupied and unoccupied bands, respectively. The gauge that $\Omega_I$ is independent of is unitary transformations $U_{nm}^{(\kv)}$ on the occupied Bloch states at each $\kv$. The gauge independence is apparent in the momentum space representation of $\Omega_I$ due to the cyclic symmetry of a trace (i.e., the trace of a matrix is invariant under similarity transformations). We note that $\Omega_I$ is equivalent~\cite{marzari1997} to the traced (over spatial DOF $i$ and $j$) integral (over $\kv$) of the quantum metric, the symmetric component of the quantum geometric tensor~\cite{provost1980,kohn1964,souza2000,resta1999,resta2017_talk},

\begin{equation}
    [\eta_\kv]_{ij} = \sum_n \bra{\partial_{k_i}u_{n,\kv}} (I - P_\kv)\ket{\partial_{k_j}u_{n,\kv}}.
\end{equation}
Furthermore, we note the finite-difference expressions,
\begin{equation}\label{app:eq:MV_variance_components_finite_difference}
\begin{split}
    \sum_j \expval{X_j^2}_n &= \frac{1}{N_\textrm{BZ}} \sum_{\kv,\bv} w_b[2 - 2\text{Re}\langle{u_{n\kv}}\ket{u_{n,\kv + \bv}}], \\
    \rv^W_n &= \frac{\ii}{N_\textrm{BZ}} \sum_{\kv,\bv} w_b \bv\left[\langle u_{n \kv} \ket{u_{n, \kv + \bv}} - 1 \right]
\end{split}
\end{equation}
from which~\eqref{app:eq:MV_variance_components} are derived via expansions of the links $M^{\kv,\bv}_{mn}$ up to second order in $|\bv|$~\cite{marzari1997}.

\subsection{Forms of the projected Resta position operator, $\tX_C^2$ and $\tX_S^2$}

We now note some relevant forms of the projected Resta position operator~\cite{resta1998}, $\tX_R=PX_RP$, a fundamental building block of periodic Spatial Localizers. Namely, we have the relations
\begin{equation}\label{app:eq:resta_as_projectors}
\begin{split}
    &\tX_R = \sum_\kv P_{\kv + \Delta k \hat{\bv}_1} P_\kv \ket{\kv + \Delta k \hat{\bv}_1}\bra{\kv}, \\
    &\tX_R^\dagger = \sum_\kv  P_\kv P_{\kv + \Delta k \hat{\bv}_1} \ket{\kv}\bra{\kv + \Delta k \hat{\mathbf{\bv}}_1}, \\
    &\tX_R^\dagger \tX_R = \sum_\kv P_\kv P_{\kv + \Delta k \hat{\bv}_1} P_\kv \ketbra{\kv}{\kv}, \\
    &\tX_R \tX_R^\dagger = \sum_\kv P_\kv P_{\kv - \Delta k \hat{\bv}_1} P_\kv \ketbra{\kv}{\kv}.
\end{split}
\end{equation}
The following operators will be used in the rest of this appendix,
\begin{align}
    L_X&=\tX_c+\ii \tX_s \nonumber\\
    \tX_C&=\frac{1}{2}(X_R+X_R^\dagger)-P\\
    \tX_S&=\frac{1}{2\ii}(X_R-X_R^\dagger)\nonumber
\end{align}
For a periodic Spatial Localizer, the operator in the lower bound of Eq.~\eqref{app:eq:final_bound_before_variance_form} amounts to

\begin{equation} \label{app:eq:resta_localizer_bound_piece}
\begin{split}
    O_X &= \tX_{C}^2 +  \tX_{S}^2 \\ &= \frac{1}{2}(L_X L_X^\dagger + L_X^\dagger L_X) \\
    &= P + \frac12 \left( \tX_R \tX_R^\dagger + \tX_R^\dagger\tX_R  \right) - 2 \text{Re} \ \tX_R,
\end{split}
\end{equation}
where we are restricting analysis to 1D (i.e., one projected Resta operator) and we have undone the transformations of the form $\texpval{\tX_R^2} = \Delta\tX_R^2 + \texpval{\tX_R}^2$ done in the derivation of the bounds. In writing \eqref{app:eq:resta_localizer_bound_piece}, we have set, without loss of generality, $\rv={\bf 0}$, and have omitted the $\rv$ label in $O_X({\bf 0})=\tX_{C}^2({\bf 0}) +  \tX_{S}^2({\bf 0})$ and other subsequent operators, e.g., $L_X$ and $\tX_R$ in lieu of $L_X({\bf 0})$ and $\tX_R({\bf 0})$ (Section~\ref{app:sec:MV_r_dependence} discusses the dependence of $O_X(\rv)$ on $\rv$ and how it enters into~\eqref{app:eq:MV_O_X_relation}). The relation in~\eqref{app:eq:resta_localizer_bound_piece} can be seen from $L_X = \tX_C + \ii \tX_S$ and/or expanding $\tX_C$ and $\tX_S$ as defined in Eq.~\eqref{app:eq_resta_real_imag}. From $P= \sum_\kv P_\kv \ket{\kv} \bra{\kv}$ and the relations in~\eqref{app:eq:resta_as_projectors}, we note that the first two terms of the last expression in~\eqref{app:eq:resta_localizer_bound_piece} can be expressed as

\begin{equation}
    \begin{split}
        &P + \frac12 \left( \tX_R \tX_R^\dagger + \tX_R^\dagger\tX_R  \right) \\ 
        = &\sum_\kv \left [ P_\kv + \frac12\left(P_\kv P_{\kv - \Delta k \hat{\bv}_1} P_\kv + P_\kv P_{\kv + \Delta k \hat{\bv}_1} P_\kv\right) \right ] \ketbra{\kv}{\kv} \\
        = &\sum_\kv \left [ 2P_\kv - \frac12\left(P_\kv Q_{\kv - \Delta k \hat{\bv}_1} P_\kv + P_\kv Q_{\kv + \Delta k \hat{\bv}_1} P_\kv\right) \right ] \ketbra{\kv}{\kv} \\
        = & \ \ 2P - \hat{\Omega}_I,
    \end{split}
\end{equation}
where, in the third line, we have written $P_\kv$ in terms of the subspace of unoccupied bands $P_\kv=I-Q_\kv$. Furthermore, we define $\hat{\Omega}_I$ as 

\begin{equation}
    \hat{\Omega}_I = \sum_{\kv,\bv}  w_bP_\kv Q_{\kv + \bv} P_\kv  \ket{\kv} \bra{\kv},
\end{equation}
where we use the hat to differentiate the operator from the quantity $\Omega_I$ of the MV localization functional. Additionally, we have $\bv\in \{\pm\Delta k \hat{b}_1\}$ as in \eqref{app:eq:MV_variance_components}, and $w_b=1/2$ for both values of $\bv$.
Returning now to the full expression of $O_X$, we have

\begin{equation}
    O_X =  2P - 2 \text{Re} \tX_R  - \hat{\Omega}_I. 
\end{equation}

\subsection{Expectation values of Wannier functions}

We now demonstrate a direct relationship between $O_X(x)$ and the MV localization functional. Specifically, we show that, when taking expectation values of $O_X(x)$ with respect to WFs, we have 

\begin{equation}\label{app:eq:MV_O_X_relation}
    \sum_n \bra{0n} O_X(x) \ket{0n} = \tilde{\Omega} + \sum_n \expval{(X - xI)}^2_n,
\end{equation}
where $\tilde{\Omega}$ is the gauge-dependent component of the MV localization functional.

WFs~\cite{Wannier1937} are defined as a Fourier transform of Bloch-like~\cite{Bloch1929,marzari1997} states,

\begin{equation}
    \ket{\transvec n} = \frac{1}{\sqrt{N_\textrm{BZ}}} \sum_\kv e^{-\ii \kv \cdot \transvec} \ket{\psi_{n, \kv}},
\end{equation}
where $\ket{\psi_{n,\kv}}$ are the Bloch eigenstates.
We now evaluate the expectation value of $O_X$ with respect to $\ket{0n}$. By the linearity of operators in expectation values, we can decompose $\bra{0n} O_X \ket{0n}$ into the two components

\begin{equation}\label{app:eq:expval_of_WF}
    \bra{0n} O_X \ket{0n} = \bra{0n} (2P - 2 \text{Re} \tX_R) \ket{0n} - \bra{0n} \hat{\Omega}_I \ket{0n}.
\end{equation}
Evaluating the first component, we find it is equal to $\expval{X^2}_n$ in \eqref{app:eq:MV_variance_components_finite_difference}. 

\begin{equation}\label{app:eq:exp_X^2_MV}
    \begin{split} 
        &\bra{0n} (2P - 2 \text{Re} \tX_R) \ket{0n} \\ &= \frac{1}{N_\textrm{BZ}} \sum_\kv [2 - 2\text{Re}\langle u_{n,\kv} \ket{u_{n,\kv+\Delta k \hat{\bv}_1}}] \\
        &=  \frac{1}{N_\textrm{BZ}} \sum_{\kv,\bv} w_b \left[2 - 2\text{Re}\langle u_{n,\kv} \ket{u_{n,\kv+\bv}}\right] \\ 
        &= \langle X^2 \rangle_n,
    \end{split}
\end{equation}
where the first two lines are connected by $\text{Re}\bra{\alpha} \beta \rangle = \text{Re} \bra{\beta} \alpha \rangle$ and $w_b=\frac12$ in the case of 1D or a cubic lattice. The connection to the third expression comes from the identification with Eq.~\eqref{app:eq:MV_variance_components_finite_difference}.
Looking to the second component of~\eqref{app:eq:expval_of_WF}, we have 

\begin{equation}\label{app:eq:omega_I_expval1}
    \begin{split}
        \bra{0n} \hat{\Omega}_I \ket{0n} = \frac{1}{N_\textrm{BZ}} \sum_{\kv, \bv} \bra{u_{n,\kv}}  P_\kv Q_{\kv + \bv} P_\kv \ket{u_{n, \kv}}.
    \end{split}
\end{equation}
From ~\eqref{app:eq:omega_I_expval1}, it is clear that summing over all occupied bands yields a relation to the gauge-invariant spread 

\begin{equation}\label{app:eq:omega_I_expval2}
    \begin{split}
        \sum_n \bra{0n} \hat{\Omega}_I \ket{0n} &= \frac{1}{N_\textrm{BZ}} \sum_{\kv, \bv} w_b \sum_n \bra{u_{n,\kv}}  P_\kv Q_{\kv + \bv} P_\kv \ket{u_{n, \kv}} \\ 
        &= \frac{1}{N_\textrm{BZ}} \sum_{\kv,\bv}w_b \Tr \left[ P_\kv Q_{\kv + \bv} \right] \\
        &= \Omega_I,
    \end{split}
\end{equation}
as it coincides with~\eqref{app:eq:MV_variance_components}.

Assuming we are considering the same number of WFs as occupied bands, one can equivalently use the expression 

\begin{equation}\label{eq:tr_omega_I}
    \Tr\left[ \hat{\Omega}_I \right] = {N_\textrm{BZ}} \Omega_I.
\end{equation}
Combining ~\eqref{app:eq:exp_X^2_MV} and ~\eqref{app:eq:omega_I_expval2}, we have

\begin{equation}
    \sum_n \bra{0n} O_X \ket{0n} = \sum_n \expval{X^2}_n - \Omega_I,
\end{equation}
which shows a close relationship between the $O_X$ expectation values of Wannier states and $\tilde{\Omega} = \Omega - \Omega_I$. The exact relationship can be seen when we consider the position $\rv$ in the localizer $L(\rv)$. 

\subsection{Including the implicit $\rv$ dependence of $O_X(\rv)$} \label{app:sec:MV_r_dependence}

The previous subsections assume $\rv=0$ in $O_X(\rv)$ for simplicity. Here we discuss the case of $\rv \neq 0$. Since we are considering 1D for simplicity, we let $\rv=x$. Recall that we define $\tX_R(x)$ in 1D as

\begin{equation}
    \tX_R(x) = P\exp\left[ \ii \Delta k (X - xI)\right]P = \tX_R(0)e^{-\ii \Delta k x}.
\end{equation}
Using this, we conclude that 

\begin{equation}
    \sum_n \bra{0n} O_X(x) \ket{0n} = \tilde{\Omega} + \sum_n \expval{(X - xI)}^2_n.
\end{equation}
That is, the expectation value of $O_X(x)$ w.r.t. the Wannier states for the occupied bands corresponds to the gauge-dependent spread, $\tilde{\Omega}$, plus the squared deviation of the means of the WFs from the point $x$.
To see this, we view the expectation value as a quadratic function of its input (the operator). Let the function of interest be

\begin{equation}
    f(X - xI) = \sum_n \bra{0n} O_X(x) \ket{0n}.
\end{equation}
For $x=0$, we have shown

\begin{equation}
    f(X) = \sum_n \expval{X^2}_n - \Omega_I.
\end{equation}
By the linearity of expectation values, we then have that 

\begin{equation}
\begin{split}
    f(X-xI) &= \sum_n \expval{(X-xI)^2}_n - \Omega_I \\
    &= \tilde{\Omega} + \sum_n \expval{(X - xI)}^2_n.
\end{split}
\end{equation}
where we utilized $\expval{(X-xI)^2}=\Delta X^2 - \expval{(X - xI)}^2$ and $\Omega - \Omega_I = \tilde{\Omega}$.

We end with a note that, while we considered expectation values of $O_X$ w.r.t. Wannier states, the bounds in Appendix~\ref{app:sec:alt_bound} correspond to expectation values of physical Schmidt vectors $\ket{\psch_1}$, which generally aren't WFs in the event that projected position operators don't commute. However, we find this section to further illustrate why the bounds derived in Appendix~\ref{app:sec:alt_bound} are relevant for periodic Spatial Localizers.

\section{Constructing localized state representations for occupied bands from the Spatial Localizer}
\label{app:localized-basis}

In this appendix, we present how to use a Spatial Localizer to construct a localized state representation. We begin by describing how to use the LIF to find favorable locations to extract localized states, and discuss how the case of commuting projected position operators $\tX_j$ results in WCs arising as zeroes of the LIF. Then, given a choice of extraction locations, we discuss how to process localizer eigenstates to extract the localized states and (if desired) orthogonalize the states to construct a localized and orthogonal basis.

Given a Spatial Localizer \(L(\mathbf r)\in\mathcal B(\mathcal H_\text{occ}\otimes\mathcal H_\Gamma)\), we outline a two–step procedure to obtain a localized orthonormal basis (WFs under translation symmetry, or Wannier-like orbitals when translation symmetry is broken).

\paragraph{Step 1: Locate centers via the LIF.}
Scan the LIF, \(\mu(\mathbf r)\), over positions \(\mathbf r\) in the system. Its (global) minima identify WCs (or generalized centers) from which we will extract localized states. For systems with discrete translation invariance, it suffices to scan a single unit cell; spatial symmetries can further reduce the scan region by reflection/rotation of \(\mu(\mathbf r)\).

In multi-band settings (e.g., more than one electron per cell), the physically relevant localized content may live in the \emph{low-lying localizer bands} and not exclusively in \(\mu(\rv)\). In such cases, examine a small set of the lowest localizer bands during the scan.

A particularly transparent situation occurs when the projected position operators commute, \([\tilde X_j,\tilde X_k]=0\) for all \(j,k\). Then the minima of \(\mu(\mathbf r)\) are zeros that coincide with the simultaneous eigenvalues \(\{\rv^W\}\) of the \(\tilde X_j\). This can be seen from the square of the localizer,
\begin{equation}
\begin{split} \label{eq:app_loc_squared}
    L^2(\rv) &= \sum_j \tX_j^2 \otimes I + \sum_{k<l} {\comm{\tX_k}{\tX_l}}\otimes \Gamma_k \Gamma_l \\
    &= Q \otimes I + \sum_{k<l} {\comm{\tX_k}{\tX_l}}\otimes \Gamma_k \Gamma_l,
\end{split}
\end{equation}
where \(Q\) is the quadratic operator~\cite{cerjan2023}. When the commutators vanish, there exists a common eigenbasis \(\{\,|\phi_W\rangle\,\}\) for all \(\tilde X_j\), with \(\tilde X_j(0)=\sum_W x^W_{j}\,|\phi_W\rangle\!\langle\phi_W|\). Evaluating
\begin{equation}
    \begin{split}
    \sum_j \tX^2_j(r_j) &= \sum_j (\tX_j(\mathbf{0}) - r_jI)^2 \\
    &= \sum_{j,W} (r^W_j - r_j)^2\ketbra{\phi_W}.
    \end{split}
\end{equation}
shows that the spectrum of \(\sum_j \tilde X_j^{\,2}(\mathbf r)\) is the set \(\{\,\sum_j(r^W_{j}-r_j)^2\,\}\). Hence \(L^2(\rv)\) (and by Hermiticity, \(L(\rv)\)) exhibits a zero in the LIF exactly at \(\mathbf r=\mathbf r^W\), and at those points the localizer eigenvectors reproduce (up to the embedding multiplicity \(g\)) the simultaneous eigenvectors \(|\phi_w\rangle\).

\paragraph{Step 2: Disentangle embedding DOFs via a Schmidt decomposition.}
At each chosen center \(\rv^W\), take one (or more) eigenvector(s) of \(L(\rv^W)\) and decompose across the physical and embedding Hilbert spaces,
\begin{equation}
|\psi^L\rangle
=\sum_i s_i\,|s_i\rangle
=\sum_i s_i\,|p_i\rangle\otimes|\xi_i\rangle,
\qquad
\sum_i s_i^2=1,
\end{equation}
where \(\{|p_i\rangle\}\in\mathcal H_\text{occ}\) and \(\{|\xi_i\rangle\}\in\mathcal H_\Gamma\) are orthonormal Schmidt vectors and \(s_i\) are Schmidt values. As a practical choice of \emph{trial WF} at \(\rv^W\), take the leading physical Schmidt vector $ |p_1\rangle $. The resulting set of trial states is not guaranteed to be orthogonal; apply Löwdin orthogonalization to obtain an orthonormal localized basis.

In periodic crystals a convenient workflow is: (i) extract one localizer eigenstate per occupied band in a single unit cell, (ii) project the corresponding trial states onto the occupied Bloch subspace, and (iii) perform \(k\)-resolved Löwdin orthogonalization, recovering a Wannier-style construction compatible with the standard projection method. In essence, we are using the projection method~\cite{marzari1997} once we have extracted trial states from a Spatial Localizer. However, we emphasize that the physical Schmidt vectors are already restricted to live in the occupied subspace, so there is no loss of norm from projecting the physical Schmidt vectors onto the Bloch states (i.e., it is more of a basis transformation than projection, but the overall method is the same).
Without translation symmetry, one needs to extract each state they are interested in from the minima of $\mu$ and perform Löwdin orthogonalization on the entire set. 

\paragraph{Remarks on degeneracies and band selection.}
Degenerate minima of \(\mu(\mathbf r)\) can arise from symmetry or from multiple occupied bands. In such cases, one can harvest the physically distinct localized content by comparing the \emph{physical} Schmidt vectors obtained from a small set of low-lying localizer bands. A systematic (localizer) band-selection criterion is an interesting direction for future work, but is not required for the examples considered here.

\section{Models}\label{app:models}

In this appendix, we discuss the models used in this work. 

In Section~\ref{app:sec:models_wse2}, we go over the 3-band model of $\mathrm{WSe_2}$ that serves as both (i) an example of an obstructed atomic limit (OAL) and (ii) a tight-binding model on a triangular lattice, which requires the inclusion of a third Resta position operator~\cite{marzari1997}. Furthermore, we analyze the robustness of the first physical Schmidt vector, $\ket{\psch_1(\rv)}$, as we vary the extraction point from the LIF minimum (Wannier center), $\rv^W$. We then look at the effectiveness of $\ket{\psch_1(\rv^W)}$ as a trial WF. We perform this analysis by comparing $\ket{\psch_1(\rv^W)}$ against two other trial WFs as inputs into Wannier90~\cite{wannier90,wannier90_v2}, a standard software for producing MLWFs. We find that the WF produced by $\ket{\psch_1(\rv^W)}$ requires no further optimization of its variance, and the other trial WFs cannot result in a WF with a variance lower than that of the WF produced by $\ket{\psch_1(\rv^W)}$. 

In Section~\ref{app:sec:models_disc}, we discuss the model on a lattice with a disclination, and present the Bloch Hamiltonian corresponding to the pristine lattice. 

In Section~\ref{app:sec:qwz_model}, we present the QWZ model, a minimal model of a Chern insulator, and the corresponding LIFs in the trivial and topological phases. Additionally, we present scatter plots of (i) $\ket{\psch_1(\rextract)}$, (ii) $\ket{\psch_2(\rextract)}$, and (iii) the optimized WF obtained using $\ket{\psch_1(\rextract)}$ and $\ket{\psch_2(\rextract)}$. We then discuss the overcompleteness relations for the coherent states $\ket{p_1(\rv)}$, which vary as a function of the center of the vortex in the Berry connection. Furthermore, we discuss the difference in behavior of $\ket{\psch_1(\rv)}$ for the trivial/topological phases of the QWZ model and the reliability of the resulting WFs for the calculation of the absolute polarization of a Chern insulator.

\subsection{3-Band Model of $\mathrm{WSe_2}$}\label{app:sec:models_wse2}

\begin{figure}
    \centering
    \includegraphics[width=1\linewidth]{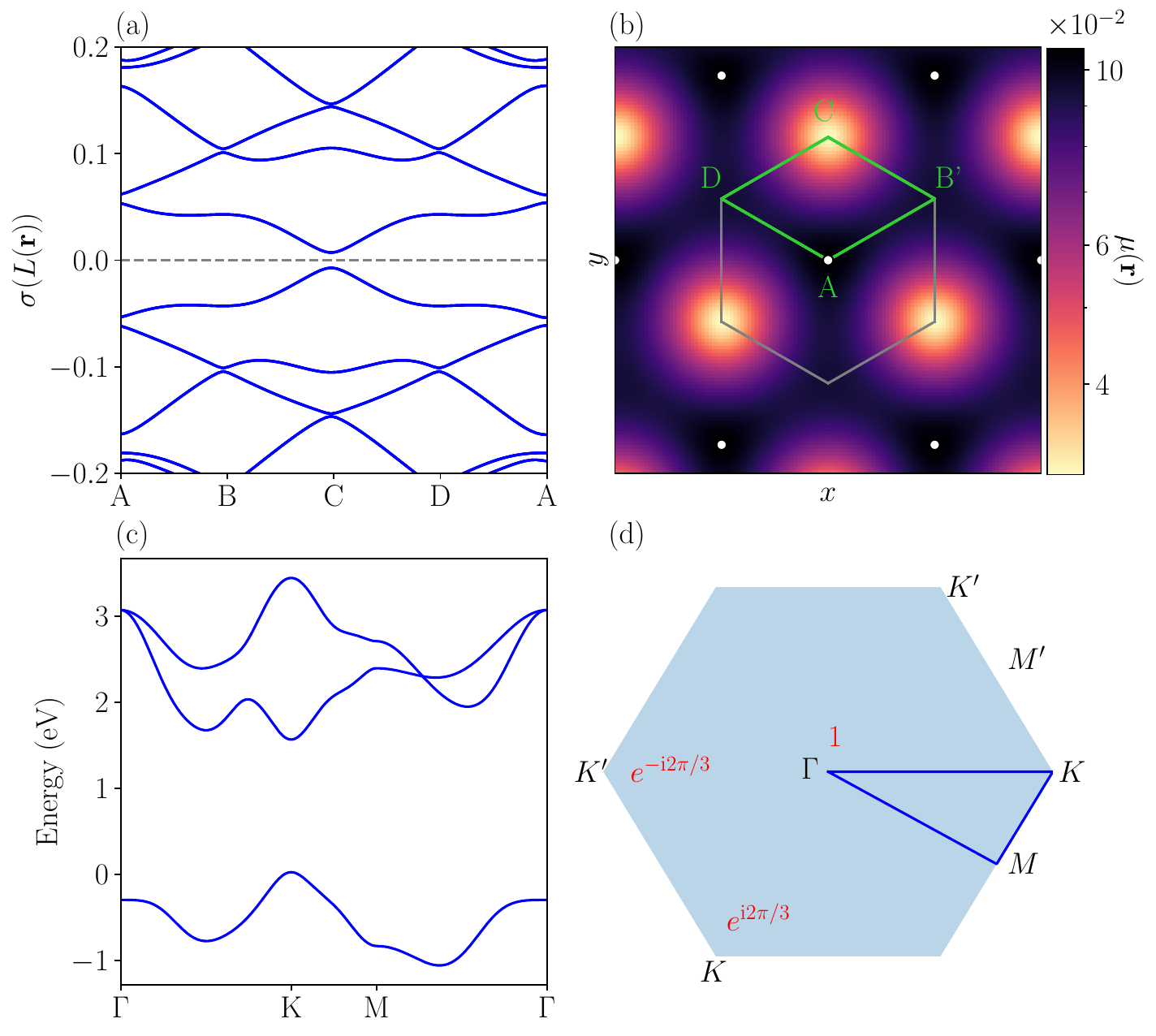}
    \caption{Near-zero spectral bands of the Spatial Localizer corresponding to the green path in (b). The lowest lying bands in (a), corresponding to the LIF, are two-fold degenerate. The dashed grey line denotes zero. The LIF (b) is a zoomed in version of the LIF presented in Fig. 1 of the main text with a green path corresponding to (a). Grey lines denote a unit cell. Energy bands (c) of the $\mathrm{WSe_2}$ Hamiltonian along the dark blue path shown in (d). Hexagon (d) corresponding to the first BZ (light blue) with a dark blue path corresponding to (c). The text in red denotes $C_3$ representations of the occupied band at the corresponding high-symmetry points.}
    \label{fig:wse2_spectra_reps}
\end{figure}

The low-energy physics of $\mathrm{WSe_2}$ can be modeled by a triangular lattice of W (metal) atoms. Specifically, we use a model~\cite{wse2_model} constructed from density-functional theory (DFT) calculations and band-fitting procedures. We consider the model with up to third nearest neighbor hoppings, referred to as the ``TNN'' model in~\cite{wse2_model}, defined as

\begin{equation} \label{eq:app_wse2_ham}
H^{\mathrm{TNN}}(\mathbf{k})
=
\begin{bmatrix}
V_0 & V_1 & V_2 \\
V_1^* & V_{11} & V_{12} \\
V_2^* & V_{12}^* & V_{22}
\end{bmatrix}
\end{equation}

where, with lattice constant $a$, $\alpha = \frac12 k_x a$, and $\beta=\frac{\sqrt{3}}{2}k_y a$ we have

\begin{equation}
\begin{split}
V_0 ={} & \epsilon_1
+ 2\,t_0\bigl(2\cos\alpha\cos\beta + \cos2\alpha\bigr)\\
&\;{}+ 2\,r_0\bigl(2\cos3\alpha\cos\beta + \cos2\beta\bigr)\\
&\;{}+ 2\,u_0\bigl(2\cos2\alpha\cos2\beta + \cos4\alpha\bigr)\,,
\end{split}
\end{equation}

\begin{equation}
\begin{split}
\text{Re}[V_1] ={} & -2\sqrt{3}\,t_2\,\sin\alpha\sin\beta
+ 2\,(r_1+r_2)\,\sin3\alpha\sin\beta\\
&\;{}- 2\sqrt{3}\,u_2\,\sin2\alpha\sin2\beta\,,
\end{split}
\end{equation}

\begin{equation}
\begin{split}
\text{Im}[V_1] ={} & 2\,t_1\,\sin\alpha\,(2\cos\alpha + \cos\beta)
+ 2\,(r_1 - r_2)\,\sin3\alpha\cos\beta\\
&\;{}+ 2\,u_1\,\sin2\alpha\,(2\cos2\alpha + \cos2\beta)\,,
\end{split}
\end{equation}

\begin{equation}
\begin{split}
\text{Re}[V_2] ={} & 2\,t_2\,\bigl(\cos2\alpha - \cos\alpha\cos\beta\bigr) \\&
- \frac{2}{\sqrt{3}}\,(r_1+r_2)\,\bigl(\cos3\alpha\cos\beta - \cos2\beta\bigr)\\
&\;{}+ 2\,u_2\,\bigl(\cos4\alpha - \cos2\alpha\cos2\beta\bigr)\,,
\end{split}
\end{equation}

\begin{equation}
\begin{split}
\text{Im}[V_2] ={} & 2\sqrt{3}\,t_1\,\cos\alpha\,\sin\beta
\\&+ \frac{2}{\sqrt{3}}\,\sin\beta\,(r_1 - r_2)\,(\cos3\alpha + 2\cos\beta)\\
&\;{}+ 2\sqrt{3}\,u_1\,\cos2\alpha\,\sin2\beta\,.
\end{split}
\end{equation}

\begin{equation}
\begin{split}
V_{11} ={} & \epsilon_2
+ \bigl(t_{11} + 3\,t_{22}\bigr)\cos\alpha\cos\beta
+ 2\,t_{11}\cos2\alpha\\
&\;{}+ 4\,r_{11}\cos3\alpha\cos\beta
+ 2\,(r_{11} + \sqrt{3}\,r_{12})\cos2\beta\\
&\;{}+ \bigl(u_{11} + 3\,u_{22}\bigr)\cos2\alpha\cos2\beta
+ 2\,u_{11}\cos4\alpha\,,  
\end{split}
\end{equation}

\begin{equation}
\begin{split}
\text{Re}[V_{12}] ={} & \sqrt{3}\,(t_{22} - t_{11})\,\sin\alpha\sin\beta
+ 4\,r_{12}\,\sin3\alpha\sin\beta\\
&\;{}+ \sqrt{3}\,(u_{22} - u_{11})\,\sin2\alpha\sin2\beta\,,  
\end{split}
\end{equation}

\begin{equation}
\begin{split}
\text{Im}[V_{12}] ={} & 4\,t_{12}\,\sin\alpha\,(\cos\alpha - \cos\beta)\\
&\;{}+ 4\,u_{12}\,\sin2\alpha\,(\cos2\alpha - \cos2\beta)\,,  
\end{split}
\end{equation}

and

\begin{equation}
\begin{split}
V_{22} ={} & \epsilon_2
+ (3\,t_{11} + t_{22})\,\cos\alpha\cos\beta
+ 2\,t_{22}\,\cos2\alpha\\
&\;{}+ 2\,r_{11}\,(2\cos3\alpha\cos\beta + \cos2\beta)\\
&\;{}+ \frac{2}{\sqrt{3}}\,r_{12}\,(4\cos3\alpha\cos\beta - \cos2\beta)\\
&\;{}+ (3\,u_{11} + u_{22})\,\cos2\alpha\cos2\beta
+ 2\,u_{22}\,\cos4\alpha\,.  
\end{split}
\end{equation}

The parameters can be found in Table III of~\cite{wse2_model} under the GGA section. $H^{\mathrm{TNN}}(\mathbf{k})$ has $D_{3h}$ point group symmetry, with band representations of $C_3$ rotation that predict WCs at the location $\frac13(\mathbf{a}_1 - \mathbf{a}_2)$ (see Fig.~\ref{fig:wse2_spectra_reps}d). 

While the Spatial Localizer has a natural construction in position space, constructing the position space Hamiltonian for a model as in Eq. \ref{eq:app_wse2_ham} is cumbersome. This motivates the use of the Spatial Localizer in (crystal) momentum space, where the Resta position operators act as translations in $\kv$ along reciprocal-lattice vectors, and projectors can be constructed from Bloch functions. 

\begin{figure}
    \centering
    \includegraphics[width=1\linewidth]{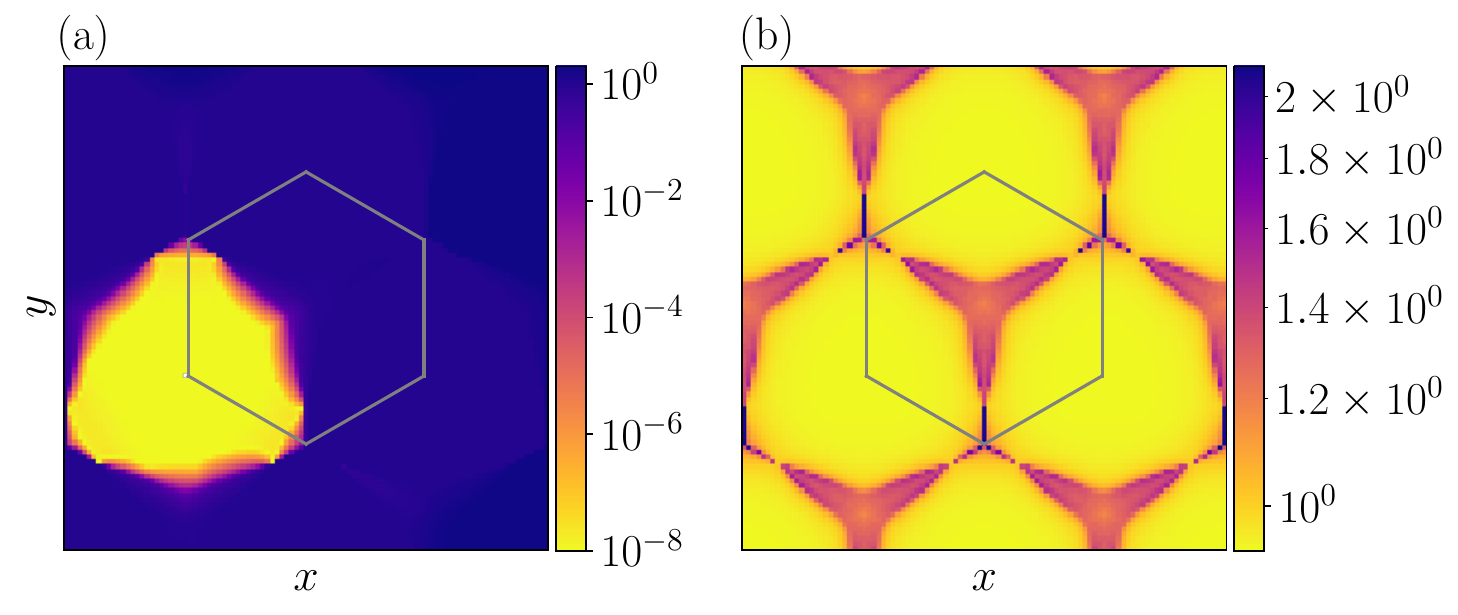}
    \caption{Absolute deviation of mean from $\rextract$ (a) and variance, $\Delta R$, (b) of WFs produced from $\ket{\psch_1(\rv)}$. In this figure, $\rextract$ corresponds to the bottom-left corner of the unit cell. In both figures, the grey hexagon denotes a unit cell as in the main text.}
    \label{fig:wse2_deviations}
\end{figure}

In this model, the minimum of the LIF, $\rv^W$, are located at symmetric points, making the ideal extraction locations easy to identify. However, we find that for producing localized WFs, the exact extraction location is not particularly important; one can choose an approximate extraction location near the minima of the LIF and still obtain WFs with almost identical center and variance. To illustrate this, we consider WFs produced as a function of the extraction point, $\rv$. Specifically, we consider the deviation of the mean of the WFs from $\rv^W$, $\abs{\texpval{\tilde{\Rv}(\rv^W)}}$, and the variance $\Delta R=\sqrt{\Delta R^2}$ in units of lattice constant. We observe exceptional robustness of both the center and variance of WFs produced with extraction points near $\rv^W$, as presented in Figure~\ref{fig:wse2_deviations}.

\subsubsection{Comparison with Wannier90}\label{app:sec:w90_comparison}

Here we illustrate the use of our Spatial Localizer framework on the three-band model of monolayer $\mathrm{WSe_2}$ discussed in the main text.  
The periodic Spatial Localizer constructed for this model inherits \emph{chiral symmetry}; as a result, its spectrum is symmetric about~$0$, so we may confine attention to the non-negative part.  
Additionally, we must consider a third Resta position operator $X_{3,R}=X_{2,R}^\dagger X_{1,R}^\dagger$ due to the triangular lattice~\cite{marzari1997}, resulting in the Clifford elements $\Gamma_j$ having dimension $g=8$. We use the Weyl-Brauer matrices~\cite{brauer1935} as the $\Gamma_j$, which correspond to the generators of the Clifford algebra $\mathrm{Cl}_{6,0}$.

Scanning the (positive) localizer bands corresponding to the LIF reveals a two-fold degenerate minimum at $\rv^W = \frac13(\mathbf{a}_1 - \mathbf{a}_2)$ as seen in Fig. \ref{fig:wse2_spectra_reps}. We denote these degenerate localizer eigenvectors by $|\psi_{1}\rangle$ and $|\psi_{2}\rangle$.  
Let  
\[
|\psi_{j}\rangle \;=\; \sum_{i} s_{i}^{(j)}\, |\psch_{i}^{\,j}\rangle\otimes|\esch_{i}^{\,j}\rangle, 
\qquad j=1,2,
\]
be their Schmidt decompositions into \emph{physical} ($|\psch_{i}^{\,j}\rangle$) and \emph{embedding} ($|\esch_{i}^{\,j}\rangle$) vectors.  
We observe that the two Schmidt spectra are identical, and thus $\ket{\psi_1}$ and $\ket{\psi_2}$ are related by a product of unitary operators acting separately on physical and embedding sectors, $U_{\text{occ}}\otimes U_{\Gamma}$, i.e. they are equivalent up to local unitary operations. 

For both eigenvectors, the leading Schmidt value $s_{1}^{(j)}$ is non-degenerate.  
In our numerics, we find
\begin{equation}
\left|\bra{p^{(1)}_1}{p^{(2)}_1}\rangle\right| =  e^{\ii \phi},
\qquad
\bra{\xi^{(1)}_1}{\xi^{(2)}_1}\rangle=0,
\end{equation}
i.e., the two degenerate localizer eigenvectors share the same leading \emph{physical} Schmidt vector (up to phase), and differ primarily in the embedding sector, despite the degenerate LIF. The role of symmetries and degeneracies of localizer spectra/eigenstates will be discussed in future work.

Although the Spatial Localizer produces localized states, it is not yet proven that it invariably yields \emph{maximally} localized WFs.  
For $\mathrm{WSe_2}$, we observe that the WFs constructed from $|\psch_{1}^{\,1}\rangle$ are \emph{real} up to a global phase, a common heuristic of having reached the global minimum of the spread functional~\cite{marzari2012} in the absence of spin-orbit coupling.  

\begin{figure}
    \centering
    \includegraphics[width=1\linewidth]{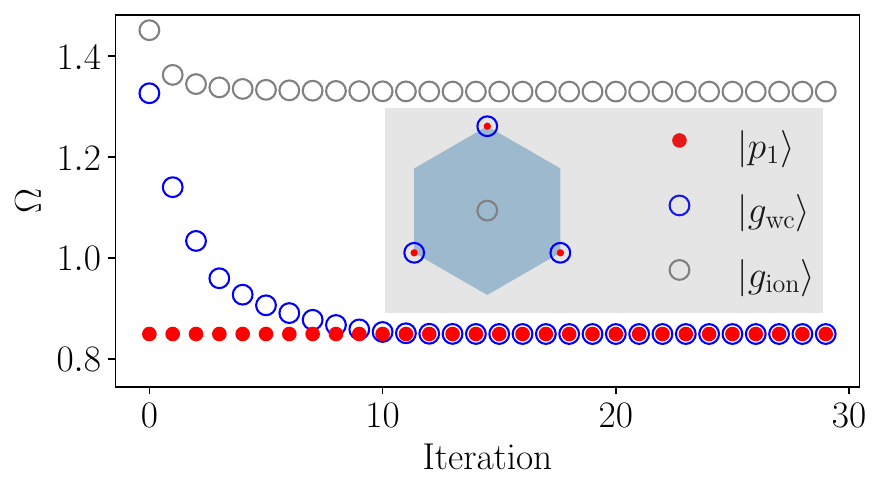}
    \caption{Comparison between physical Schmidt vector, $\ket{\psch_1}$, and two ansätze, $\ket{g_{\mathrm{wc}}}$ and $\ket{g_{\mathrm{ion}}}$, for WFs of the effective $\mathrm{WSe_2}$ model considered in the main text. Main plot : Marzari-Vanderbilt localization functional~\cite{marzari1997}, $\Omega$, throughout 30 iterations of the Wannier90~\cite{wannier90} optimization process. Inset : Legend and WCs of the WFs on a unit cell (blue) at the end of the optimization process. We note that the WCs do not change appreciably during the optimization process.} 
    \label{fig:wannier_90_comparison}
\end{figure}

To test optimality, we use Wannier90~\cite{wannier90,wannier90_v2,marzari1997}, a standard software package for finding maximally localized WFs by optimizing the gauge of Bloch functions. In this method, one needs (i) the Bloch functions for the occupied bands and (ii) an ansatz state to serve as an initial gauge choice.  
Besides the physical Schmidt vector $|\psch_{1}^{\,1}\rangle$, we supplied two intuitive ansatz states:
(i) a delta-function centered on a W atom, and  
(ii) a delta-function centered on $\rv^W=\tfrac13 (\mathbf{a}_{1}-\mathbf{a}_{2}$), which is the WC predicted by band representation theory.
Both trials are symmetrized to the trivial $C_{3}$ irrep, yielding

\begin{equation}
    \begin{split}
        |g_{\text{ion}}\rangle = \frac{1}{\sqrt3}\bigl(C_{3}^{2}+C_{3}+I\bigr)\,|0,0\rangle, \\
        |g_{\text{wc}}\rangle  = \frac{1}{\sqrt3}\bigl(C_{3}^{2}(\rv^W)+C_{3}(\rv^W)+I\bigr)\,|1,0\rangle,
    \end{split}
\end{equation}

where $|n,m\rangle$ is the basis state located at the unit cell $n\mathbf{a}_{1}+m\mathbf{a}_{2}$ with equal phase and weight on internal degrees of freedom.  
After 200 total iterations of spread minimization, we find that the $\ket{\psch_1}$ WF spread did not change, while the $|g_{\text{wc}}\rangle$ WF spread converges to the spread of the $\ket{\psch_1}$ WF (see Figure~\ref{fig:wannier_90_comparison}). Furthermore, the $|g_{\text{ion}}\rangle$ WF spread became trapped in a higher-spread local minimum.

The $\mathrm{WSe_2}$ test therefore supports, but does not yet prove, the conjecture that Spatial Localizers furnish a gauge of maximally localized WFs without external seeding or knowledge of the symmetry representations present in the occupied bands.  Establishing general conditions for this property and clarifying the role of symmetries, such as space group and anti-unitary symmetries, remain directions for future work.

\subsection{Space group F222 Model}
As a three-dimensional example, we consider a space group F222 model~\cite{cano2022}. This model realizes two obstructed atomic limits whose Wannier centers lie at the inequivalent Wyckoff positions $4a$ and $4b$, located at $\mathbf{r}_\text{4a}=(0,0,0)$ and $\mathbf{r}_\text{4b}=(0,0,\tfrac{1}{2})$ in conventional coordinates. The lattice is face-centered orthorhombic, with primitive direct-lattice vectors
\begin{align}
\mathbf{a}_1 &= \frac{1}{2}(0,b,c),\\
\mathbf{a}_2 &= \frac{1}{2}(a,0,c),\\
\mathbf{a}_3 &= \frac{1}{2}(a,b,0),
\end{align}
and corresponding reciprocal vectors
\begin{align}
\mathbf{b}_1 &= 2\pi\left(-\frac{1}{a},\frac{1}{b},\frac{1}{c}\right),\\
\mathbf{b}_2 &= 2\pi\left(\frac{1}{a},-\frac{1}{b},\frac{1}{c}\right),\\
\mathbf{b}_3 &= 2\pi\left(\frac{1}{a},\frac{1}{b},-\frac{1}{c}\right).
\end{align}
For simplicity, we set $a=b=c=1$. In the minimal two-band description, the Bloch Hamiltonian takes the stacked Rice--Mele form
\begin{equation}
H_0(\mathbf{k}) = -\frac{\epsilon}{2}(1+\sigma_z) - \frac{t}{2}\left(1+\sigma_z\cos k_z+\sigma_y\sin k_z\right),
\end{equation}
which transitions between the two obstructed atomic limits as the model parameters are varied.

To move away from this effectively stacked one-dimensional limit while preserving the symmetry of the model, we add the perturbation
\begin{equation}
\Delta H(\mathbf{k}) = (\cos k_x - \sin k_y)\sin k_z\,\sigma_x.
\end{equation}
This term introduces explicit dependence on all three crystal-momentum components and mixes the $k_z$ chain structure with the transverse directions. The full Hamiltonian is therefore
\begin{equation}
H(\mathbf{k}) = H_0(\mathbf{k}) + \Delta H(\mathbf{k}).
\end{equation}
In the calculations shown in the main text, this perturbation is included so that the model is no longer reducible to decoupled Rice--Mele-like chains, while the two insulating phases remain associated with Wannier centers at $4a$ and $4b$.

For visualization, we evaluate the LIF along the boundary of a conventional F222 unit cell with one octant removed. This choice makes the inequivalent candidate Wannier-center locations visible within a single three-dimensional rendering and allows the minima associated with the two phases to be read off directly from the boundary data. In the phase connected to the $4a$ limit, the minima occur at the corner-type positions of the cell, whereas in the phase connected to the $4b$ limit they occur at the corresponding half-translated positions. The octant cut therefore provides a convenient way to display, in a single figure, the distinct real-space center structure of the two obstructed atomic phases.

Regarding the construction of the corresponding Spatial Localizer, we note that one needs to include a 4th Resta position operator, $X_{4,R}= X_{3,R}^\dagger X_{2,R}^\dagger X_{1,R}^\dagger$ to ensure isotropic localization.

\subsection{4-band lattice with disclination}\label{app:sec:models_disc}

\begin{figure}
    \centering
    \includegraphics[width=1\linewidth]{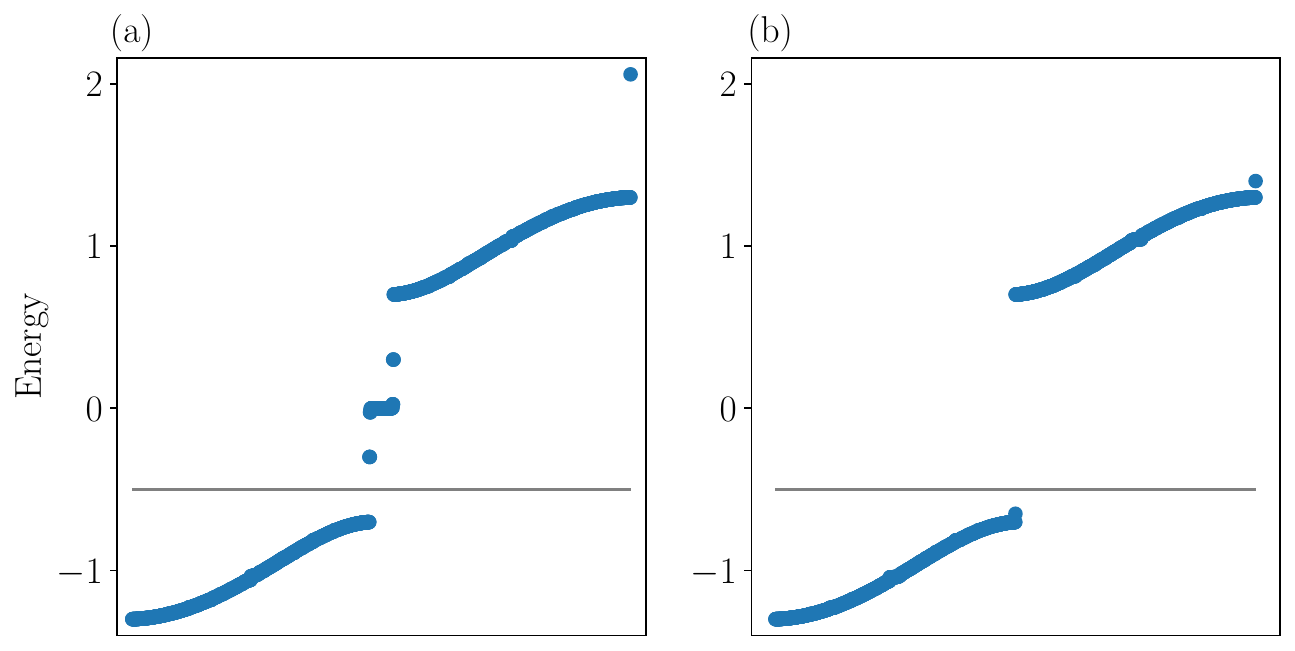}
    \caption{Spectrum of disclination model in topological (a) and trivial (b) phases. Grey line denotes the Fermi Energy $E_F=-0.5$.}
    \label{fig:disc_ham_spectrums}
\end{figure}

\begin{figure}
    \centering
    \includegraphics[width=1\linewidth]{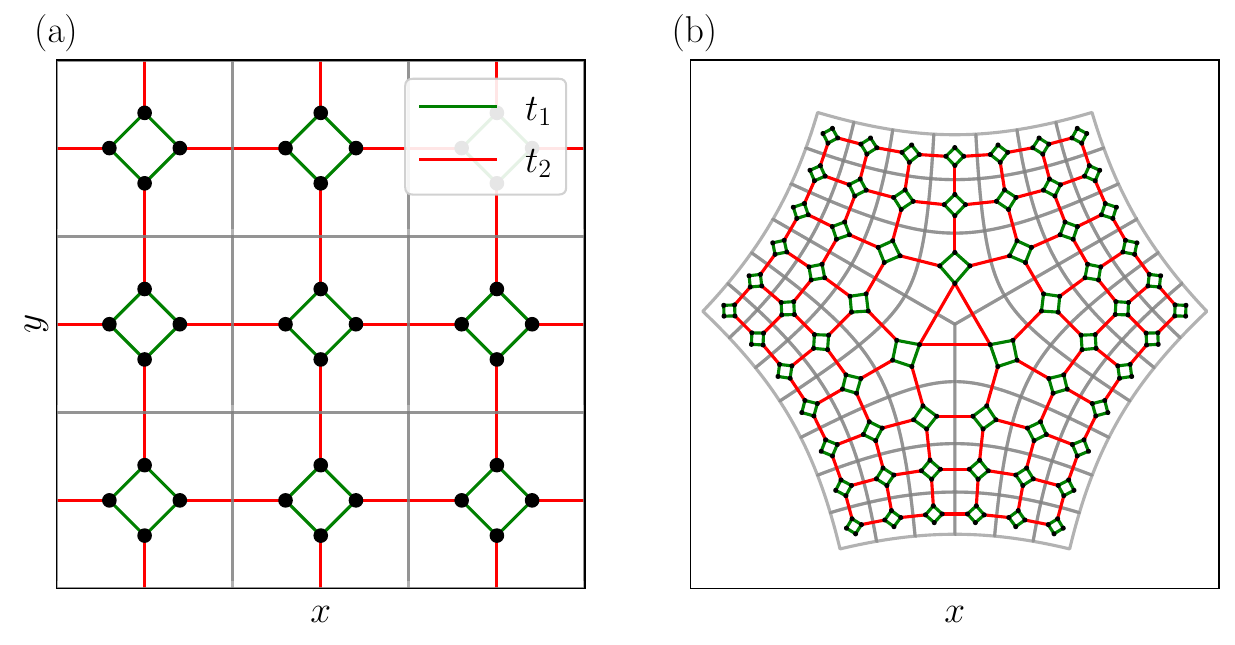}
    \caption{Graphical representation of the pristine lattice (a) and lattice with a disclination (b). The red and green lines correspond, respectively, to intra-cell ($t_1$) and inter-cell ($t_2$) hoppings.}
    \label{fig:disc_graphical}
\end{figure}

The Bloch Hamiltonian for the pristine lattice reads as 

\begin{equation}\label{eq:disc_pristine_ham}
    \begin{split}
    h(\kv) = t_1 \sigma_1 \otimes (\sigma_1 + I_{\text{2x2}}) &+ \bigg[t_2\left(\cos(k_y)\sigma_1 + \sin(k_y) \sigma_2\right) \\ &\oplus t_2\left(\cos(k_x)\sigma_1 + \sin(k_x) \sigma_2\right) \bigg]
    \end{split}
\end{equation}

where $t_1$ and $t_2$ respectively denote intra- and inter-cell hopping amplitudes. For the trivial phase, we have $t_1=1$ and $t_2=0.3$. For the topological phase, we have $t_1=0.3$ and $t_2=1$. Such a choice of parameters allows us to define a Fermi energy $E_F=-0.5$ for both phases (See Fig.~\ref{fig:disc_ham_spectrums}). This corresponds to 126 and 107 occupied states, respectively, in the trivial and topological phases for the lattice of unit cells depicted in Fig. \ref{fig:disc_graphical}.

\subsection{Chern Insulator}\label{app:sec:qwz_model}

\begin{figure}
    \centering
    \includegraphics[width=1\linewidth]{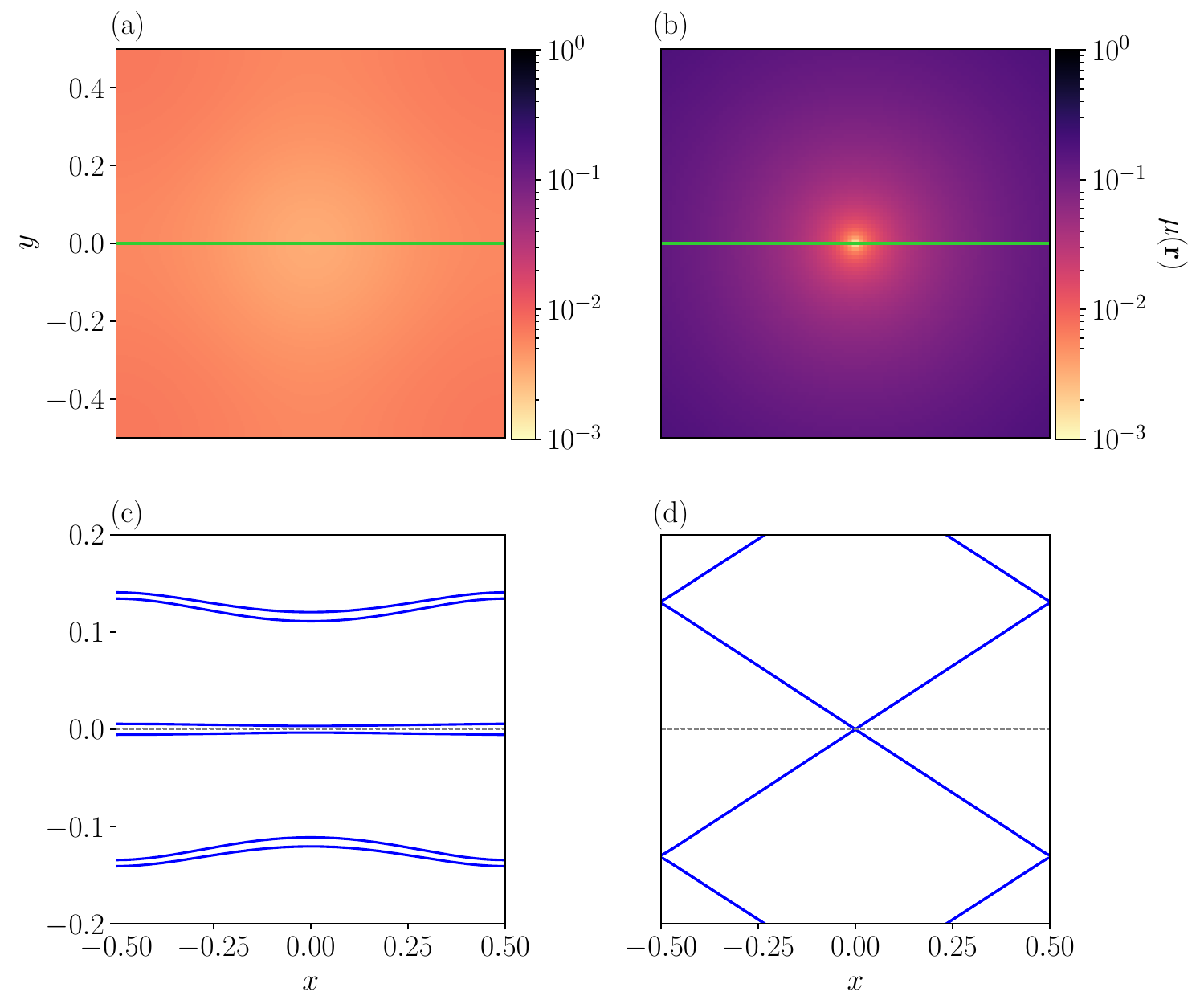}
    \caption{LIF ((a) and (b)) and slices of the near-zero localizer spectrum ((c) and (d)) of the topological (left) and trivial (right) phases of the QWZ model over a unit cell. The green lines in (a) and (b) correspond to the slices in (c) and (d).}
    \label{fig:qwz_gap_and_cuts}
\end{figure}

\begin{figure}
    \centering
    \includegraphics[width=1\linewidth]{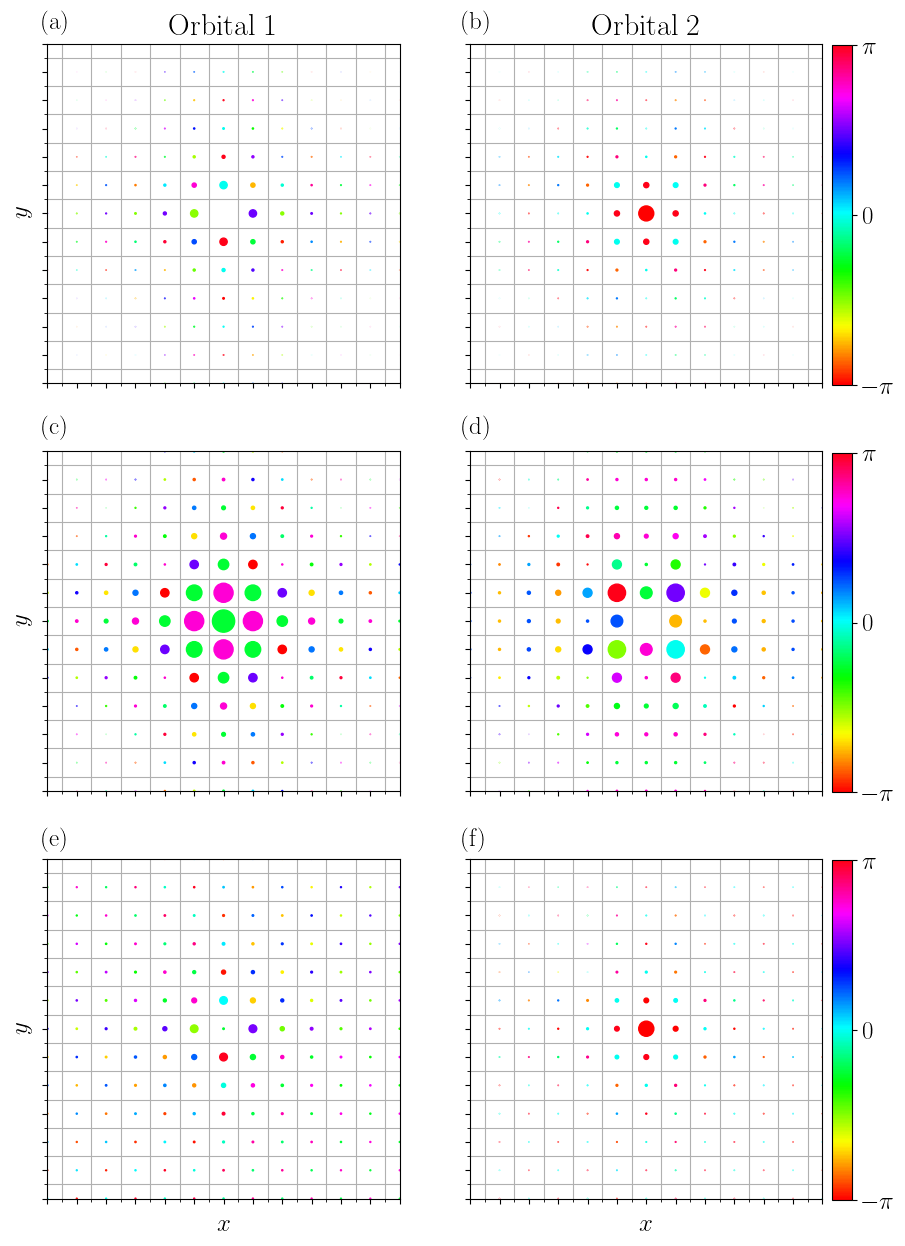}
    \caption{Scatter plots of $\ket{\psch_1(\rextract)}$ ((a) and (b)), $\ket{\psch_2(\rextract)}$ ((c) and (d)), and the resulting WF ((e) and (f)) for the topological phase of the QWZ model with $N=50$. The left and right sides respectively denote different intracell DOF, as shown in~\eqref{eq:ham_QWZ}. Note that the relative magnitude of weights (sizes of scatter points) is only meaningful here for different orbitals of the same state, i.e., the relationship between scatter point size and magnitude is not the same for different states. }
    \label{fig:qwz_psch_wf_scatter}
\end{figure}

We consider the QWZ~\cite{qi2006} model on a square lattice as a minimal example of a Chern insulator, with Bloch Hamiltonian,

\begin{equation}\label{eq:ham_QWZ}
    h(\kv) = (m + \cos k_x + \cos k_y)\sigma_3 + \sin k_x \sigma_1 + \sin k_y \sigma_2,
\end{equation}

The spectrum of the QWZ model under PBC is gapless when $m=\pm2,0$. For the topological and trivial phases, we set $m=1$ and $m=3$, respectively. In both phases, we occupy the lower band in the construction of a Spatial Localizer. 

As discussed in the main text, we observe a flat LIF in the topological phase, yet a consistent minimum. For obtaining localized states, we have chosen the lattice site as the extraction point $\rv^*$ as it both minimizes the LIF (see Fig.~\ref{fig:qwz_gap_and_cuts}) and maximizes the first Schmidt value of the localizer eigenvector, providing a minimal value of $\mathcal{S}$ in Eq. (9) of the main text. Furthermore, we find that the first physical Schmidt vector and its lattice translated counterparts, $\{\ket{\psch_1(\rextract + \transvec)}\}$ for direct lattice vectors $\transvec$, form an overcomplete, exponentially localized set of states as a result of $\ket{\psch_1(\rextract + \transvec)}$ lacking support at the $M$ point in the BZ (see overcompleteness relations below).
To complete the basis and construct WFs, we use a linear combination of the first two physical Schmidt vectors to produce WFs since $\ket{\psch_2(\rextract)}$ has support at the $M$ point (see Figure ~\ref{app:fig:p1_p2_bloch_overlaps}). Specifically, we use $\ket{\psi^T}= \alpha_1 \ket{\psch_1(\rextract)} + \alpha_2 \ket{\psch_2(\rextract)}$ as a trial WF where $|\alpha_2/\alpha_1| \approx 0.023$. A mesh-based search of relative magnitudes and phases was used to determine the optimal linear combination of $\ket{\psch_1(\rextract)}$ and $\ket{\psch_2(\rextract)}$. The mean of the corresponding WF is slightly shifted from the lattice site, with mean $\texpval{\hat{\Rv}}\approx(0.01,-0.01)$ in units of the lattice constant. For completeness, the Schmidt spectrum of $\ket{\psi^L(\rextract)}$ used to construct $\ket{\psi^T}$ is $s_1=0.99991, s_2=0.0116, s_3=0.00646,$ and $s_4=0.0004445$. We present scatter plots of $\ket{\psch_1(\rextract)}$, $\ket{\psch_2(\rextract)}$, and the resulting WF in Figure~\ref{fig:qwz_psch_wf_scatter}.

We note that $\ket{\psch_1(\rextract)}$ and its lattice-translated counterparts result in an overcomplete set of states only if the corresponding vortex is centered at a point of the k-mesh used. Thus, at finite system sizes, it is possible to construct a set of WFs from $\ket{\psch_1(\rv)}$ alone by shifting the extraction point such that the vortex is between points of the k-mesh. However, we expect $\ket{\psch_1(\rv)}$ to correspond to an overcomplete set of states in the thermodynamic limit, regardless of the extraction point. 

\subsubsection{Overcompleteness relations}\label{app:sec:overcompleteness_relations}

\begin{figure}
    \centering
    \includegraphics[width=1\linewidth]{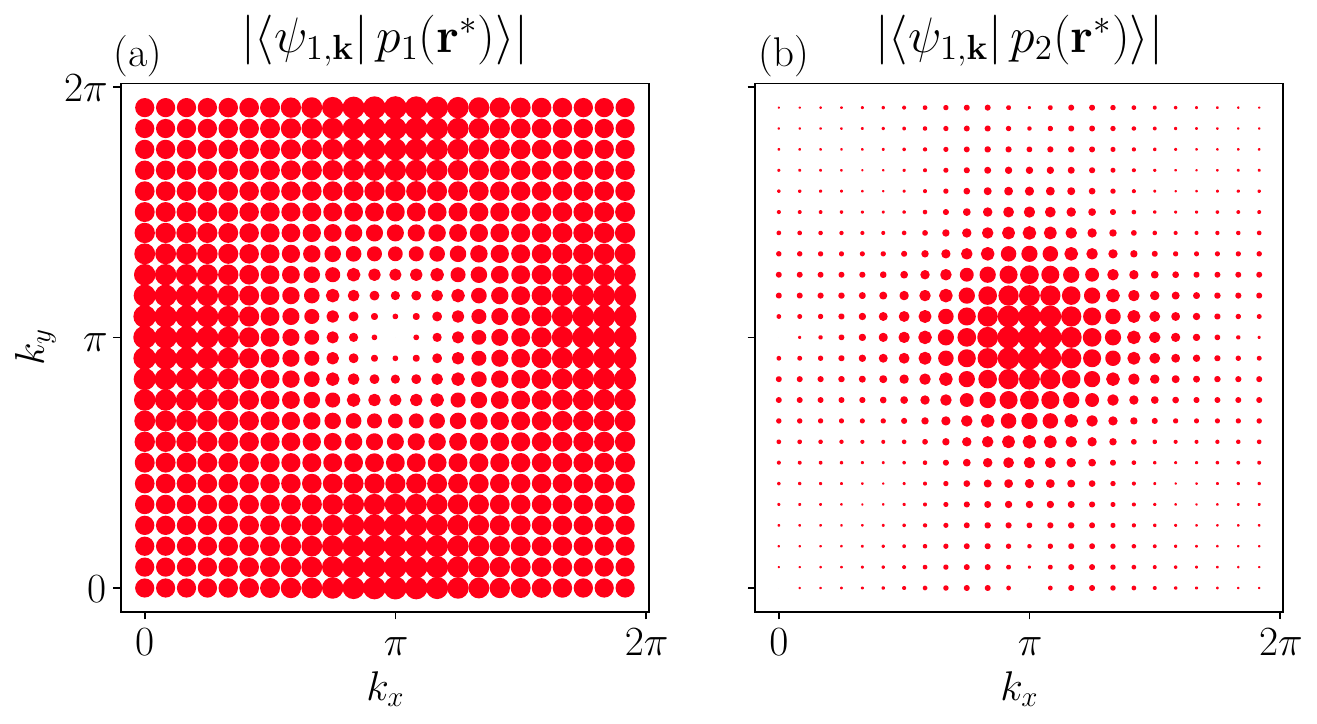}
    \includegraphics[width=1\linewidth]{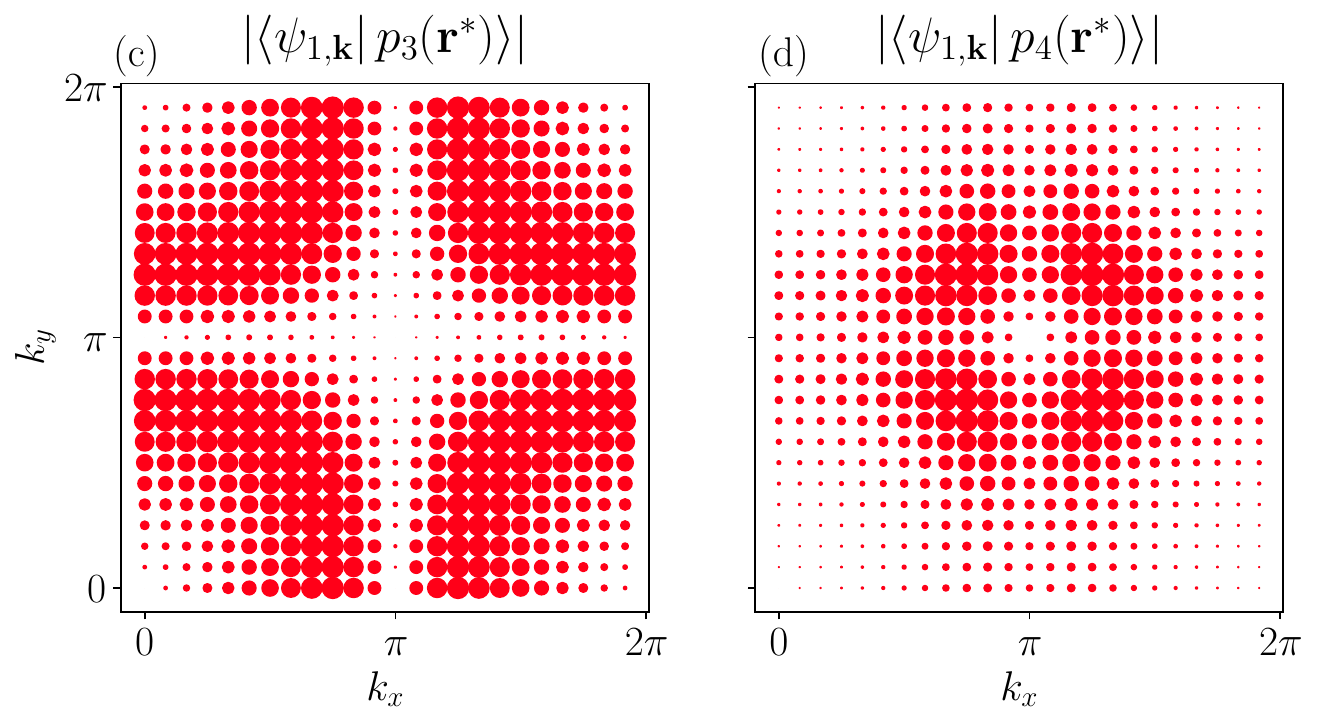}
    \caption{Magnitude of overlaps of occupied Bloch states $\ket{\psi_{1,\kv}}$ with physical Schmidt vectors $\ket{\psch_i(\rextract)}$ for the Chern ($C=1$) phase of the QWZ model. The radius of the scatter points are proportional to the magnitudes. Note that the overlap for $\ket{\psch_1(\rextract)}$ vanishes at $\kv_\mathbf{v}=(\pi,\pi)$, indicating the overcompleteness. We emphasize that the first physical Schmidt vector holds the dominant contribution for the localizer eigenvector, with Schmidt value $s_1 \approx0.9996$.}
    \label{app:fig:p1_p2_bloch_overlaps}
\end{figure}

Given a coherent state $\ket{\psch_1(\rextract)}$ with Berry connection vortex centered at $\kv_\mathbf{v}$, an overcompleteness relation (an equation showing the linear dependence of a set of states) can be made for the set of states comprised of $\ket{\psch_1(\rextract)}$ translated to every unit cell in the system. 
Let $\ket{\psch_1(\rextract + \transvec)}$ denote $\ket{p_1(\rextract)}$ centered at the unit cell located at $\transvec = n_1\mathbf{a}_1 + n_2\mathbf{a}_2$ for primitive lattice vectors $\mathbf{a}_i$ and $n_1,n_2\in\mathbb{Z}$. $\ket{\psch_1(\rextract+\transvec)}$ can then be described as 

\begin{equation}
    \ket{\psch_1(\rextract+\transvec)} = T_\transvec\ket{\psch_1(\rextract)} = \sum_\kv e^{\ii \kv \cdot \transvec} \alpha_\kv \ket{\psi_{\kv}},
\end{equation}

where $T_\transvec$ is the translation operator, $\ket{\psi_\kv}$ are Bloch states of the occupied Chern band, and $\alpha_\kv$ are complex coefficients. Furthermore, we have $\alpha_{\kv_\mathbf{v}}=0$ (the coherent state has no support at the center of the vortex) and we have made use of the relation $T_\transvec\ket{\psi_\kv}=e^{\ii \kv \cdot \transvec}\ket{\psi_{\kv}}$ for position-space translations on Bloch states. For a vortex center $\kv_\mathbf{v}$, we then have the overcompleteness relation 

\begin{equation}
    \begin{split}
        \sum_\transvec e^{-\ii \kv_\mathbf{v} \cdot \transvec} \ket{p_1(\rextract + \transvec)} 
        &= \sum_{\transvec, \kv} e^{-\ii \kv_\mathbf{v} \cdot \transvec} e^{\ii \kv \cdot \transvec} \alpha_\kv \ket{\psi_{\kv}} \\
        &= \sum_\kv \alpha_\kv \sum_\transvec e^{\ii (\kv - \kv_\mathbf{v}) \cdot \transvec} \ket{\psi_{\kv}} \\
        &\propto \sum_\kv \alpha_\kv \delta_{\kv,\kv_\mathbf{v}} \ket{\psi_{\kv}} \\ 
        &= 0 ,
    \end{split}
\end{equation}

where the last line is zero due to the combination of $\delta_{\kv,\kv_\mathbf{v}}$ and $\alpha_{\kv_\mathbf{v}}=0$. Such an overcompleteness relation has been discussed~\cite{li2024} in the context of a generalized Perelomov identity~\cite{Perelomov1971} for Chern bands. In our case for the QWZ model with $C=1$, $\ket{p_1(\rextract)}$ has a vortex located at the $M$ point ($\kv_\mathbf{v}=(\pi,\pi)$), yielding the overcompleteness relation

\begin{equation}
\begin{split}
    \sum_\transvec e^{-\ii \kv_\mathbf{v} \cdot \transvec} \ket{p_1(\rextract + \transvec)} 
    &= \sum_{n_1,n_2} e^{-\ii \pi (n_1 + n_2)} \ket{p_1(\rextract + \transvec)} \\
    &= \sum_{n_1,n_2} (-1)^{(n_1 + n_2)} \ket{p_1(\rextract + \transvec)} \\&= 0,
\end{split}
\end{equation}
which has a striking similarity to the overcompleteness relation given by Thouless~\cite{thouless1984_WF_OC} for the coherent states of the lowest Landau level, where the only difference is a factor of $(-1)^{n_1n_2}$ due to the use of magnetic translations in the symmetric gauge of the Quantum Hall Effect (QHE)~\cite{fradkin2013field,okuma2024} as opposed to lattice translations in a Chern insulator. We note that the overcompleteness can be seen directly from the magnitude of the overlaps between $\ket{\psch_1(\rextract)}$ and the occupied Bloch states $\ket{\psi_{1,\kv}}$, given by $\abs{\langle \psi_{1,\kv} \ket{\psch_1(\rextract)}}$. We see that $\abs{\langle \psi_{1,\kv'} \ket{\psch_1(\rextract)}}=0$. Furthermore, we note that the overlap $\abs{\langle \psi_{1,\kv'} \ket{\psch_2(\rextract)}}\neq0$, which motivated the use of $\ket{\psch_1(\rextract)}$ and $\ket{\psch_2(\rextract)}$ in the construction of WFs for the Chern phase. We present the overlaps $\abs{\langle \psi_{1,\kv} \ket{\psch_i(\rextract)}}$ for all physical Schmidt vectors in Figure~\ref{app:fig:p1_p2_bloch_overlaps}.

\subsubsection{$\ket{\psch_1(\rv)}$ in Chern vs OAL phases}

\begin{figure*}
    \centering
    \includegraphics[width=1\linewidth]{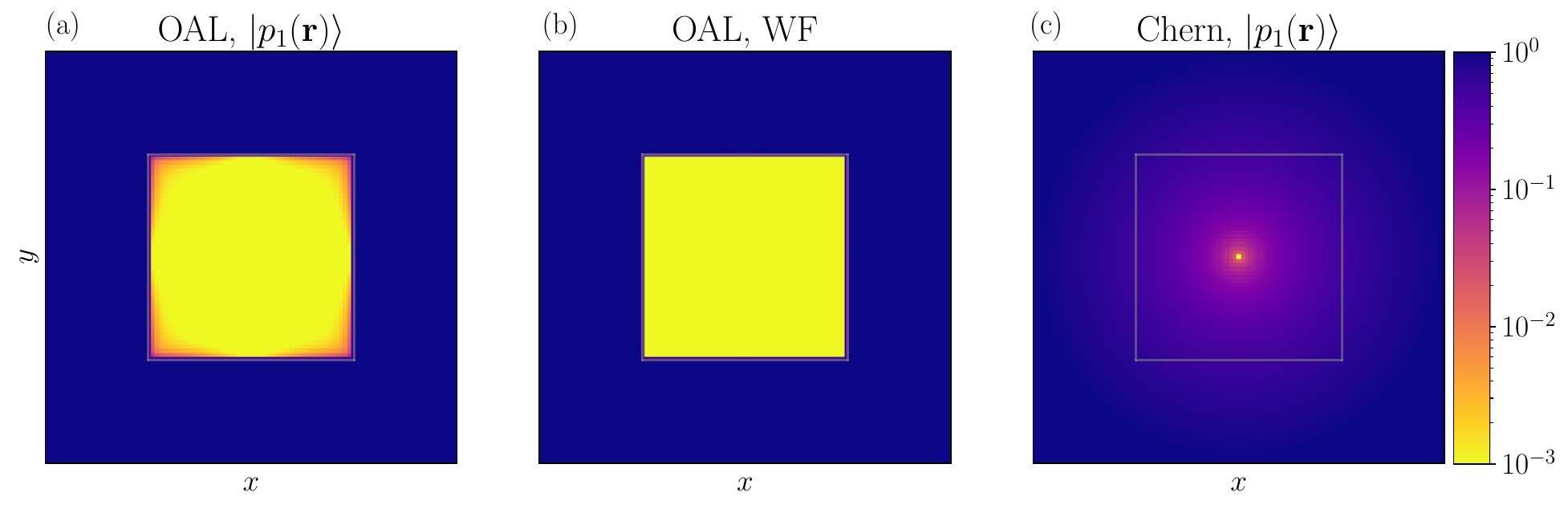}
    \caption{The deviation of mean position from the LIF minimum, $\rextract=\mathbf{0}$ (center of unit cell), given by $\abs{\langle\hat{\Rv}\rangle - \rextract}$, for states extracted from a Spatial Localizer for the  QWZ model. (a) [(b)] corresponds to $\ket{\psch_1(\rv)}$ [the WF produced by $\ket{\psch_1(\rv)}$] for the trivial phase. (c) corresponds to $\ket{\psch_1(\rv)}$ for the topological phase. All heatmaps are in units of lattice constant and were produced with $N=24$ as the system size. The grey boxes denote a unit cell.}
    \label{fig:qwz_deviations_rextract}
\end{figure*}

\begin{figure*}
    \centering
    \includegraphics[width=1\linewidth]{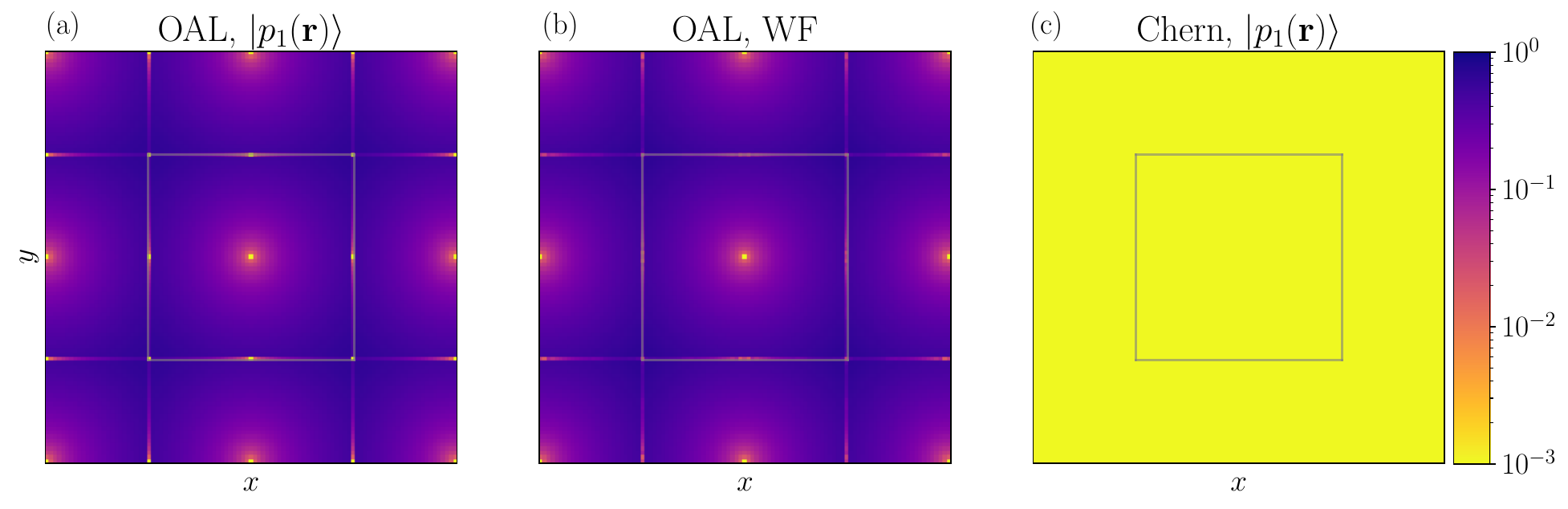}
    \caption{The deviation of mean position from extraction point $\rv$, given by $\abs{\langle\hat{\Rv}\rangle - \rv}$, for states extracted from a Spatial Localizer for the QWZ model. (a) [(b)] corresponds to $\ket{\psch_1(\rv)}$ [the WF produced by $\ket{\psch_1(\rv)}$] for the trivial phase. (c) corresponds to $\ket{\psch_1(\rv)}$ for the topological phase. All heatmaps are in units of lattice constant and were produced with $N=24$ as the system size. The grey boxes denote a unit cell. 
    }
    \label{fig:qwz_deviations_rv}
\end{figure*}

\begin{figure*}
    \centering
    \includegraphics[width=1\linewidth]{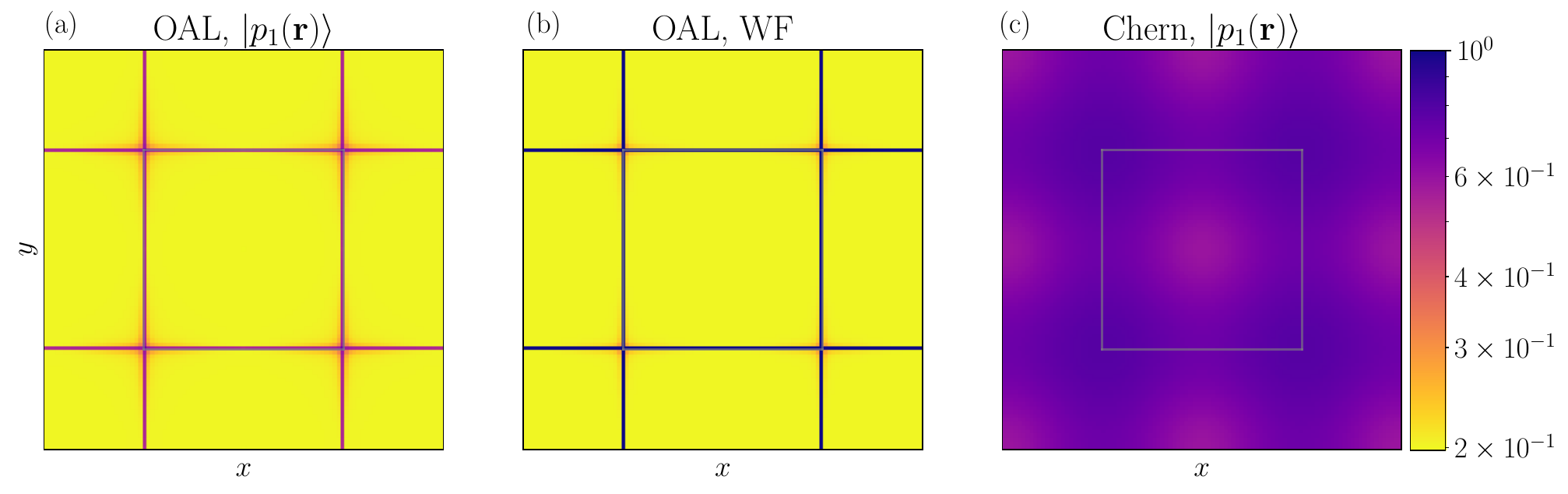}
    \caption{The variance, $\Delta R=\sqrt{\Delta R^2}$, of states extracted from a Spatial Localizer for the QWZ model. (a) [(b)] corresponds to $\ket{\psch_1(\rv)}$ [the WF produced by $\ket{\psch_1(\rv)}$] for the trivial phase. (c) corresponds to $\ket{\psch_1(\rv)}$ for the topological phase. All heatmaps are in units of lattice constant and were produced with $N=24$ as the system size. The grey boxes denote a unit cell. }
    \label{fig:qwz_var_both}
\end{figure*}

\begin{figure*}
    \centering
    \includegraphics[width=1\linewidth]{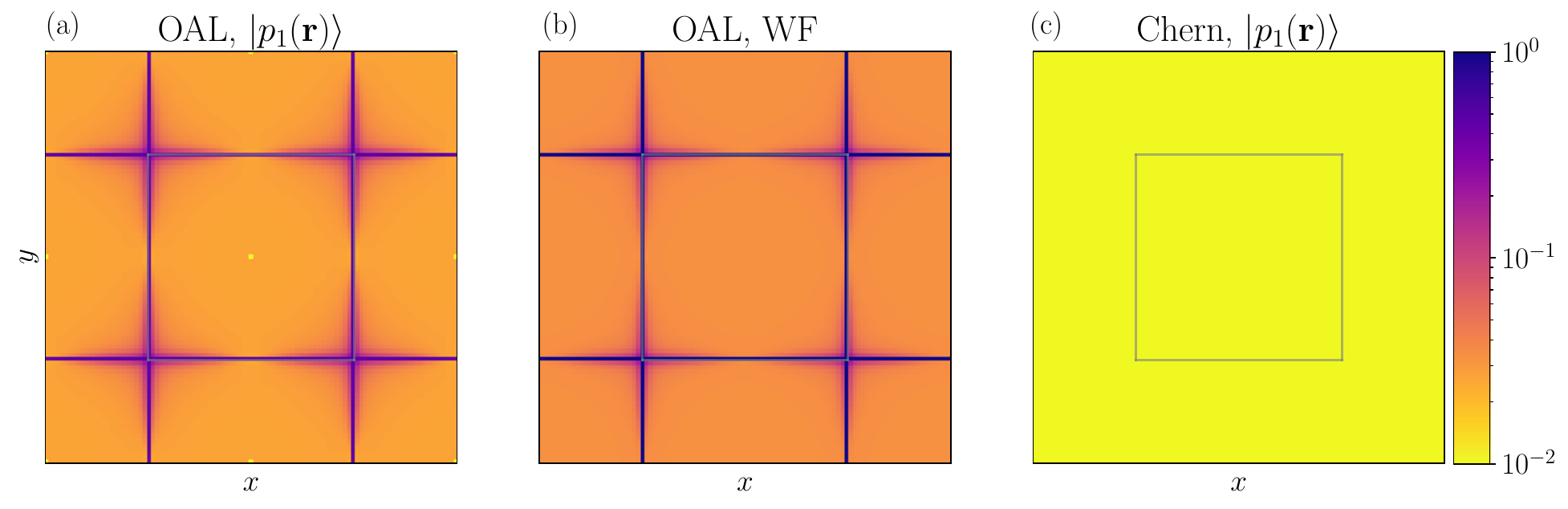}
    \caption{Coherence, measured by $\sqrt{\Delta R^2} - \sqrt{\abs{\comm{\tX}{\tY}}}$, of states extracted from a Spatial Localizer for the QWZ model. (a) [(b)] corresponds to $\ket{\psch_1(\rv)}$ [the WF produced by $\ket{\psch_1(\rv)}$] for the trivial phase. (c) corresponds to $\ket{\psch_1(\rv)}$ for the topological phase. All heatmaps are in units of lattice constant and were produced with $N=24$ as the system size. The grey boxes denote a unit cell. 
    }
    \label{fig:qwz_coherence}
\end{figure*}

As discussed in Section~\ref{app:sec:models_wse2}, the choice of extraction point $\rv$ for producing localized WFs in an OAL phase is robust with respect to perturbations  $\rv=\rextract$. I.e., the mean position and variance of the resulting WF are stable and minimally changed upon such perturbations. This same behavior is present in the trivial/OAL ($C=0$) phase of the QWZ model. Furthermore, other properties, such as the coherence of the WFs, are robust against perturbations of the extraction point.

However, in the Chern ($C=1$) phase, we observe a completely different behavior as $\ket{\psch_1(\rv)}$ now corresponds to the coherent and overcomplete states. Specifically, we find that the mean position of $\ket{\psch_1}$ tracks the extraction point $\rv$, rather than remaining near $\rextract$ as in the OAL phase. This behavior strongly mirrors the QHE, where one has a continuum of coherent states centered anywhere in the system, and they are quasi-zero modes of the annihilation operator $\tZ^\dagger(\rv)=\tX(x) - \ii \tY(y)$. We note that the corresponding vortex in the Berry connection moves in a fashion that preserves the relationship between position and vortex centers regarding the calculation of polarization (see below). Furthermore, the coherence of $\ket{\psch_1(\rv)}$ is $\order{10^{-3}}$ across a unit cell (Figure~\ref{fig:qwz_coherence}(c)), indicating a saturation of the HRUR.

For states from both the trivial/OAL and Chern phases, we present (i) deviation of mean position from the LIF minimum, $\rextract=\mathbf{0}$, in Figure~\ref{fig:qwz_deviations_rextract}, (ii) the deviation of mean position from extraction point in Figure~\ref{fig:qwz_deviations_rv}, (iii) the variance in units of lattice constant in Figure~\ref{fig:qwz_var_both}, and the coherence in Figure~\ref{fig:qwz_coherence}. 
The exact expressions used to calculate these quantities are provided in the figure captions.
We note that we do not present data regarding the WFs in the Chern phase as a general method for constructing WFs from the coherent states is beyond the scope of this work and we do not generally expect that $\ket{\psch_2(\rv)}$ will be optimal for constructing a WF in superposition with $\ket{\psch_1(\rextract)}$, as was done in the main text. We do note the use of ``bump'' functions proposed in~\cite{li2024} as a possibly general method for constructing optimal WFs from coherent states.

\subsubsection{Vortices and Polarization}

\begin{figure}
    \centering
    \includegraphics[width=1\linewidth]{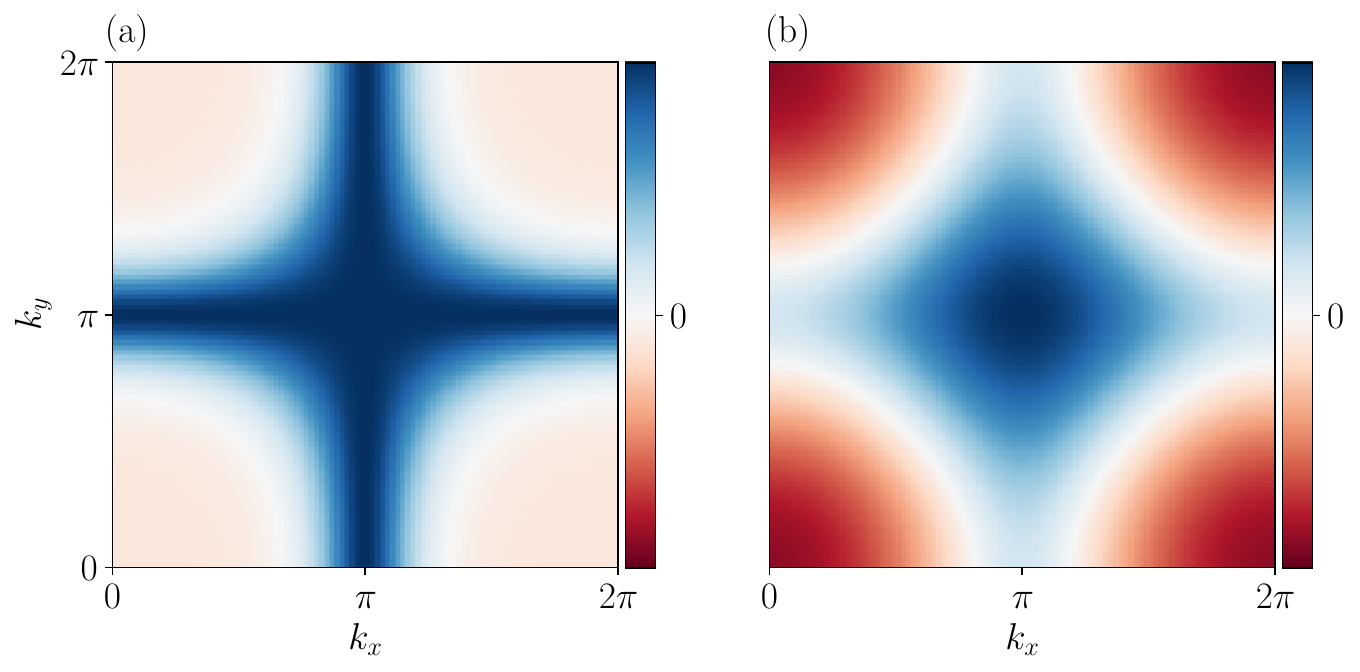}

    \caption{Berry curvature $\mathcal{F}({\mathbf{k}})$ (a) and corresponding smooth potential $\phi_s(\kv)$ (b) of the QWZ model in the Chern phase ($C=1$, $m=1$).}
    \label{fig:bcurv_and_potential}
\end{figure}

Recently, Gunawardana et al.~\cite{gunawardana2025} proposed a gauge-invariant method for calculating the polarization of a Chern insulator, provided one is in the vortex gauge~\cite{gunawardana2024}, corresponding to a vortex (or vortices in the case of $\abs{C}>1$) in a divergence-free Berry connection. The polarization for a single band is calculated as

\begin{equation}\label{app:eq:gunawardana_polarization}
    \frac{\mathbf{P}}{e} = -\frac{\texpval{\hat{\Rv}}}{V_c} - \frac{C}{2\pi} \hat{z} \times \mathbf{k_v},
\end{equation}
where $\texpval{\hat{\Rv}}$ is the WC, $V_c$ is the volume (area in 2D) of the unit cell, $C$ is the Chern number, and $\mathbf{k_v}=k_{vx}\hat{x} + k_{vy}\hat{y}$ is the location of the vortex in the Brillouin zone. We note that the corresponding expression in~\cite{gunawardana2025} differs by a minus sign on the second term. This is due to a difference of sign in the definition of the Berry connection, and thus a difference of sign in the calculation of the Chern number.

For the WF of the $C=1$ QWZ model presented in the main text, we approximate the vortex location as the average momentum of the magnitude of the Berry connection in the Brillouin zone, e.g., 

\begin{equation}
    \begin{split}
        k_{vx} &= \frac{\int_\mathrm{BZ} dk_x dk_y \ \ k_x |\mathcal{A}(k_x,k_y)|}{\int_\mathrm{BZ} dk_x dk_y \ \ |\mathcal{A}(k_x,k_y)|},  \\
        k_{vy} &= \frac{\int_\mathrm{BZ} dk_x dk_y \ \ k_y |\mathcal{A}(k_x,k_y)|}{\int_\mathrm{BZ} dk_x dk_y \ \  |\mathcal{A}(k_x,k_y)|}.
    \end{split}
\end{equation}
With these (approximate) vortex locations, we calculate a polarization equal to the value predicted in~\cite{gunawardana2025} up to $\order{10^{-2}}$. We note that the work in~\cite{gunawardana2025} uses a spin-rotated QWZ model. However, such a spin rotation will have no effect on polarization as all intra-unit-cell degrees of freedom lie at the same point in position space (i.e., the spin rotation operator commutes with the position operator and has no effect on the Berry connection). 

Furthermore, ~\cite{gunawardana2024} predicts the optimal vortex location, i.e., the $\kv_\mathbf{v}$ that minimizes the variance of the corresponding WF, to be located at the maximum of the potential, $\phi_s(\kv)$, which can be calculated as 

\begin{equation}
    \phi_s(\kv) = \sum_{\transvec \neq \mathbf{0}}\frac{\mathcal{F}({\transvec})}{\norm{\transvec}^2} e^{-i\mathbf{k}\cdot\transvec},
\end{equation}
where $\mathcal{F}({\transvec})$ are the Fourier components of the Berry curvature. We observe that our vortex center is located at the predicted optimal location, namely the $M$ point for the Chern insulator in this work, with an error of $\order{10^{-2}}$. We present both the Berry curvature, $\mathcal{F}({\mathbf{k}})$ and the smooth potential $\phi_s(\kv)$ in Figure~\ref{fig:bcurv_and_potential}.

\end{document}